\documentclass[11pt,a4paper]{article}

\usepackage{amsmath}
\usepackage{amssymb}
\usepackage[usenames]{color} 
\usepackage{multirow}
\usepackage{graphicx}
\usepackage{epstopdf}
\usepackage{array}
\usepackage{hhline}
\usepackage{cite}
\usepackage{microtype,colortbl}
\usepackage[bf]{caption}
\usepackage{color}
\usepackage[usenames,dvipsnames]{xcolor}
\usepackage{subcaption}
\usepackage{pstool}
\usepackage{bbold}
\usepackage{tikz}
\usepackage{ytableau}
\usepackage{latexsym,stmaryrd}
\usepackage[bookmarks=false]{hyperref}
 
\usepackage{ifpdf}
\def\hybrid{\topmargin -20pt    \oddsidemargin 0pt
        \headheight 0pt \headsep 0pt
        \textwidth 6.25in       
        \textheight 9 in       
        \marginparwidth .875in
        \parskip 5pt plus 1pt 
          \jot = 1.5ex
   }
\hybrid
\numberwithin{equation}{section}
\numberwithin{table}{section}

\newcommand{\beq}{\begin{equation}}  \newcommand{\eeq}{\end{equation}}
\newcommand{\bal}{\begin{aligned}}   \newcommand{\eal}{\end{aligned}}
\newcommand{\bea}{\begin{eqnarray}}  \newcommand{\eea}{\end{eqnarray}}

\newcommand{\bmat}{\left(\begin{array}}
\newcommand{\emat}{\end{array}\right)}

\newcommand{\bbC}{\mathbb{C}}
\newcommand{\bbR}{\mathbb{R}}
\newcommand{\bbP}{\mathbb{P}}

\newcommand{\nn}{\nonumber}

\newcommand{\cO}{\mathcal{O}}

\newcommand{\cE}{\mathcal{E}}

\newcommand{\cN}{\mathcal{N}}

\newcommand{\cH}{\mathcal{H}}

\newcommand{\cI}{\mathcal{I}}
\newcommand{\cJ}{\mathcal{J}}
\newcommand{\cR}{\mathcal{R}}

\newcommand{\cM}{\mathcal M}

\renewcommand{\Im}{\mathrm{Im}\,}
\renewcommand{\Re}{\mathrm{Re}\,}
\newcommand{\bqz}{\mathbf{q}_0}
\newcommand{\bQ}{\mathbf{Q}}

\newcommand{\I}{\text{Im}}
\newcommand{\R}{\text{Re}}

\usepackage{cancel}

\newcommand{\be}{\begin{equation}}
\newcommand{\ee}{\end{equation}}

\newcommand{\im}{\mathbf{i}}
\newcommand{\slt}{\mathfrak{sl}(2)}
\newcommand{\SLt}{\mathrm{Sl}(2)}
\newcommand{\rI}{\mathrm{I}}
\newcommand{\rII}{\mathrm{II}}
\newcommand{\rIII}{\mathrm{III}}
\newcommand{\rIV}{\mathrm{IV}}

\newcommand{\bbZ}{\mathbb{Z}}
\newcommand{\subs}{\subset}
\newcommand{\spanR}[1]{\mathrm{span}_\bbR\{#1\}}
\newcommand{\spanC}[1]{\mathrm{span}_\bbC\{#1\}}
\newcommand{\Gr}[2]{\mathrm{Gr}^{#1}_{#2}}
\newcommand{\conj}[1]{\overline{#1}}
\newcommand{\HD}{\Diamond}
\newcommand{\taE}{\ensuremath{\mathbf{\tilde{a}}_0^{(n_\mathcal{E})}}}
\newcommand{\sumN}[1]{\ensuremath{N^-_{(#1)}}}
\newcommand{\sumNsqr}[1]{\ensuremath{\big(N^-_{(#1)}\big)^2}}
\newcommand{\oneCh}[1]{\ensuremath{\boxed{#1} \to \cdots}}
\newcommand{\twoCh}[2]{\ensuremath{#1 \to \boxed{#2} \to \cdots}}
\newcommand{\threeCh}[3]{\ensuremath{#1 \to #2 \to \boxed{#3} \to \cdots}}
\newcommand{\fourCh}[4]{\ensuremath{#1 \to #2 \to #3 \to \boxed{#4} \to \cdots}}

\newcommand{\shdots}{\ensuremath{.\hspace{-.01cm} . \hspace{-.01cm}.}}

\definecolor{Gray}{gray}{0.95}

\begin{document}

\baselineskip=14pt
\parskip 5pt plus 1pt

\vspace*{-1.5cm}
\begin{flushright}    
  {\small 
  MPP-2018-260
  }
\end{flushright}

\vspace{2cm}
\begin{center}        
  {\LARGE Infinite Distance Networks in Field Space and Charge Orbits}
\end{center}

\vspace{0.5cm}
\begin{center}        
{\large  Thomas W.~Grimm$^{1}$, Chongchuo Li$^1$, Eran Palti$^2$}
\end{center}

\vspace{0.15cm}
\begin{center}        
\emph{$^1$ Institute for Theoretical Physics \\
Utrecht University, Princetonplein 5, 3584 CE Utrecht, The Netherlands}\\[.3cm]
  \emph{$^2$Max-Planck-Institut f\"ur Physik (Werner-Heisenberg-Institut), \\
Fohringer Ring 6, 80805 Munchen, Germany}
             \\[0.15cm]
 
\end{center}

\vspace{2cm}


\begin{abstract}
\noindent
The Swampland Distance Conjecture proposes that approaching infinite distances in field space an infinite tower of states becomes exponentially light. We study this conjecture for the complex structure moduli space of Calabi-Yau manifolds. In this context, we uncover significant structure within the proposal by showing that there is a rich spectrum of different infinite distance loci that can be classified by certain topological data derived from an associated discrete symmetry. We show how this data also determines the rules for how the different infinite distance loci can intersect and form an infinite distance network. We study the properties of the intersections in detail and, in particular, propose an identification of the infinite tower of states near such intersections in terms of what we term charge orbits. These orbits have the property that they are not completely local, but depend on data within a finite patch around the intersection, thereby forming an initial step towards understanding global aspects of the distance conjecture in field spaces. 
Our results follow from a deep mathematical structure captured by the so-called orbit theorems, which gives a handle on singularities in the moduli space through mixed Hodge structures, and is related to a local notion of mirror symmetry thereby allowing us to apply it also to the large volume setting. These theorems are general and apply far beyond Calabi-Yau moduli spaces, leading us to propose that similarly the infinite distance structures we uncover are also more general.
\end{abstract}

\thispagestyle{empty}
\clearpage

\setcounter{page}{1}


\newpage

\begingroup
  \flushbottom
  \setlength{\parskip}{0pt plus 1fil} 
  \tableofcontents
  \newpage
\endgroup

\section{Introduction}
\label{sec:intro}

There are a number of proposed consistency constraints on effective quantum field theories that could potentially arise from string theory. One of them, the Swampland Distance Conjecture (SDC), states that infinite distances in moduli space lead to an infinite tower of states becoming massless exponentially fast in the proper field distance \cite{Ooguri:2006in}. So if we consider two points in field space $P$ and $Q$, with a geodesic proper distance between them of $d\left(P,Q\right)$, then upon approaching the point $P$ there should exist an infinite tower of states with characteristic mass scale $m$ such that 
\be
\label{SC}
m\left(P\right) \sim m\left(Q\right)e^{-\gamma d\left(P,Q\right)} \mathrm{\;as\;} d\left(P,Q\right) \rightarrow \infty \;.
\ee
Here $\gamma$ is some positive constant which depends on the choice of $P$ and $Q$ but which is not specified in generality. The distance conjecture has been studied in a number of different settings and utilising different approaches \cite{Cecotti:2015wqa,Baume:2016psm,Klaewer:2016kiy,Palti:2015xra,Valenzuela:2016yny,Blumenhagen:2017cxt,Palti:2017elp,Lust:2017wrl,Hebecker:2017lxm,Cicoli:2018tcq,Grimm:2018ohb,Heidenreich:2018kpg,Blumenhagen:2018nts,Landete:2018kqf,Lee:2018urn,Lee:2018spm}. It has also been generalised and refined to a proposal which should hold for fields with a potential and for any super-Planckian variations in field space \cite{Baume:2016psm,Klaewer:2016kiy}. It therefore has potentially important phenomenological implications within cosmology, particularly in the context of large field inflation, see \cite{Blumenhagen:2018hsh} for a review. More recently, it has also been utilised in \cite{Ooguri:2018wrx} in relation to the Swampland de Sitter conjecture \cite{Obied:2018sgi,Ooguri:2018wrx,Garg:2018reu}. 

The distance conjecture, as stated in (\ref{SC}) is rather coarse. It does not say anything about properties of the tower of states beyond their mass, and in particular, about what is the overall structure of different infinite distances in the field space. In order to build up intuition about these questions, and evidence for the conjecture, it is useful to study large rich classes of field spaces in string theory. In \cite{Grimm:2018ohb} such a systematic study was initiated for the complex structure moduli space of Calabi-Yau manifolds in compactifications of type IIB string theory to four dimensions. We will retain this setting in this paper.\footnote{See \cite{Lee:2018urn,Lee:2018spm} for a general analysis of weak gauge coupling limits in compactifications of F-theory to six-dimensions.} The conjecture was shown to hold for a large class of infinite distances without referring to any specific example. The reason such a general approach is possible is because infinite distance loci in moduli space have some very general properties. In particular, they have a discrete set of data associated to monodromies when circling them, and this data determined the local form of the moduli space as well as the spectrum of charged states. In this paper we will build on these ideas and uncover more of the structure contained in this discrete data.
In terms of the distance conjecture, this structure will `resolve' the infinite distance divergence into a fine classification of different types of infinite distances, and begin to shed light on how such infinite distance types can intersect and form a complex network of infinite distance loci. It will also determine how the towers of states can arise and be inter-related within such a network.

\begin{figure}
\centering
 \includegraphics[width=0.8\textwidth]{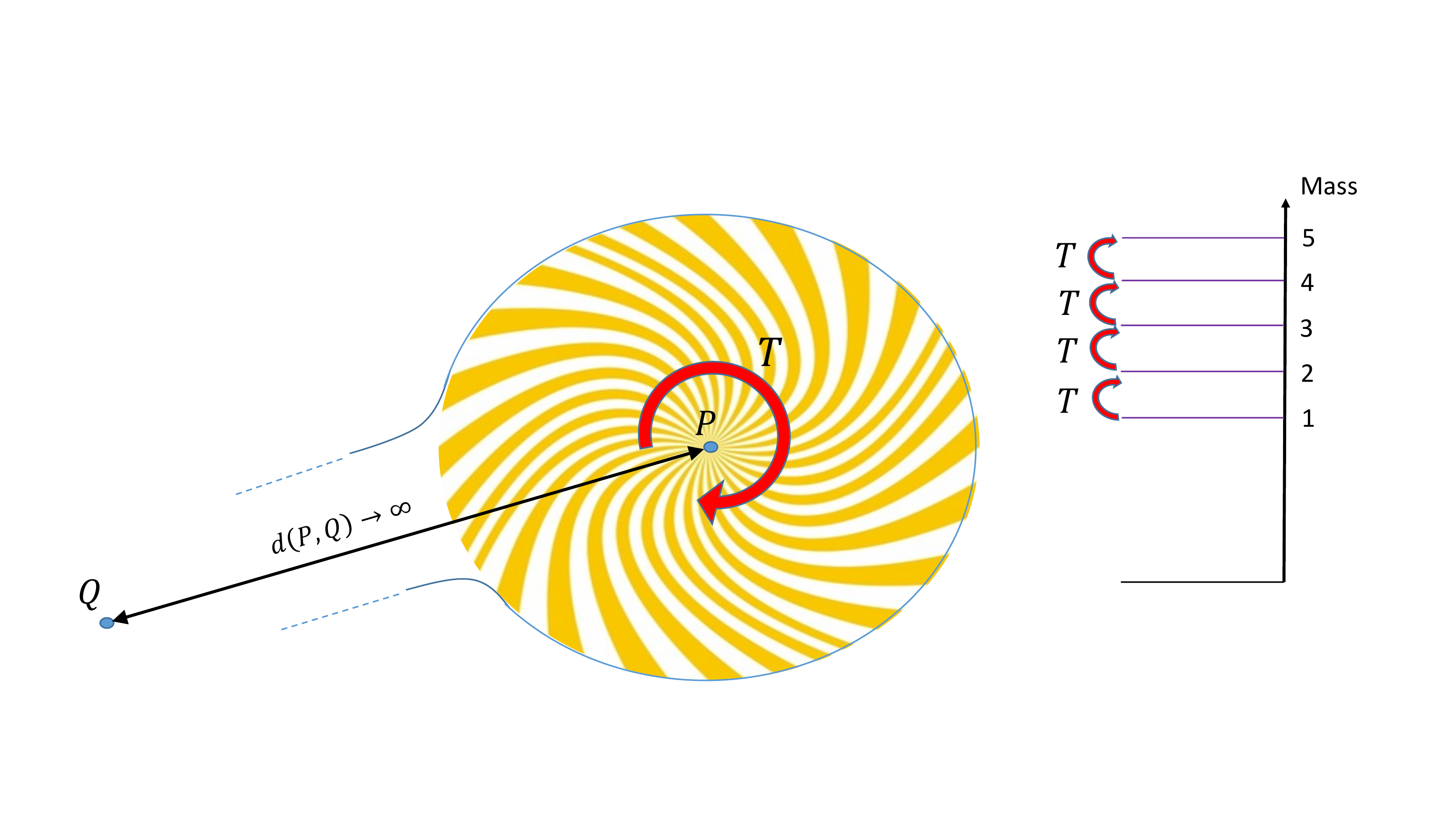}
\caption{Figure illustrating the relation between the distance conjecture and monodromy. The point $P$ is at infinite distance and the monodromy about it is denoted by T. The monodromy determines the local singular geometry of the moduli space, which leads to the exponential behaviour of the mass of the tower of states. The monodromy also acts on the spectrum of states picking out a specific infinite set of states.}
\label{intmon}
\end{figure}

First, we recall the local aspect of the data. The results of \cite{Grimm:2018ohb} showed that infinite distance loci are singular loci in the moduli space and have an associated discrete monodromy transformation, denoted by $T$. This transformation determines the local geometry of the moduli space. It also picks out an infinite tower of states where it acts as the transformation moving one step up the tower. This general picture is illustrated in figure \ref{intmon}.
The presence of such a universal structure allowed for a very general analysis and so to proofs of very general results. 
It was also proposed that the infinite distance is itself induced by integrating out the tower of states. In this sense, it is quite natural that the same object $T$ controls both the tower of states and the infinite distance behaviour.

\begin{figure}[h]
\centering
 \includegraphics[width=0.8\textwidth]{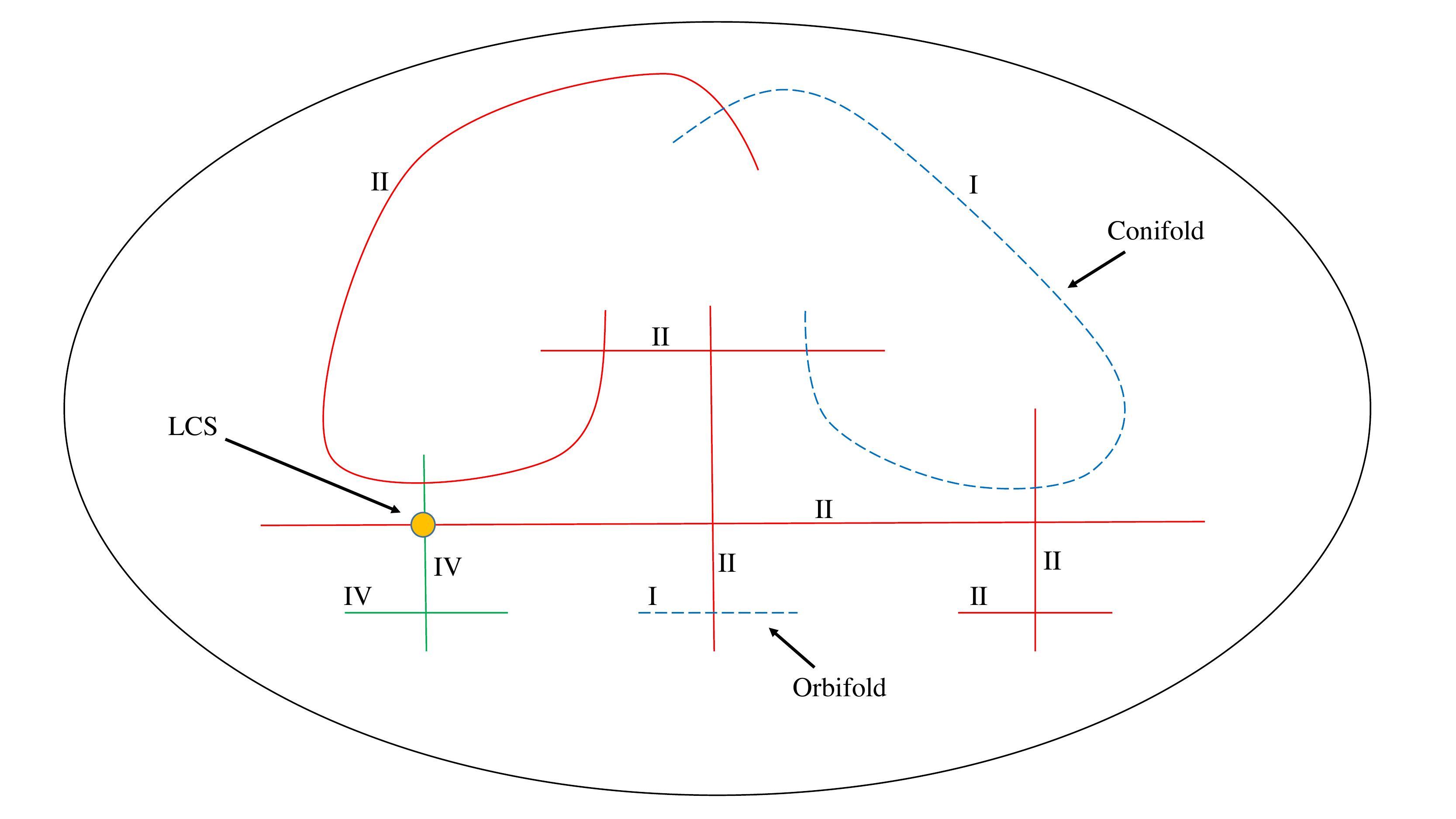}
\caption{Figure showing an example field space with multiple infinite distance loci. The example is the (resolved) complex structure moduli space of the (mirror of the) two parameter Calabi-Yau $\mathbb{P}^{1,1,2,2,2}[8]$ as studied in \cite{Candelas:1993dm}. Each infinite distance locus is denoted by a solid line and assigned a type labelled by $\mathrm{II}$, $\mathrm{III}$, or $\mathrm{IV}$. We also show special finite distance loci with dashed lines, and these are associated to type I. Some well-known loci are labelled explicitly, the finite distance conifold and orbifold loci, and the infinite distance large complex-structure point.}
\label{fig:modCY}
\end{figure}

So far we have only considered a single point $P$ at infinite distance. But the moduli space is a high-dimensional space, and $P$ actually belongs to a continuous set of points which together form an infinite distance locus. This full locus can be characterised by discrete data related to $T$. The locus can also intersect other similar infinite distance loci. Together, all these loci form a network of infinite distances. This structure is perhaps best illustrated with an example. In figure \ref{fig:modCY} we present an example field space, the complex structure moduli space of a particular Calabi-Yau manifold. Each locus of infinite distance in the moduli space is denoted by a solid line, and the full structure of the network is manifest.
The loci in figure \ref{fig:modCY} are labelled by a type, which (for Calabi-Yau threefolds) can be I, II, III, or IV. Type I loci are at finite distance in moduli space. Type II, III or IV loci are at infinite distance and the increasing type denotes a sense of increasingly strongly divergent distances. In \cite{Grimm:2018ohb} a generic point $P$ on one of the infinite distance loci was assigned a type inherited from the locus type.\footnote{The notation in \cite{Grimm:2018ohb} is that types I, II, III, IV are labelled by $d=0,1,2,3$ respectively.} This was done away from the intersection points and is in this sense a purely local analysis.

\begin{figure}
\centering
 \includegraphics[width=0.8\textwidth]{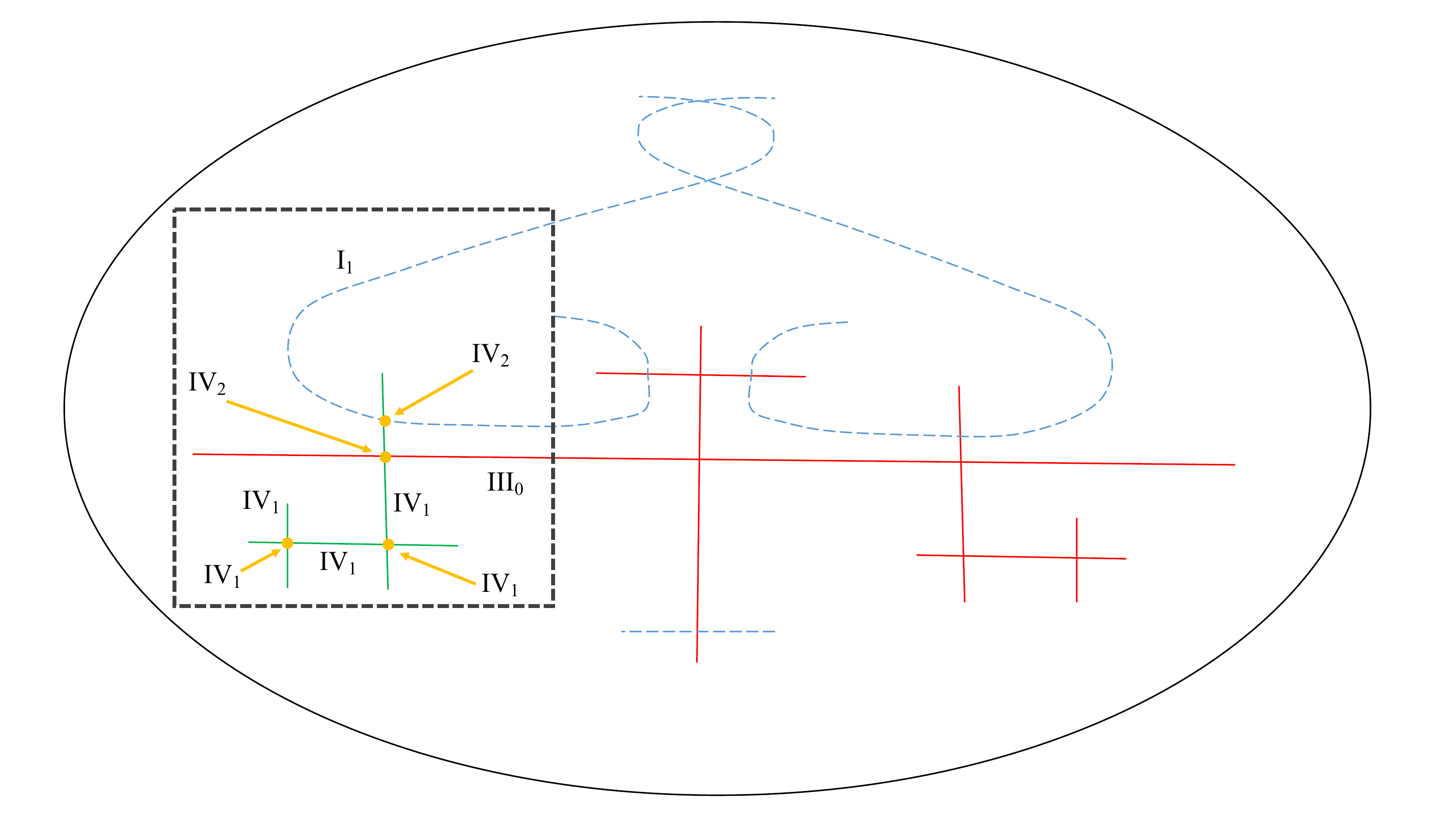}
\caption{Figure showing an example intersecting network for the (mirror of the) Calabi-Yau $\mathbb{P}^{1,1,1,6,9}[18]$ as studied in \cite{Candelas:1994hw}. In this case we focus in one a particular region of the network, within the box, and show the more refined data for each locus including the sub-index. At the points of intersections the type of a locus can be modified. We show the types associated to each intersection point in the focused region.}
\label{fig:modCY2}
\end{figure}

In this paper we will begin to explore the global structure of the infinite distance network. The first thing we will introduce is a more refined classification of the infinite distance loci which takes into account important additional data. The type will now be supplements by a numerical sub-index, so for example, will take the form II$_2$. This more refined type can then change, or enhance, at points where the loci intersect. In figure \ref{fig:modCY2} we give a different example of an infinite distance network where we now focus in on the intersection structure in a particular region. We see that the loci are assigned a more refined data type and also each intersection locus has an associated type which may differ from the generic point on the locus. 
We will explain what the more refined data captures, and how it can be calculated from the monodromy $T$. 

The next step will be to understand the distance conjecture when approaching the intersection points themselves. The whole notion of the nature of the infinite distance is vastly more complicated at the intersection points. In particular, the finiteness of the distance itself, as well as the masses of states, become path dependent questions. So whether a state becomes massless or not at the intersection loci depends on how one approaches them. We will show how to incorporate this path dependence into the formalism. 
 
\begin{figure}[h]
    \begin{minipage}{6cm}
    \includegraphics[width=5.5cm]{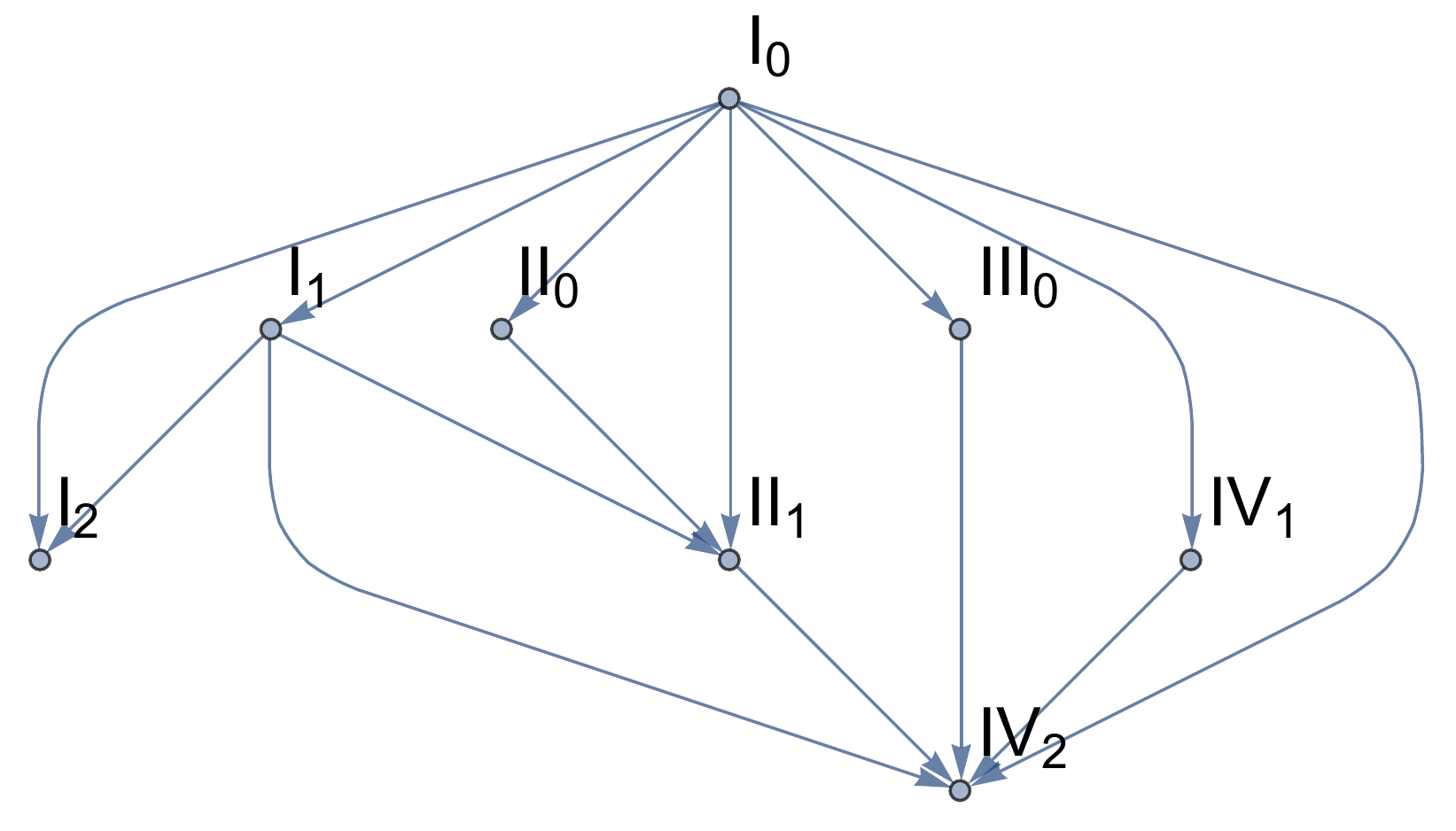}\end{minipage} 
     \begin{minipage}{9.0cm} \includegraphics[width=10cm]{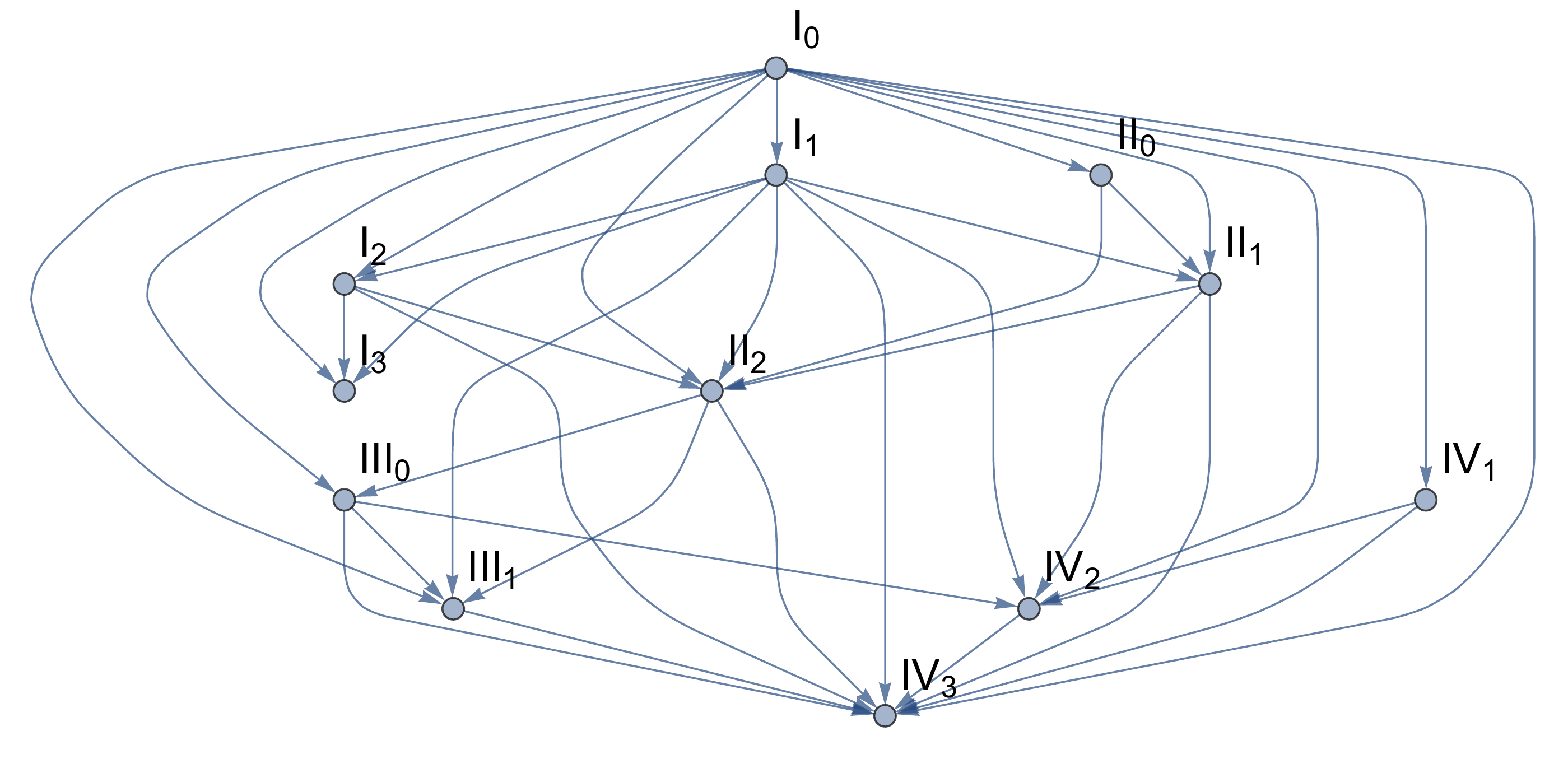} \end{minipage}

\begin{picture}(0,0)
\put(10,30){$h^{2,1}=2$}
\put(160,30){$h^{2,1}=3$}
\end{picture}
\caption{Graphs of allowed type enhancements for field spaces with $h^{2,1}$ complex fields. In terms of Calabi-Yau geometries, $h^{2,1}$ is the associated Hodge number. An arrow denotes that a starting type of locus may enhance over a sub-locus, corresponding to an intersection, to the end type. Note that the enhancement relations are not transitive. For example,  in the $h^{2, 1} = 2$ case, there is a chain of $\rII_0 \to \rII_1 \to \rIV_2$ enhancements, but there is no direct enhancement from $\rII_0$ to $\rIV_2$.} 
\label{Graphs23in}
\end{figure}

The refined discrete data not only gives the properties of the infinite distance loci but also the rules for which types of infinite distance loci can intersect each other and what are the possible types to which they could enhance on the intersection points. We therefore find rules for what type of infinite distance networks could be built. These intersection rules have deep mathematics behind them, as initially developed in \cite{CKS} and studied recently in \cite{Kerr2017}.  The rules can be expressed in terms of which types of infinite distance loci can enhance to which types over certain sub-loci corresponding to intersections. Expressed this way the intersection, or enhancement, rules for two example classes of networks are shown in figure \ref{Graphs23in}. 
The example network in figure \ref{fig:modCY2} falls into the type $h^{2,1}=2$. One can then readily check that the enhancement of the locus types at the intersections indeed follows the general rules. 

In \cite{Grimm:2018ohb} the tower of states was identified as generated by an infinite action of the monodromy matrix $T$ on some BPS state charge. In this work we will introduce a more general notion of such a tower that is associated to the monodromy action, which we term a {\it charge orbit}. A crucial aspect of the charge orbit is that it will not be associated to a point on an infinite distance locus, but to a patch, which means that it can capture the structure of intersections. This will therefore form a first step towards connecting the towers of the different infinite distance loci into a network. A non-trivial result which we will be able to prove already is that if the type of the infinite distance increases at the intersection, then there is an infinite charge orbit of states which become massless approaching the locus even away from the intersection point itself. We call this an inheritance of a charge orbit by a locus from its intersection point. It is important to note, however, that in \cite{Grimm:2018ohb} the monodromy induced tower was shown to be populated by BPS states, while in this paper we will identify the charge orbit but will be unable to prove that it is populated by BPS states. Nonetheless, we propose that it indeed captures the tower of states of the distance conjecture, while leaving a proof in terms of BPS states for future work. 

The paper is structured as follows. In section \ref{math_background} we introduce the formalism and underlying theorems which we will use in the paper. In section \ref{sec:classification_results} we show how the data of the type of infinite distance locus can be used to form a complete classification of such loci, and how this type can be extracted from the discrete monodromy. In section \ref{sec:infinite_towers} we utilise these results to define the charge orbits at intersections of infinite distance loci. We summarise our results, and discuss extensions and interpretations of them in section \ref{sec:sum}. In the appendix we present a detailed analysis of some example intersection loci as well as collect some of the more technical formalism.

\section{Monodromy and Orbit Theorems in Calabi-Yau Moduli Spaces} \label{math_background}

In this section we introduce, and develop in a way adapted to our needs, the crucial mathematical theorems and structures associated to so-called orbits. The central elements are the nilpotent orbit theorem, the Sl(2)-Orbit theorem and the growth theorems. The theorems lead to a detailed and powerful description of the moduli space locally around any singular loci. In particular, we will utilise their multi-variable versions which will allow for a description of a patch of moduli space that can include intersections of infinite distance loci. 

\subsection{Complex Structure Moduli Spaces and Monodromy} \label{sec:CS_moduli_space}

The focus of this paper lies on a particular sector of Type II string compactifications 
on Calabi-Yau threefolds. More precisely, we will investigate the geometry 
of the field space spanned by the scalars in the $\cN=2$ vector multiplets
arising in these compactifications. These scalars correspond to complex structure 
deformations of the Calabi-Yau threefold in Type IIB string theory and complexified 
K\"ahler structure deformations in Type IIA. Since these two compactifications are 
deeply linked via mirror symmetry, it will often suffice to address only one of the 
two sides. In particular, it is important to recall that the complex structure side 
captures the more general perspective and hence will be the focus for the 
first part of our exposition. Later on, we will address aspects of the K\"ahler structure 
side by discussing large volume compactifications.  

To begin with, let us denote the complex structure moduli space by $\cM_{\rm cs}$ 
and introduce the Weil-Petersson metric $g_{\rm WP}$ that arises in the Type IIB
string theory comactification. The space $\cM_{\rm cs}$ has complex dimension $h^{2,1}$, 
where $h^{p, q} = \text{dim}_\bbC(H^{p,q}(Y_3))$ are the Hodge numbers of the Calabi-Yau 
threefold $Y_3$. In a local patch we can thus introduce  complex coordinates 
$z^I$, $I=1,\ldots,h^{2,1} $, which are called the complex structure moduli. 
The metric  $g_{\rm WP}$ on $\cM_{\rm cs}$ is special K\"ahler 
and determined by the complex structure variations of the holomorphic 
$(3,0)$-form $\Omega$ on $Y_3$ \cite{Andrianopoli:1996cm,Craps:1997gp,Alim:2012gq}.
Its components $g_{I \bar J} = \partial_{z^I} \partial_{\bar z^J} K$  can locally be obtained from the K\"ahler potential 
\beq \label{Kpot_cs}
    K(z,\bar z) = - \log \Big[\im \int_{Y_3} \Omega \wedge \bar \Omega \Big] \equiv  - \log \Big[ \im \, \bar  \Pi^\cI \eta_{\cI\cJ}  \Pi^\cJ \Big]  \ .
\eeq
In the second equality we have expanded $\Omega$ into a real integral basis $\gamma_\cI$, $\cI = 1, \ldots ,2h^{2,1}+2$ spanning 
$H^{3}(Y_3,\mathbb{Z})$. More precisely, we introduced 
\beq   \label{Omega-exp}
    \Omega = \Pi^\cI \, \gamma_\cI   \ ,  \qquad \eta_{\cI\cJ} =- \int_{Y_3} \gamma_\cI \wedge \gamma_\cJ \ .
\eeq
In order to simplify notation we will introduce bold-faced letters to denote coefficient vectors
in the three-form basis $\gamma_\cI$, i.e.
\beq
  \mathbf{\Pi} \equiv \big(\Pi^1, \ldots ,\Pi^{2h^{2,1}+2} \big)^\mathrm{T}\ .
\eeq 
The complex coefficients $\Pi^\cI$ can be shown to be holomorphic function and are called 
the \textit{periods} of $\Omega$. 
Let us stress that $z^I$, $\Pi^\cI$, 
and $\gamma_\cI$ are adapted to the considered patch in 
$\cM_{\rm cs}$ and can very non-trivially 
change when moving to different patches in $\cM_{\rm cs}$.\footnote{Furthermore, there is 
the freedom to rescale the whole vector $\mathbf{\Pi}$ with a holomorphic function $f(z)$, which corresponds 
to a K\"ahler transformation of \eqref{Kpot_cs}. While one should keep this freedom in mind, we will not mention it again.}

It is important to discuss the possible transformations preserving the above structure. 
To begin with, we note that $\eta = (\eta_{\cI \cJ})$ is an anti-symmetric matrix. It defines 
an anti-symmetric bilinear form 
\beq \label{def-S}
   S( v,w ) \equiv S(\mathbf{v},\mathbf{w})= \mathbf{v}^\mathrm{T} \, \eta \, \mathbf{w} \equiv - \int_{Y_3} v \wedge w\ , 
\eeq
where $v,w$ are three-forms in $H^{3}(Y_3,\bbC)$ and $\mathbf{v},\ \mathbf{w}$ are their 
coefficient vectors in the integral basis $\gamma_\cI$. We will use the notations $S(v, w)$ and $S(\mathbf{v}, \mathbf{w})$ interchangeably.
One shows that the group preserving 
$\eta$  is the real symplectic group $\mathrm{Sp}(2h^{2,1}+2,\bbR)$ acting as 
\beq \label{preserve_eta}
  M^\mathrm{T} \eta M= \eta \ ,\quad M \in \mathrm{Sp}(2h^{2,1}+2, \bbR)\ .
\eeq 
The action of this group thus corresponds to actions on the basis that preserve $S(\mathbf{v},\mathbf{w})= S(M \mathbf{v}, M \mathbf{w})$. Crucially, we stress 
that they do not correspond to a symmetry of the effective theory, but rather to a choice of frame in which 
to consider the fields. The true symmetry of the effective theory is encoded by the so-called \textit{monodromy 
group} $\Gamma \subset \mathrm{Sp}(2h^{2,1}+2)$, which we will discuss next.  

\begin{figure}[h!]
\vspace*{.4cm}
\begin{center} 
\includegraphics[width=8cm]{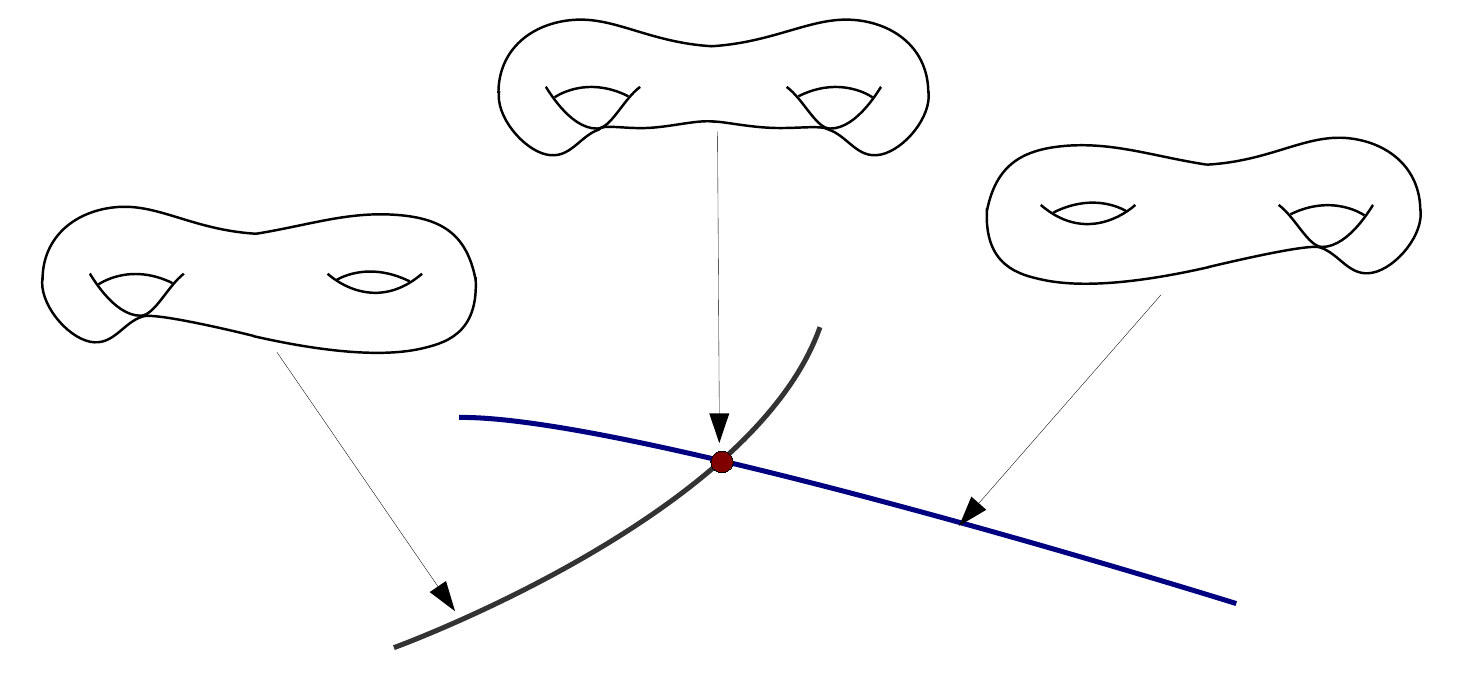} 
\vspace*{-.4cm}
\end{center}
\caption{Two normally intersecting divisors of the discriminant locus $\Delta$. The singularity of the Calabi-Yau threefold, here depicted as 
genus-two Riemann surface, worsens at the intersection.} \label{enhanced_sing}
\end{figure}

A crucial fact about the complex structure moduli space $\cM_{\rm cs}$ is that 
it is neither smooth nor compact. It generally admits points at which the Calabi-Yau manifold 
becomes singular. These form the so-called 
discriminant locus. Clearly, it is non-trivial to show general results about these discriminant loci 
and we first summarize some of the main abstract results. Later on we will give a more detailed 
classification of what actually can happen at this locus. Firstly, we note that the moduli space 
of smooth Calabi-Yau threefolds is quasi-projective \cite{Viehweg}, which roughly implies that 
as long as one removes a divisor $\Delta_s$ corresponding to singular Calabi-Yau manifolds it can be embedded 
into a projective space. The discriminant locus $\Delta_s$ can have a very non-trivial structure, since it will 
generically consist of many intersecting components. Crucially the singularities of the Calabi-Yau manifolds 
can get worse when moving along $\Delta_s$. A cartoon picture of this is shown in figure \ref{enhanced_sing} and 
we already gave a more realistic description of an actually occurring moduli space in the introduction, see figures \ref{fig:modCY}
and \ref{fig:modCY2}.  
It was also shown \cite{Hironaka,Viehweg}
that one can resolve $\Delta_s$ to $\Delta = \cup_k \Delta_k$ such that it consists of divisors $\Delta_k$ that intersect normally. 
This result is crucial to justify the local model that we employ to describe the individual patches of the moduli 
space. Hence, in the following we will always work with the desingularized discriminant locus $\Delta$.
It will also be convenient to introduce a shorthand notation for the intersection of $l$ divisors we define
\beq \label{intersection_locus}
    \Delta_{k_1 \ldots k_l} = \Delta_{k_1} \cap \ldots \cap \Delta_{k_l}\ .
\eeq

Another important aspect of the above description of $\cM_{\rm cs}$ is the fact that 
$\mathbf{\Pi}$ can be understood as being multi-valued and experience 
monodromies along paths encircling the divisor components $\Delta_k$ of $\Delta$. 
To make this more precise, let us introduce local coordinates $z^I$, such that the divisor $\Delta_k$ is given 
by $z^k = 0$ for some $k \in \{1,\ldots , h^{2,1}\}$. 
The intersection of divisors $\Delta_k$ and $\Delta_{l}$ can be parametrized if one introduces several vanishing 
local coordinates $z^k = z^{l}=0$.
We encircle $\Delta_k$ by sending $z^k \rightarrow e^{2\pi \im} z^k$. In general the 
periods will non-trivially transform with a matrix $T_k$.
When defining the monodromy, especially when writing $T_k$ as a matrix, there is a choice between whether the $T_k$ is defined to act on the homology $3$-cycles or the cohomology $3$-forms. Our convention in this paper is to let the monodromy act on the integral basis of the $3$-forms. Explicitly, with a multi-valued integral basis of $3$-forms chosen to be $\{\gamma_\cI\}$, the monodromy operator $T_k$ induced by the loop $z^k \rightarrow e^{2\pi \im} z^k$ is defined by
\begin{equation}
  \gamma_\cI(\ldots, e^{2\pi\im} z^k, \ldots) = \gamma_\cJ(\ldots, z^k, \ldots) \left(T_k\right)^{\cJ}_{\cI}, \textrm{ for all } \cI.
\end{equation}
In terms of the period vector $\mathbf{\Pi}$, under our convention, we have
\beq \label{monodromy-trafo}
   \mathbf{\Pi}(  \ldots  ,e^{2\pi \im} z^k, \ldots  ) = T_{k}^{-1} \ \mathbf{\Pi}( \ldots ,z^k, \ldots )\ .   
\eeq
The monodromy matrices are shown to be quasi-unipotent \cite{landman,Schmid},
i.e.~they satisfy an equation of the form $ (T^{m} - \text{Id} )^{n+1} = 0$ for some positive integers $m,n$. 
Furthermore, the monodromies arising from intersecting divisors $\Delta_k, \Delta_{l}$ commute
$ \left[T_{k} , T_{l}\right] = 0$. This fact remains true for each pair of monodromy matrices if one considers higher intersections.
Collecting all $T_k$ from all components of $\Delta$ one obtains a group $\Gamma$ known
as the monodromy group. It preserves the pairing $\eta$, such that by \eqref{preserve_eta} we 
have 
\beq
   \Gamma \subset \mathrm{Sp}(2h^{2,1} + 2,\bbR)\ . 
\eeq
More abstractly, the monodromy group can be defined by considering representations 
of the fundamental group $\pi_1(\cM_{\rm cs}) $ acting on the period vectors. In general, the elements of $\Gamma$ will not commute. However, in this work we will restrict ourselves to the commuting monodromies arising at intersections of divisors $\Delta_k$. 

In the next section we will have a closer look at the singularities occurring along the $\Delta_k$ and their intersections.
In order to do that it will be important to extract the unipotent part $T_k^{(u)}$ of each $T_k$. We define 
\beq \label{def-Nk}
   N_k = \frac{1}{m_k} \log(T_k^{m_k}) \equiv \log (T_{k}^{(u)})\ ,
\eeq 
where $m_k$ is the smallest integer that satisfies $ (T^{m_k}_k - \text{Id} )^{n_k+1} = 0$. 
This implies that the $N_k$ are nilpotent, i.e.~that there exist integers $n_k$ such that 
\beq
     N_k^{n_k + 1} = 0 \ . 
\eeq
Since each $T_k$ preserves the bilinear form $S$ introduced in \eqref{def-S}, i.e.~$S(T_k\cdot,T_k\cdot)=S(\cdot,\cdot)$
one finds 
\beq \label{Npres_S}
    S(N_k \mathbf{v}, \mathbf{w}) = - S(\mathbf{v},N_k \mathbf{w})\ , 
\eeq
and since $T_k^{(u)} \in \mathrm{Sp}(2h^{2,1} + 2,\bbR)$ we have $N_k \in \frak{sp}(2h^{2,1}+2,\bbR)$, where $\frak{sp}(n,\bbR)$ is the Lie algebra of $\mathrm{Sp}(n,\bbR)$.
The nilpotent elements $N_k$ will be the key players in much of 
the following discussion. Therefore, it is convenient to make a base transformation 
and pick right away coordinates for which the monodromies are unipotent. This can be achieved 
by sending $z^k \rightarrow (z^k)^{m_k}$. We should stress that this implies that we lose information 
about certain types of singularities, such as orbifold singularities. We will see below 
that it is the unipotent part of $T_k$ that encodes whether or not a point on $\Delta$ is at 
finite or infinite distance. In fact, one checks that the above coordinate change does not alter 
the discussion relevant to this work.

\subsection{Approximating the periods: Nilpotent orbits} \label{sec:period_approx_nil}

In this section we discuss the first important tool which is used in establishing the mathematical 
structure that we will explore throughout this work. The general 
important question one wants to address is: Are there simpler functions that approximate the 
periods $\mathbf{\Pi}$ introduced in \eqref{Omega-exp} and capture some of their key features? 
In the following we will introduce a set of such functions known as \textit{nilpotent orbits} following \cite{Schmid}. 
These not only approximate the periods, but also share their transformation behaviour \eqref{monodromy-trafo}
under local monodromy transformations. We will also comment on the importance of nilpotent orbits 
in the context of variations of Hodge structures.

To begin with, let us note that the periods $\mathbf{\Pi}$ of $\Omega$ are in general 
very complicated functions on the moduli space $\cM_{\rm cs}$. This can be already expected 
from the figure \ref{fig:modCY}. Hence, at best one can hope to approximate the $\mathbf{\Pi}$ locally. 
The nilpotent orbits approximate $\mathbf{\Pi}$ in a local patch denoted by $\cE$ 
containing points of the discriminant locus $\Delta$.  The local patch is chosen to be of the form 
\beq 
   \cE = (\mathbb{D}^*)^{n_\cE} \times \mathbb{D}^{h^{2, 1} - n_\cE}\ ,
\eeq 
i.e.~a product of punctured disks $\mathbb{D}^* = \left\{z\in\bbC \mid 0 < |z| < 1\right\}$ and unit disks $\mathbb{D} = \left\{\zeta\in\bbC \mid |\zeta| < 1\right\}$ so that the singular point ``lies in the puncture''.
In other words, we approximate the periods near points at the intersection of $n_\cE$ discriminant divisors 
$\Delta_i$, $i=1,\ldots,n_{\cE}$, but away from any further intersection. 
The introduced local coordinates $z^I = (z^i,\zeta^\kappa)$ parametrize the 
 $n_\cE$ intersecting discriminant 
divisors $\Delta_i$ given by $z^i=0$.
The coordinates $\zeta^\kappa$ parametrize additional complex directions and 
do not play an important role in the following discussion.
We have introduced the nilpotent matrices $N_i$ 
in \eqref{def-Nk}. It was then shown 
by Schmid \cite{Schmid} that locally around the point $P$ with $z^i=0$ the periods take the form
\bea \label{period_orbit}
\quad \mathbf{\Pi}(z,\zeta) &=&  \text{exp} \Big[ \sum_{j=1}^{n_\cE} -\frac{1}{2\pi \im} (\log z^j ) N_j  \Big] \mathbf{A}(z,\zeta) \;, \\
  &\equiv& \text{exp} \Big[ \sum_{j=1}^{n_\cE} -t^j N_j  \Big] \mathbf{A}\big(e^{2\pi \im t},\zeta \big) \;, \nn 
\eea
with $\mathbf{A}$ being holomorphic in $z^i,\zeta^\kappa$ near $P$. Here we have also expressed the result in the coordinates
\beq \label{t-def}
   t^j \equiv x^j + \im\, y^j= \frac{1}{2\pi \im} \log z^j\ .
\eeq
This implies that crucial information about the \textit{singular behaviour} of the periods $\mathbf{\Pi}$ near 
the point $P$ is in the matrices $N_j$. Furthermore, the second essential information is the leading term in the 
 vector $\mathbf{A}(z,\zeta)$. Since it is holomorphic it admits an expansion
\beq
    \mathbf{A}(z,\zeta) = \mathbf{a}_0(\zeta) +  \mathbf{a}_{j} (\zeta)  z^j +  \mathbf{a}_{jl} (\zeta)  z^j z^l+  \mathbf{a}_{jlm} (\zeta)  z^j z^l z^m + \ldots \ , 
\eeq
with the $ \mathbf{a}_0(\zeta),\ \mathbf{a}_{j} (\zeta) \ , \ldots$ being holomorphic functions of $\zeta^\kappa$. 
The nilpotent orbit theorem underlies the statement \eqref{period_orbit}. Namely, it establishes the 
fact that the periods $\mathbf{\Pi}$ are well-approximated by the nilpotent orbit 
\beq \label{nilp-orbit}
  \mathbf{\Pi}_{\text{nil}} =  \text{exp} \Big[ \sum_{j=1}^{n_\cE} -\frac{1}{2\pi \im} (\log z^j ) N_j  \Big] \mathbf{a}_0(\zeta) \equiv 
    \text{exp} \Big[ \sum_{j=1}^{n_\cE} -t^j N_j  \Big] \mathbf{a}_0(\zeta) \;,
\eeq
where an estimate how well the orbit \eqref{nilp-orbit} approximates the actual period $\mathbf{\Pi}$ was given in \cite{Schmid} and \cite{CKS}. 
We stress that the nilpotent orbit drops the exponential corrections in the coordinates $t$, i.e.
\beq
   \mathbf{\Pi}(t,\zeta) =   \underbrace{ \text{exp} \Big[{\mathrm \sum_{j=1}^{n_\cE}} \ -t^j N_j  \Big] \Big( \mathbf{a}_0(\zeta)}_{\text{nilpotent orbit}\ \mathbf{\Pi}_{\rm nil}}  + \cO(e^{2\pi \im t}) \Big)\ .
\eeq
This result is crucial, for example, in evaluating the leading form of the K\"ahler potential \eqref{Kpot_cs}. 

Having defined the nilpotent orbit, one immediately sees that it shares the transformation behaviour of the periods under the shifts $t^i \rightarrow t^i - \delta_k^i$, i.e.
\beq
    \mathbf{\Pi}_{\text{nil}}( \ldots , t^k - 1, \ldots ) = e^{N_k}\,  \mathbf{\Pi}_{\text{nil}}( \ldots , t^k, \ldots )  = T^{\rm (u)}_{k} \ \mathbf{\Pi}_{\text{nil}}( \ldots , t^k, \ldots )\ .
\eeq
Here we stress again that $N_k$ defined via \eqref{def-Nk} only captures the unipotent part of the monodromy transformation,
which is the only relevant part since we assume a coordinate transformation $t^k \rightarrow m_kt^k$ as 
at the end of subsection \ref{sec:CS_moduli_space} have been performed. 

Let us close this section by recalling some basic facts about Hodge structures and Hodge filtrations and their 
relation to nilpotent orbit. Recall, that the third cohomology group splits for a given complex structure as 
\beq \label{H3-cohom-split}
     H^{3}(Y_3,\mathbb{C}) = H^{3,0} \oplus H^{2,1} \oplus H^{1,2} \oplus H^{0,3}\ .
\eeq
This $(p,q)$-split for a smooth geometry $Y_3$ defines a so-called \textit{pure Hodge structure} of weight $3$ (see appendix \ref{app:MWF_MHS}, for some additional details). The changes of this split as one moves in complex structure moduli space 
are captured by the study of variations of Hodge structures. In order to make this more explicit, we first combine the $H^{p,q}$
as 
\bea \label{Hodge-filtration}
    F^3 &=& H^{3,0}\, ,\qquad F^2 = H^{3,0} \oplus H^{2,1} \, , \qquad F^1 = H^{3,0} \oplus H^{2,1} \oplus H^{1,2}\, ,  
    \\ F^0 &=& H^{3,0} \oplus H^{2,1} \oplus H^{1,2} \oplus H^{0,3}\, . \nn 
\eea
These complex spaces vary holomorphically with the complex structure moduli $z^I$. 
Introducing a flat connection $\nabla_I \equiv \nabla_{\partial/\partial z^I}$, known as the Gauss-Manin connection, one has $\nabla_I F^p \subset F^{p-1} $.
For Calabi-Yau threefolds one furthermore finds that all elements of the lower $F^p$, $p<3$ are obtained as derivatives 
of $F^3$ spanned by the holomorphic $(3,0)$-form. Roughly speaking this implies that all information about the 
filtration $F \equiv (F^3,F^2,F^1,F^0)$ is encoded by $\Omega$. 

Since the periods of $\Omega$ are approximated by the nilpotent orbit given in \eqref{nilp-orbit}, 
we can also obtain a filtration by taking derivatives of $\mathbf{\Pi}_{\rm nil}$ when $\mathbf{\Pi}_{\rm nil}$ is represented in a flat frame. Concretely, we 
evaluate 
\beq \label{Pi-der}
   \mathbf{\Pi}_{\rm nil} \xrightarrow{\quad \partial_{t^i}\quad }  N_i  \mathbf{\Pi}_{\rm nil} \xrightarrow{\quad \partial_{t^j}\quad }      N_i  N_j \mathbf{\Pi}_{\rm nil} \ \rightarrow \ \ldots\ ,
\eeq
and note that the derivatives with respect to $\zeta^\kappa$ are encoded by $\nabla_\kappa \mathbf{a}_{0}, \nabla_\kappa  \nabla_\lambda \mathbf{a}_{0},$ etc. 
Due to the nilpotent orbit theorem the derivatives of $\mathbf{\Pi}_{\rm nil}$ approximate 
the elements in spaces $F^2,F^1,F^0$ up to corrections proportional to $z^j = e^{2\pi \im t^j}$. 
Clearly, when moving to the points on $\Delta$ by sending $t^i \rightarrow \im \infty$ 
the elements \eqref{Pi-der} are singular. However, this singularity arises in $\mathbf{\Pi}_{\rm nil}$
and all its derivatives only via the exponential prefactor exp$(\sum_i t^i N_i)$. As we discuss in 
the next subsection, we can characterize singularities after dropping the singular prefactor, e.g.~by
replacing \eqref{Pi-der} with 
\beq \label{Pi-der_2}
   \mathbf{a}_{0} \xrightarrow{\quad \quad}  N_i  \mathbf{a}_{0} \xrightarrow{\quad \quad }      N_i  N_j \mathbf{a}_{0} \ \rightarrow \ \ldots\ ,
\eeq
and considering in the $\zeta^\kappa$-directions the derivatives  $\nabla_\kappa \mathbf{a}_{0},\nabla_\lambda \mathbf{a}_{0},$ etc. 
The limiting Hodge filtrations $F^p_{\Delta}$ spanned by these vectors will be discussed in more detail in the next subsection.

\subsection{Characterizing Singularities in Calabi-Yau Threefolds} \label{sec:charact_sing}

We now 
have a closer look at the arising singularities at the divisors $\Delta_i$ and their intersections. 
In subsections~\ref{sec:sing_class} and \ref{sec:class_enh} we summarize a recent classification of singularities 
and allowed enhancements carried out in \cite{Kerr2017}. This work builds on many important 
and deep mathematical results about so-called limiting mixed Hodge structures. This 
subsection aims to give the reader a somewhat condensed summary of the underlying mathematical 
tools with some additional details deferred to appendix \ref{app:MWF_MHS}. 

The basic object that one associates to the points on $\Delta$ is a limiting mixed Hodge structure. 
For our purposes, rather then introducing in detail the concept of a mixed Hodge structure, it  
turns out to be useful to directly work with the so-called \textit{Deligne splitting}. We will introduce this 
splitting in the following. Roughly speaking it captures a finer split $I^{p,q}$, $p,q=0, \ldots ,3$ 
of the third cohomology group $H^{3}(Y_3,\bbC)$
as one moves to a singularity of $Y_3$.
In other words the $(p,q)$-split  \eqref{H3-cohom-split}  for a smooth 
geometry $Y_3$ splits into this finer Deligne splitting schematically depicted as
\beq \label{movetoDelta}
(  H^{3,0},\  H^{2,1},\ H^{1,2},\ H^{0,3})\quad \xrightarrow{\text{\ \ move to $\Delta$\ \ }} \quad \begin{array}{ccccccccc} &&& I^{3,3}&&\\ 
&& I^{3,2} \hspace*{-.2cm}& & \hspace*{-.2cm} I^{2,3} \\ 
& I^{3,1}\hspace*{-.2cm} & & \hspace*{-.2cm} I^{2,2} \hspace*{-.2cm} & & \hspace*{-.2cm} I^{1,3}\\
 I^{3,0} \hspace*{-.2cm} & & I^{2,1} \hspace*{-.2cm} & & \hspace*{-.2cm} I^{1,2} &&\hspace*{-.2cm} I^{0,3}\\
&I^{2,0} \hspace*{-.2cm} & & \hspace*{-.2cm} I^{1,1} \hspace*{-.2cm} & & \hspace*{-.2cm} I^{0,2}\\
&&I^{1,0} \hspace*{-.2cm} & &  \hspace*{-.2cm} I^{0,1} \\
&&&I^{0,0}&& \\  \end{array}\ .
\eeq 
To introduce this splitting we follow the 
filtration $F \equiv (F^3,F^2,F^1,F^0)$ given in \eqref{Hodge-filtration} to a point in $\Delta$. As pointed 
out already in the previous subsection the form will become singular in this limit. However, 
we can remove these singularities as we discuss in the following. 

We begin our consideration with the simplest situation, namely consider points on a divisor $\Delta_1$ 
that are not elements of any other $\Delta_{l}$, i.e.~we are away from any intersection locus $\Delta_{1 l}=\Delta_{1} \cap \Delta_{l}$. 
We denote this set of points by $ \Delta_1^\circ$, generally setting
\beq \label{def-Deltacirc1}
    \Delta_k^\circ = \Delta_k - \bigcup_{l\neq k} \Delta_{kl}\ .
\eeq
To reach the locus $\Delta_1$ we have to send $z^{1} \rightarrow 0$, which by \eqref{t-def} is equivalent to $t^1\rightarrow \im \infty$. For points on $\Delta_1^\circ$
one shows that 
\beq \label{Finfdef_onediv}
F^p(\Delta_1^\circ) =\lim_{t^{1}\rightarrow \im \infty}\ \text{exp} \left[-t^1 N_{1}  \right] F^p\ ,
\eeq
is well-behaved. In this expression we let $N_k$ act on the basis $\gamma_\cI$ in which all elements of $F^p$ can be expanded. 
Clearly, $F^p(\Delta_1^\circ)$ is defined on $\Delta_1^\circ$ and still depends homomorphically
on the other $h^{2,1} - 1$ complex structure moduli. 

\begin{figure}[h!]
\vspace*{.5cm}
\begin{center} 
\includegraphics[width=7cm]{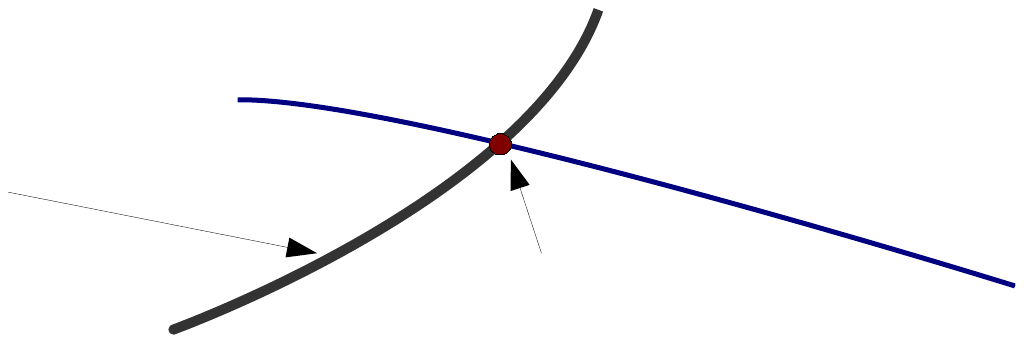} 
\vspace*{-1cm}
\end{center}
\begin{picture}(0,0)
\put(230,10){$F^p(\Delta^\circ_{12})$}
\put(105,30){$F^p(\Delta^\circ_{1})$}
\put(245,65){$ \Delta_{1}$}
\put(312,10){$ \Delta_{2}$}
\end{picture}
\caption{Association of a limiting $F^p$ to the points on the discriminant locus.} \label{limitingF_pic}
\end{figure}

Let us next move towards the intersection of $\Delta_1$ with another divisor, say $\Delta_{2}$ in $\Delta$, i.e.~let us consider the surface
$\Delta_{1 2} = \Delta_{1} \cap \Delta_{2}$. This requires to send both $z^1,z^{2} \rightarrow 0$ or $t^1,t^2\rightarrow \im \infty$ and one shows that the 
spaces $F^p(\Delta^\circ_1)$ are also not generally well-behaved in this limit. To remedy this problem, we consider the locus $\Delta_{12}^\circ$, generally defining
defined as
\beq
    \Delta_{kl}^\circ = \Delta_{kl} - \bigcup_{m\neq k, l} \Delta_{klm}\ .
\eeq
Hence, $\Delta_{12}^\circ$ consists of points on $\Delta_{12}$ away 
from any further intersection. On this locus one considers 
 \beq \label{Finfdef_twodiv}
F^p(\Delta_{12}^\circ) =\lim_{t^{1},t^{2}\rightarrow \im \infty}\ \text{exp} \left[-t^{1} N_{1} - t^{2} N_{2}  \right] F^p\ .
\eeq
The $F^p(\Delta_{12}^\circ) $ now depend on the remaining $h^{2,1}-2$ coordinates and are non-singular. We have depicted the assignment of the $F^p(\Delta_1^\circ)$ and $F^p(\Delta_{12}^\circ)$ to the points in $\Delta$ 
in figure \ref{limitingF_pic}. From this discussion it should be clear that one can proceed in a similar fashion for higher intersections. 

Let us now turn to the finer split arising at the points of the discriminate locus $\Delta$. 
This is known as the \textit{Deligne splitting} and encoded by complex vector 
spaces $I^{r,s}$ with $r+s \in \{0, \ldots ,6\}$. 
The data defining the splitting at each point of $\Delta$ are a limiting $F^p$, such as $F^p(\Delta^\circ_{1})$ and $F^p(\Delta^\circ_{12})$
introduced in \eqref{Finfdef_onediv} and \eqref{Finfdef_twodiv}, and an associated nilpotent matrix. The simplest case 
are again points that are on $\Delta^\circ_1$ defined in \eqref{def-Deltacirc1}. The associated nilpotent 
element is simply $N_1$. In other words, one associates 
\beq \label{associateDegline_1}
   (F(\Delta^\circ_1), N_1) \quad \mapsto \quad  \{ I^{p,q}(\Delta^\circ_1) \}_{p,q=0, \ldots ,3}\ ,
\eeq  
where we denote $F(\Delta^\circ_1) \equiv (F^3(\Delta^\circ_1), \ldots ,F^0(\Delta^\circ_{1}))$
More involved are points that lie on the intersection locus $\Delta^\circ_{12}$ of two divisors, since here the immediate 
question for the associated nilpotent matrix arises. It turns out \cite{Schmid} that one is actually free to choose 
any $N_{12}$ in the cone  
\beq \label{postive-sum}
    \sigma(N_1,N_{2}) =\left\{ a_1 N_1 + a_2 N_{2}  \mid a_i>0 \right\}\ . 
\eeq

It is crucial to note that each choice of $a_1, a_2$ in \eqref{postive-sum} yields the same $I^{p,q}(\Delta^\circ_{12})$ and 
we can pick the most convenient combination, such as $N_1 + N_2$. In summary, 
at the intersection $\Delta^\circ_{12}$ and away from any further intersection, one associates 
\beq \label{associateDegline_2}
  (F(\Delta^\circ_{12}), N_1+N_{2}) \quad \mapsto \quad  \{ I^{p,q}(\Delta^\circ_{12}) \}_{p,q=0, \ldots ,3}\ .
\eeq 
It should be clear how to generalize this discussion to even higher intersection loci $\Delta_{k_1 \ldots k_l}$ introduced 
in \eqref{intersection_locus}. The associated nilpotent element are now elements 
of the cone 
\beq \label{sigma_gen}
    \sigma(N_{k_1}, \ldots ,N_{k_l}) = \left\{ a_{k_1} N_{k_1} +  \ldots + a_{k_l} N_{k_l} \mid a_i>0 \right\}\ .
\eeq 
For example, let us consider the intersection of $\Delta_i,\, i=1, \ldots ,l$, away from any further intersection and 
denote this space by $\Delta^\circ_{1 \ldots l}$.
By an appropriate relabelling this is the general situation. The limiting Hodge filtration for points on 
such intersections are given by 
\beq \label{Fp-gen}
   F^p(\Delta_{1 \ldots l}^\circ) =\lim_{t^{1}, \ldots ,t^{l}\rightarrow \im \infty}\ \text{exp} \left[-\begin{array}{c}\sum_{i=1}^l t^{i} N_{i} \end{array} \right] F^p\ .
\eeq
Then 
the map to the Deligne splitting is 
\beq
   (F(\Delta^\circ_{1 \ldots l}), N_{(l)}) \quad \mapsto \quad  \{ I^{p,q}(\Delta^\circ_{1 \ldots l}) \}_{p,q=0, \ldots ,3}\ .
   \label{dsda}
\eeq
Here $N_{(l)}$ is an element of \eqref{sigma_gen} and we have chosen a simple representative by picking  
\beq
    N_{(l)} = \sum_{i=1}^l N_i \ .
    \label{lbr} 
\eeq
This also allows us to introduce a notation which will be used throughout the paper, namely that an index $(l)$ in brackets 
on a matrix indicates that we add the first $l$ elements of an ordered set, i.e.~$(N_1, \ldots ,N_l, \ldots )$. Indeed, we will often denote (\ref{dsda}) this way
\be
 I^{p,q}_{(l)} \equiv I^{p,q}(\Delta^\circ_{1 \ldots l})  \;.
\label{dsl}
\ee

With this information at hand we are now in the position to introduce the Deligne splitting $I^{p,q}$
and discuss its properties. To keep the notation simple we will study the map 
\beq \label{simpleFI}
(F_\Delta, N) \quad \mapsto \quad  \{ I^{p,q} \}_{p,q=0, \ldots ,3}\ ,
\eeq
with $F_\Delta = (F^3_\Delta, \ldots ,F^0_\Delta)$.
The $F^p_\Delta$ is the limiting $F^p$ and 
the $N$ stands for the nilpotent element associated to the considered point on $\Delta$. In other words 
\eqref{simpleFI} can correspond to the cases \eqref{associateDegline_1} and \eqref{associateDegline_2} or any higher intersection. 
In order to define the $I^{p,q}$ we first note that there is a natural set of vector spaces associated 
to a nilpotent $N$ known as the \textit{monodromy filtration}
$W_i,\ i=0, \ldots ,6$ . The most natural spaces associated to a given $N$ acting on $H^3(Y_3,\bbC)$ 
are constructed from the images $\mathrm{Im}\,N^p$ and kernels $\mathrm{Ker}\,N^q$. 
These allow us to define 
\begin{equation} \label{def-Wi}
\begin{array}{ccl}
  W_6  & = & V,\\
  \cup &   &\\
  W_5  & = & \mathrm{Ker}\,N^3,\\
  \cup &   &\\
  W_4  & = & \mathrm{Ker}\,N^2  + \mathrm{Im}\,N,\\
  \cup &   &\\
  W_3  & = & \mathrm{Ker}\,N  + \mathrm{Im}\,N \cap\mathrm{Ker}\,N^2,\\
  \cup &   &\\
  W_2  & = & \mathrm{Im}\,N \cap\mathrm{Ker}\,N  + \mathrm{Im}\,N^2,\\
  \cup &   &\\
  W_1  & = & \mathrm{Im}\,N^2 \cap\mathrm{Ker}\,N,\\
  \cup &   &\\
  W_0  & = & \mathrm{Im}\,N^3.
\end{array}
\end{equation}
The properties of the so-defined $W_i$ will be discussed in more detail in appendix \ref{app:MWF_MHS}. 
It is a crucial fact that this filtration $W_i$ associated to higher intersections does not depend 
on the precise element chosen among the $\sum_i a_i N_i $ with $a_i>0$ \cite{CattaniKaplan}. For example, 
on $\Delta_{12}^\circ$, the associated spaces $W$ do not depend on which element $N_{12}$ in \eqref{postive-sum} one picks.

We now have all the required information to define the 
Deligne splitting 
\beq \label{def-Ipq}
   I^{p,q} = F^p_\Delta \cap W_{p+q} \cap \Big( \bar F^{q}_\Delta \cap W_{p+q} + \sum_{j \geq 1} \bar F^{q-j}_\Delta \cap W_{p+q-j-1}  \Big)\ .
\eeq
At first, this definition looks rather involved and somewhat arbitrary. However, it has many remarkable features, such as being the unique 
definition satisfying 
\beq  \label{Fp-Wi_split} 
  F^{p}_\Delta  = \bigoplus_{r\geq p} \bigoplus_s I^{r,s} \ ,\qquad W_{l} = \bigoplus_{p+q \leq l} I^{p,q} \ ,  \qquad
     \overline{I^{p,q}} = I^{q,p} \ \text{mod}\ \bigoplus_{r < q,s<p} I^{r,s}\ .
\eeq
While the details of this definition are important in our explicit constructions, within 
this section it often suffices to view $I^{p,q}$ as spaces obtained from $F^{p}_\Delta,\ N$ and use the features that we will discuss next. 
Let us note that it is often convenient to use the shorthand notation $(F,W)$ to summarize the relevant data for 
the map \eqref{def-Ipq}. Here $F$ is a vector containing the spaces $F^3,\ldots,F^0$ relevant at the point in $\Delta$, and $W$ 
is the weight filtration relevant at this point. This data $(F,W)$ also determines a limiting mixed Hodge structure as
described in appendix \ref{app:MWF_MHS}. However, it will be more convenient in the following to 
work with the Deligne splitting.

As a first important property of \eqref{def-Ipq} one checks that the spaces indeed define 
a splitting of the total vector space. In fact, at any point of $\Delta$ one needs to replace the split \eqref{H3-cohom-split} by 
\beq \label{H3split}
      H^3(Y_3,\bbC) = \bigoplus^3_{p, q=0} I^{p, q} \ ,
\eeq
where we remind the reader that the $I^{p,q}$ crucially depend on the location of the point, as indicated in \eqref{associateDegline_1} and \eqref{associateDegline_2}. 
One of the most important features of the Deligne splitting 
arises from the fact that $N$ acts as $N F^p_\Delta \subset F^{p-1}_\Delta$ and $N W_i \subset W_{i-2}$. 
Applied to \eqref{def-Ipq} we find 
\beq \label{NI=I}
    N I^{p,q} \subset I^{p-1,q-1}\ . 
\eeq 
We note that this does not mean that the whole lower $(p,q)$-spaces are obtained by 
acting with $N$. In fact, there is a natural way to split each $I^{p,q}$ into 
a primitive part $P^{p,q}$ that is not obtained by acting with $N$ on a $(p+i,q+i)$-element
and a non-primitive part. Explicitly one defines the \textit{primitive} parts to be 
\beq \label{def-Ppq}
    {P}^{p,q}  =  I^{p,q}  \cap \mathrm{Ker}\,N^{p+q-2}\ .
\eeq 
Clearly, the primitive part $P^{p,q}$ of $I^{p,q}$ contains the core information in the Deligne splitting, 
since all other elements are obtained by the action of $N_k$. One shows that 
\beq \label{I=NP}
I^{p,q} = \bigoplus_{i \geq 0} \, N^i (P^{p+i,q+i})\ .
\eeq 
The primitive elements satisfy another remarkable feature, namely, their norm is positive and non-vanishing
for non-vanishing elements. 
More explicitly, one has 
\bea
  S(P^{p,q}, N^l P^{r,s}) &=& 0  \qquad \text{for}\ p + q = r + s = l + 3\ \text{and}\  (p,q) \neq (s,r)\ , \label{Sj-orthogonality1-I}\\
  \im^{p-q} S (v ,N^{p+q-3} \bar v) &>& 0 \qquad \text{for} \ v \in P^{p,q}\ , \ v\neq 0\ ,  \label{polarizingSj}
\eea
where we use the bilinear form introduced in \eqref{def-S}. These properties give us a powerful tool 
to analyse positivity and vanishing properties of forms at $\Delta$. As we will discuss in the next subsection 
they are actually key in systematically classifying the allowed singularities and enhancement patterns. 

In summary, we have now explained the following picture. As we change the complex structure 
moduli from a smooth Calabi-Yau threefold to a singular threefold on
the discriminant locus on $\Delta$, we need to replace the splitting of $H^{3}(Y_3,\bbC)$ as in 
\eqref{movetoDelta} with the $I^{p,q}$ defined via \eqref{def-Ipq} or \eqref{Fp-Wi_split}. The 
occurring splits characterize the singularity at $P \in \Delta$. 
In subsection \ref{sec:sing_class} we will focus in detail on such splits and explain how these 
can be classified systematically for Calabi-Yau threefolds. 

From the above construction it is clear that the precise form of $I^{p,q}$ depends on the 
considered point on $\Delta$ and will generally differ for 
points, for example, on $\Delta_1^\circ$ and points on the intersection $\Delta_{12}^\circ$. 
This implies that we could also move from a generic point on $\Delta_{1}^\circ$ to the intersection locus 
$\Delta_{12}^\circ$. In this case we expect that the 
$ I^{p,q}(\Delta_1^\circ)$ change to the $I^{p,q}(\Delta_{12}^\circ)$. We write this as 
\beq
    I^{p,q}(\Delta_1^\circ) \quad \longrightarrow \quad I^{p,q}(\Delta_{12}^\circ)\ ,  
\eeq
with an arrow indicating the enhancement direction. 
It is crucial in this step to ensure that the polarization conditions \eqref{Sj-orthogonality1-I}, \eqref{polarizingSj} are transmitted correctly, which 
in fact imposes severe constraints on the form of the enhancement. 
As stressed in the introduction it is crucial for us not only to classify all the allowed splittings 
$I^{p,q}$, but also all the allowed enhancement. This 
formidable task was carried out in \cite{Kerr2017} and will be the subject of the next two subsections.

\subsection{Commuting $\slt$s and the Sl(2)-orbit} \label{sec:sl2-orbit}

While the nilpotent orbits are useful, for example, in approximating the periods they are, in general,
not a simple representation encoding the information about the asymptotic limit when approaching $\Delta$.
However, there is 
a foundational result of Cattani, Kaplan, and Schmid \cite{CKS}, which shows that there is asymptotically 
 a special representation of the data contained in the nilpotent orbit \eqref{nilp-orbit}.
Roughly speaking, one is able to replace $(N_i,\mathbf{a}_0)$ with $(N^-_i,\mathbf{\tilde a}_0)$ such 
that the $N^-_i$ are part of commuting $\mathfrak{sl}(2,\bbC)$ algebras and $\mathbf{\tilde a}_0$ splits into 
subvectors affected by the action of these $\mathfrak{sl}(2,\bbC)$. In this representation many of the 
asymptotic properties of the setting are rather easy to show and can then be translated back into 
the representation $(N_i,\mathbf{a}_0)$. For example, the growth properties of the Hodge norm 
discussed in subsection \ref{sec:growth_theorems} are proved in this way. For us the existence of 
the commuting $\mathfrak{sl}(2,\bbC)$ algebras will be of crucial importance when 
constructing the infinite charge orbit relevant for the Swampland Distance Conjecture 
as we describe in detail in section \ref{sec:infinite_towers}. 

We begin our exposition by introducing the commuting $\mathfrak{sl}(2,\bbC)$ algebras in more detail and 
introduce the steps required to explicitly construct them. In order to do that we consider 
a local patch $\cE$ of the complex structure moduli space that intersects $n_\cE$ discriminant 
divisors $\Delta_i$, $i= 1, \ldots ,n_\cE$, which non-trivially intersect each other. In other words, 
we assume that the highest intersection in $\cE$ is $\Delta_{1 \ldots n_\cE}$ which is obtained by intersecting 
all $n_\cE$ divisors. Clearly, all other intersections of a smaller number of $\Delta_i$ can also be 
considered. As usual we denote the monodromy logarithms associated to $\Delta_i$ by $N_i$. 
Crucially, we will choose an ordering of the $N_i$: $(N_1, \ldots ,N_{n_\cE})$, and all the results 
presented below will depend on this ordering. Clearly, one still is free to pick any other ordering, 
but then has to adjust the statements below accordingly. Furthermore, we will assume that 
the patch $\cE$ is chosen such that the nilpotent orbit 
\beq \label{gen-nilp_start}
     \mathbf{\Pi}_{\rm nil} = \text{exp} \Big[ \sum_{j=1}^{n_\cE} -t^j N_j  \Big] \mathbf{a}_0^{(n_\cE)}\ ,
\eeq
approximates the periods in $\cE$. Starting from this data we want to construct associated 
commuting $\mathfrak{sl}(2,\bbC)$ algebras. Each of these algebras $\mathfrak{sl}(2,\bbC)_i$ is generated 
by three elements, and we denote these triples by 
\beq \label{Sl2-triples}
   \text{commuting $\mathfrak{sl}(2,\bbC)_i$ triple:}\quad (N_i^-,N_i^+,Y_i) \ .
\eeq
These elements satisfy the standard $\mathfrak{sl}$(2)-commutation relations $[Y_i,N_i^\pm]=\pm 2 N_i^\pm$ and $[N_i^+,N^-_i]=Y_i$.
Furthermore, the triples are pairwise commuting, i.e.~all elements 
in the $i$th triple commute with all elements of the $j$th triple for $i\neq j$. Together these triples define a 
Lie algebra homomorphism  
\beq
  \rho_* :\  \bigoplus_{i} \mathfrak{sl}(2,\bbC)_i \ \longrightarrow \ \mathfrak{sp}(2h^{2,1}+2,\bbC)\ ,
\eeq 
where $\rho$ gives the map of the corresponding Lie groups. 
The Sl(2)-orbit theorem of \cite{CKS} states the properties of the triples 
in relation to a given nilpotent orbit. 

Given a nilpotent orbit \eqref{gen-nilp_start} around the highest intersection $\Delta_{1 \ldots n_\cE}$ 
one can read off the filtrations $(F_\Delta,W)$ with $F^p_{\Delta}$ 
defined in \eqref{Fp-gen} and $W_i(N_{(n_\cE)})$ discussed in \eqref{def-Wi}. Here we recall that 
the $W_i$ are induced by $N_{(n_\cE)} = \sum_{i=1}^{n_\cE} N_i$ or any other positive linear 
combination of the $N_i$. The corresponding 
Deligne splitting $I^{p,q}(\Delta_{1 \ldots n_\cE})$ is determined via \eqref{def-Ipq} or \eqref{Fp-Wi_split}. 
A splitting $I^{p,q}$ is called $\bbR$-split, if it obeys 
\beq 
    \overline{I^{p,q}}=I^{q,p}, \textrm{ for all } p, q\ . 
 \eeq
 It is crucial
 that the limiting $F^p_{\Delta}$ do not generally induce an $\bbR$-split Deligne splitting.
The $\SLt$-orbit theorem of \cite{CKS}\footnote{
More precisely Proposition 2.20 and Theorem 3.25 of \cite{CKS}.} remedies this problem by 
assigning two matrices $\delta, \zeta$ and a Hodge filtration $ \hat{F} = e^\zeta e^{-\im\delta} F$
to $(F, W)$ such that the new Deligne splitting $\tilde I^{p,q}$ derived from $(\hat F,W)$ is $\bbR$-split. This new structure $(\hat{F}, W)$ is 
called the Sl(2)-splitting of $(F, W)$.  We will review its construction in appendix \ref{app:Sl2-splitting}. 
The Sl(2)-splitting is central to the construction of commuting $\slt$-triples as we discuss in  appendix \ref{app:constructSl2s}. In particular, both are linked 
via the relation 
\beq \label{Yk-action}
   Y_{(k)}\, \tilde I^{p,q}(\Delta^\circ_{1 \ldots k}) = (p+q-3) \,  \tilde I^{p,q}(\Delta^\circ_{1 \ldots k})\ ,
\eeq
where $Y_{(k)} = Y_1 +  \ldots  + Y_k$ and $\tilde I^{p,q}(\Delta^\circ_{1 \ldots k}) $ is the Sl(2)-splitting associated 
to $\Delta_{1 \ldots k}^\circ$. 

Note that for Calabi-Yau threefolds we have discussed after \eqref{Pi-der} that all information contained in $F_\Delta^p$ 
can be inferred from $\mathbf{a}_0$ and its $\zeta^\kappa$-derivatives and the nilpotent elements. 
Hence, the existence of an Sl(2)-splitting 
can be reformulated to the statement that there exists a special 
\beq
   \mathbf{\tilde a}_0 = e^\zeta e^{-\im\delta}\mathbf{a}_0\ .
\eeq
The $ \mathbf{\tilde a}_0$ for the highest point of intersection $\Delta_{1 \ldots n_\cE}$ will serve as a starting 
point for the construction of the $\slt$-triples \eqref{Sl2-triples}. Let us denote this by 
\beq
    \mathbf{\tilde a}^{(n_\cE)}_0 \equiv  \mathbf{\tilde a}_0(\Delta^\circ_{1 \ldots n_\cE})\ .
\eeq
One then constructs the $\mathbf{\tilde a}_0$ relevant at the lower intersections stepwise as we 
summarize in appendix \ref{app:constructSl2s}. The crucial point 
for our later discussion is the fact that the initial step for constructing the commuting 
Sl(2)-triples always requires to start at the highest intersection. The $\mathbf{\tilde a}_0^{(n_P)}\equiv  \mathbf{\tilde a}_0(\Delta^\circ_{1 \ldots n_P})$ relevant 
for a point $P \in \Delta^\circ_{1 \ldots n_P}$ is given by \cite{CKS} \footnote{Note that there is an additional minus sign in 
the exponent compared to (4.56) of \cite{CKS}. This arises from the fact that we let $N_i^-$ act on the 
coefficients in an integer basis, rather then the basis itself.}
\beq \label{a0relation}
\mathbf{\tilde a}_0^{(n_P)} =\text{exp}\!\! \begin{array}{c}\left( - \im  N^-_{n_P+1} \right)\end{array} \mathbf{\tilde a}_0^{(n_P+1)} =\ \ldots\ = \text{exp}\!\! \begin{array}{c}\left(- \im \sum_{i=n_P+1}^{n_\cE}  N^-_i \right)\end{array} \mathbf{\tilde a}_0^{(n_\cE)} \ .
\eeq
This implies that considering such a point $P$ on a lower intersection also the data of the highest intersection $\Delta^\circ_{1 \ldots n_\cE}$
is relevant. This non-local information will be crucial in section \ref{sec:infinite_towers} when constructing an infinite charge orbit. 
An explicit construction of the triples $(N_i^-,N_i^+,Y_i)$ for a two-parameter example is presented in appendix 
\ref{sec:exa}. 

Another important statement of the Sl(2)-orbit theorem is that the nilpotent orbit can be written in terms of yet another orbit, namely 
the Sl(2)-orbit $\mathbf{\Pi}_{\text{Sl(2)}}$. However, in contrast to the above discussion of the nilpotent orbit 
approximating the periods, the Sl(2)-orbit should be viewed as an alternative description capturing the main structure of 
the limiting variation of Hodge structure. Explicitly the relation between the nilpotent orbits and the Sl(2)-orbit are given by 
\beq \label{rel_Sl2-nil}
   \mathbf{\Pi}_{\text{nil}} \equiv 
    \text{exp} \Big[ \sum_{j=1}^{n_\cE} -t^j N_j  \Big] \mathbf{a}_0(\zeta) = \text{exp} \Big[ \sum_{j=1}^{n_\cE} -x^j N_j  \Big]\cdot M(y)\cdot  \mathbf{\Pi}_{\text{Sl(2)}} 
 \eeq
 where the Sl(2)-orbit is given by
 \beq \label{Sl(2)-orbit}
    \mathbf{\Pi}_{\text{Sl(2)}} \equiv  \text{exp} \Big[ \sum_{j=1}^{n_\cE} -\im\, y^j N_j^-  \Big] \mathbf{\tilde a}^{(n_\cE)}_0(\zeta) \ ,
 \eeq
 and we recall the notation $t^i = x^i + \im y^i$.
It is crucial here to introduce the $y^i$-dependent matrix $M(y)$. The Sl(2)-orbit theorem 
states that $M(y)$ can be written as 
\beq 
    M(y) = \prod_{r=n}^1 g_r \left( \frac{y^1}{y^{r+1}},\ldots ,\frac{y^r}{y^{r+1}}\right) \ ,
\eeq
where $g_r(y^1,\ldots,y^r)$ are $\mathrm{Sp}(2h^{2,1}+2,\bbR)$-valued functions. 
Furthermore, $g_r(y^1,\ldots,y^r)$
and $g^{-1}_r(y^1,\ldots,y^r)$ have power-series expansions 
in non-positive powers of $y^1/y^2$, $y^2/y^3$,  \ldots , $y^r$ with constant term $1$ and 
convergent in a region 
\beq \label{sector_Sl2orbit}
    \mathcal{\hat R}_{1 \ldots r} =\left\{ \frac{y^1}{y^2}>\lambda\ ,\quad \frac{y^2}{y^3} > \lambda\ ,  \ldots , \quad y^r > \lambda \right\}\ ,
\eeq
for some $\lambda > 0$. 
In other words, writing $ \mathbf{\Pi}_{\text{nil}}$ in terms of an Sl(2)-orbit $\mathbf{\Pi}_{\text{Sl(2)}}$ depends on the considered region $ \mathcal{\hat R}_{1 \ldots r}$ in 
moduli space. Of course, we can always reorder the $y^i$ sending $y^i \rightarrow y^{\sigma(i)}$ 
to be in a region $ \mathcal{\hat R}_{\sigma(1) \ldots \sigma(r)}$ that satisfies the above conditions. 
This implies that the Sl(2)-orbit will then be adapted to this ordering.

\subsection{Growth of the Hodge norm} \label{sec:growth_theorems} 

In this subsection we introduce an important result that follows from the correspondence between 
nilpotent orbits and commuting Sl(2)s. Namely, we discuss the asymptotic behaviour of the Hodge norm 
of general three-forms near the discriminant locus $\Delta$. The Hodge norm on a smooth space $Y_3$ 
is defined as 
\beq
\label{hodgenorm}
    || v||^2 \equiv || \mathbf{v}||^2 = \int_{Y_3} v \wedge *  \bar v  \equiv  S(C \mathbf{v} ,  \mathbf{\bar v} )\ ,
\eeq 
where $v$ is a complex 3-form, $*$ is the Hodge star operator, and 
$\mathbf{v}$ are the components of $v$ in the integral basis $\gamma_I$. In the pure Hodge decomposition \eqref{H3-cohom-split} the 
Hodge operator acts as $* v = \im^{p-q} v$ for $v \in H^{p,q}(Y_3)$. Let us note that the Hodge norm can also be written in terms of the bilinear form 
$S$ defined in \eqref{def-S} and the Weil operator $C$. The Weil operator acts as $\im^{p-q}$ on $(p,q)$-forms 
and is used in \cite{Schmid,CKS}.  The definition \eqref{hodgenorm} implies, for example, that the Hodge norm of the unique $(3,0)$-form 
$\Omega$ on $Y_3$ is given by 
\beq \label{Omega_norm}
   || \Omega ||^2 \equiv || \mathbf{\Pi}||^2 = \im \int_{Y_3} \Omega \wedge   \bar \Omega = e^{-K}\ ,
\eeq 
where we have expressed the result using the K\"ahler potential \eqref{Kpot_cs} on the 
complex structure moduli space.

\begin{figure}[h!]
\vspace*{.5cm}
\begin{center} 
\includegraphics[width=7.5cm]{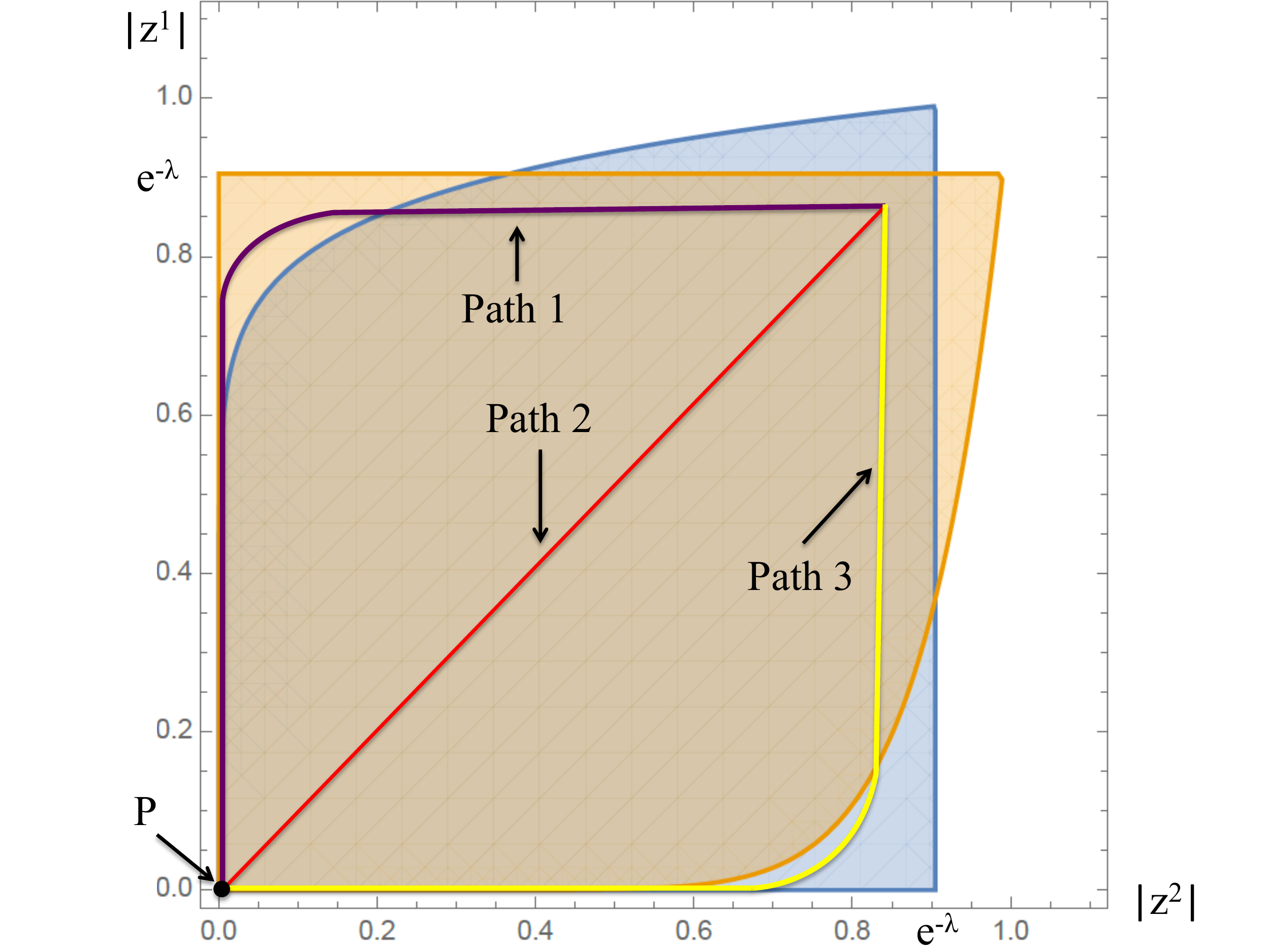} 
\end{center}
\caption{Real slice through two intersecting divisors located at $z^1= e^{2\pi\im t^1} =0$ and $z^{2}=e^{2\pi\im t^2} =0$. The 
intersection point is the origin $P=\{|z^1|=|z^{2}|=0\}$. The shaded  areas show the two overlapping regions  \eqref{(1)growthbound},  \eqref{(2)vt-growth_2} a 
path to the singularity at the origin can pass through, in order that the growth can be 
evaluated using \eqref{(1)vt-growth_2}, \eqref{(2)vt-growth_2} for the constant $\lambda$ set to $\lambda=0.1$. Three paths of different nature with respect to this are shown.} \label{growthsectors}
\end{figure}

Extracting the behaviour of $||v||^2$ when approaching a point 
on $\Delta$ is, of course, a very non-trivial task. In fact, at first, it seems impossible 
that this can be done at all, since it appears to be a highly \textit{path-dependent} question. 
To highlight this point further, let us consider 
a two-dimensional moduli space, locally parameterized by two coordinates $z^1,z^2$. 
We consider two divisors $\Delta_1$ and $\Delta_2$ intersecting in a point (see also subsection \ref{sec:charact_sing}). 
Clearly an intersection point $P = \Delta_1 \cap \Delta_2$ can be approached on many different paths, 
as indicated in figure \ref{growthsectors}. Recalling the discussion of subsection \ref{sec:class_infinitedistanceP} the points on 
$\Delta_1^\circ$ and $\Delta_2^\circ$ can be at finite or infinite distance, and one expects that 
the growth of the norm $|| v ||^2$ can differ greatly when approaching a finite or an infinite distance 
point. Considering, for example, $|| \Omega||^2$ the growth of the Hodge norm corresponds 
to the growth of the K\"ahler potential which clearly is connected to the 
distance of the point. The issue becomes particularly eminent 
when $P$ is at this intersection of divisors with $\Delta_1^\circ$ of type I and $\Delta_2^\circ$ of type IV, 
i.e.~using \eqref{one-modulus} one at finite distance and one at infinite distance. The growth of the Hodge norm 
along the paths in figure \ref{growthsectors} then should differ significantly. Remarkably, the growth theorem 
proven in \cite{CKS} takes into account this path dependence and nevertheless gives 
a universal result. We present this results for $v$ being 
a flat section under the Gauss-Manin connection $\nabla$ briefly discussed at the end of 
subsection \ref{sec:period_approx_nil} and briefly comment on generalizations in \eqref{growthequiv}.

To begin with let us state the growth theorem for the case that we consider points at 
a single divisor $\Delta_1^\circ$ at $t^1 = \im \infty$, i.e.~a point on $\Delta_1$ away from any intersection. 
We consider a three-form $v$ that satisfies 
\beq  \label{location_1}
  v \ \in \ W_j(N_1)\ , \qquad W_j(N_1) = \bigoplus_{p+q \leq j} I^{p,q}(\Delta_1^\circ)\ ,
\eeq
where we recalled that $W_j$, defined in \eqref{def-Wi}, can be decomposed into the Deligne splitting $I^{p,q}$
associated to the locus $\Delta_1^\circ$ (see \eqref{Fp-Wi_split}). Here $j$ is corresponding to the smallest 
value $0,1, \ldots ,6$ such that \eqref{location_1} holds. This is relevant due to the fact that we have $W_{j+1}\subset W_j$.
Then the growth theorem  \cite{Schmid}  states, that for $\I\, t^1  > \lambda$ and $\R \, t^1 < \delta$, with $\lambda,\delta$ being sufficiently large constants, one finds the dominant growth 
\beq \label{vt-growth}
     || v||^2 \sim c\, (\I\, t^1)^{j-3} \ , \qquad c>0\ . 
\eeq
Here and below the $\sim$ indicates that there are generally 
terms that grow slower than this leading term. In particular, one can have terms
proportional to $ (\I\, t^1)^{j-3-k}$ for $k>0$ or exponentially suppressed terms proportional to $e^{- \I\, t^1}$.
Clearly, in this one-parameter case, path dependence is not an issue. 

Let us next consider the two-parameter situation, i.e.~we consider a point $P$ on the 
intersection of $\Delta_1$ and $\Delta_2$, located at $t^1=\im \infty$ and $t^2=\im \infty$, but 
away from any further intersection within $\Delta$. Then the growth theorem 
depends on the path taken towards the point $P$ at $t^1=t^2=\im \infty$.
We can think of this as corresponding to the two ways we can reach the singularity 
type at point $P$. Namely, we can first enhance to the singularity at $\Delta_1^\circ$  
and then move to $\Delta_{12}^\circ$ or we can first enhance to the singularity at 
$\Delta_2^\circ$ and then to $\Delta_{12}^\circ$. This corresponds to paths 1 and 3 in figure \ref{growthsectors}.
The relevant nilpotent elements are then 
\bea
   (1) & \Delta_1^\circ \ \rightarrow \ \Delta_{12}^\circ:\qquad (N_1, N_1+N_2)\ ,\\
   (2) & \Delta_2^\circ \ \rightarrow\  \Delta_{12}^\circ:\qquad (N_2, N_1+N_2)\ .\nn
\eea
Let us start with the case $(1)$ and consider a three-form $v$ satisfying
\beq  \label{(1)vlocation}
  v \ \in \ W_{l_1}(N_1) \cap W_{l_2}(N_1+N_2) \ ,
\eeq
where $W_{l_1}(N_1)$ can be split as in \eqref{location_1}, while 
$W_{l_2}(N_1+N_2)$ is associated to $N_1+N_2$ and hence splits into 
the Deligne splitting on $\Delta_{12}^\circ$ as $W_{l_2}(N_1+N_2) = \bigoplus_{p+q \leq l_2} I^{p,q}(\Delta_{12}^\circ)$.
Note that here $l_1$ and $l_2$ are the lowest values for which \eqref{(1)vlocation} is satisfied.  
The growth theorem of \cite{CKS} now states that this $v$ has the leading growth 
\beq \label{(1)vt-growth_2}
     || v||^2 \sim c\, \left( \frac{\I\, t^1 }{\I \, t^2 }\right)^{l_1-3} (\I \, t^2)^{l_2-3} \ , \qquad c>0\ ,
\eeq
when approaching the intersection point $t^1=t^2=\im\infty$. In order for this to be true, however, one has 
to restrict to paths in the region 
\beq \label{(1)growthbound}
 \cR_{12} = \left\{ \ \frac{\I\, t^1 }{\I \, t^2 } > \lambda \ , \ \ \I \, t^2 > \lambda \ \right\}\ ,
\eeq
for any constant $\lambda>0$ and demand that $\R\, t^1,\R\, t^2$ are bounded by some constant. We will denote such a restriction as a {\it growth sector}, so that all paths in $\cR_{12}$ are in the same growth sector.
We have depicted this condition in figure \ref{growthsectors}. Let us stress that 
the growth in \eqref{(1)vt-growth_2} is polynomial as long as $l_1\leq l_2$. This will always happen, for example, for the 
growth of the K\"ahler potential $e^{-K}$ as we will see below.

Clearly, in order to discuss the case $(2)$ we simply have to exchange $N_1$ and $N_2$ and $t^1$ and $t^2$ 
in all formulas. One thus finds that for 
\beq 
  v \ \in \ W_{l_1}(N_2) \cap W_{l_2}(N_1+N_2) \ ,
\eeq
one has the leading growth of the Hodge norm 
\beq
\label{(2)vt-growth_2}
     || v||^2 \sim c\, \left( \frac{\I\, t^2 }{\I \, t^1 }\right)^{l_1-3} (\I \, t^1)^{l_2-3} \quad \text{in}\quad 
 \cR_{21}= \left\{\  \frac{\I\, t^2 }{\I \, t^1 } > \lambda \, , \  \I \, t^1 > \lambda \ \right\} \ ,
\eeq
for any constant $\lambda>0$ and bounded $\R\, t^1,\R\, t^2$.

While having only discussed the two-parameter case, the reader might anticipate 
the form of the general growth theorem for any number of intersecting divisors. 
We summarize this important result of Cattani, Kaplan, and Schmid \cite{CKS} and Kashiwara \cite{Kashiwara} in the 
following. 
Let us consider the leading growth when approaching a point $P$ on the intersection of 
$n_P$ divisors $\Delta_1,\ldots,\Delta_{n_P}$ in $\Delta$ located at $t^1 =\ldots=t^{n_P} = \im \infty$. 
To simplify the notation we recall that we introduced in \eqref{t-def} that $t^i = x^i + \im y^i$.
The sectors are specified for 
fixed $\lambda,\delta>0 $. The ${n_P}!$ orderings give different overlapping sectors. 
We pick for the $N_i$ the ordering 
\beq \label{ordering_gen}
    \text{chosen ordering:}  \qquad N_1,\, N_2,\,  \ldots  ,\ N_{n_P}\ ,
 \eeq   
with all other orderings obtained by exchanging $N_i$ and $t^i$ in the following formulas. 
Next we consider a $v$ with 
\beq \label{general_growth-loc}
 v \ \in \ W_{l_1}\big(N_{(1)} \big) \cap W_{l_2} \big(N_{(2)} \big) \cap \ldots \cap W_{l_{n_P}}\big(N_{({n_P})} \big) 
 \eeq 
 where $N_{(j)}=\sum_{i=1}^j N_i $ and  $l_i$ are the lowest values for which this is true. The 
 leading growth of the Hodge norm is then
\beq \label{general_growth-res}
   ||v||^2 \sim c\, \left( \frac{y^1 }{y^2 }\right)^{l_1-3}\cdots \left( \frac{y^{{n_P}-1} }{y^{n_P} }\right)^{l_{{n_P}-1}-3} (y^{n_P})^{l_{n_P}-3} 
 \eeq 
for some $c>0$. Associated to the ordering \eqref{ordering_gen} the growth sector 
for the allowed paths takes the form
\beq \label{sector_gen}
   \cR_{1 \ldots {n_P}} = \left\{t^i: \quad \frac{y^{1} }{y^2 } > \lambda \, ,\ldots , \ 
\frac{y^{{n_P}-1}}{y^{{n_P}} } > \lambda  \,  ,\   y^{n_P} > \lambda\, ,
\quad x^i < \delta\, \right\} .
\eeq 
It might be interesting to stress that the proof in \cite{CKS} of this theorem uses fundamentally 
the  Sl(2)-orbit theorem briefly discussed in subsection \ref{sec:sl2-orbit} and much of the technology reviewed in this 
section. In particular, the relevant sector \eqref{sector_gen} for allowed paths arises due to the convergence properties of 
the Sl(2)-orbit and agrees with \eqref{sector_Sl2orbit} in its $y^i$-part.

As an application of this growth theorem, let us evaluate the growth of $||\Omega||^2$ 
and hence via \eqref{Omega_norm} of the K\"ahler potential $e^{-K}$. The first step is 
to approximate the periods $\mathbf{\Pi}$ by the nilpotent orbit $\mathbf{\Pi}_{\rm nil}$. 
The nilpotent orbit $\mathbf{\Pi}_{\rm nil}$ then can be approximated by the Sl(2)-orbit
as in \eqref{rel_Sl2-nil}, when restricting to the appropriate sector \eqref{sector_Sl2orbit}. The latter is defined using the $\mathbf{\tilde a}^{(n_\cE)}_0$ introduced in 
subsection \ref{sec:sl2-orbit}. While the relation between the nilpotent and Sl(2)-orbit contains 
non-trivial $y^i$-dependent terms, one can show that they are bounded and do not alter the 
growth. 
In fact, one has that the growth of both $\mathbf{\Pi}_{\rm nil}$ and $\mathbf{\tilde a}_0^{(n_\cE)}$ agree
such that  \cite{CKS}\footnote{In fact,  
it was shown generally in \cite{CKS} that the growth result \eqref{general_growth-res} also hold if one multiplies 
$v$ by either exp$(\sum_i x^i N_i)$, exp$(\sum_i t^i N_i)$, or even the matrix relating $\mathbf{\Pi}$ and $\mathbf{a}_0$.} 
\beq \label{growthequiv}
     ||\mathbf{\Pi}||^2 \ \sim\   ||\mathbf{\Pi}_{\rm nil}||^2 \ \sim \ \big| \big|\mathbf{\tilde a}^{(n_\cE)}_0\big| \big|^2 \ , 
\eeq
where the symbol $\sim$ as above indicates that we are only considering the leading growth near the 
point $P$ on the discriminant locus. We can now infer the growth by 
using the location of $\mathbf{a}_0$ in the filtrations $W_{l}(N_{(k)}^-)$, where 
$N_{(k)}^- = \sum_{i=1}^k N_{i}^-$ in analogy to \eqref{lbr}. We note that \footnote{This can be
inferred by using \eqref{a0relation} extended to all $\mathbf{a}_0^{(i)}$. The $\mathbf{a}_0^{(i)}$  
are the vectors spanning the Sl(2)-split $\hat F^3 = \tilde I^{3,d_i}$ on the intersection loci $\Delta^\circ_{1 \ldots i}$.
The fact, that the location of $\mathbf{a}_0^{(i)}$ and $\mathbf{a}_0^{(n_\cE)}$ follows from the 
commutativity of the $\slt$-triples, as we will discuss in a slightly different context when we study the charge orbit below.}
\beq \label{locationa0}
  \mathbf{\tilde a}^{(n_\cE)}_0 \ \in \ W_{d_1+3}\big(N^-_{(1)} \big) \cap W_{d_2+3} \big(N^-_{(2)} \big) \cap \ldots \cap W_{d_{n_P}+3}\big(N^-_{(n_P)} \big) \ .
\eeq
The integer $d_i$ is defined by 
\beq \label{N_on_a0}
   \big(N^-_{(i)} \big)^{d_i} \mathbf{\tilde a}^{(n_\cE)}_0 \neq 0\ , \qquad   \big(N^-_{(i)}  \big)^{d_i+1} \mathbf{\tilde a}^{(n_\cE)}_0 =0 \ .
\eeq
In other words it labels in which $\tilde I^{p,q}$ the $\mathbf{\tilde a}^{(n_\cE)}_0$ resides at the various intersection 
loci. Denoting the Sl(2)-split Deligne splitting on $\Delta^\circ_{1 \ldots k}$ by $\tilde I^{p,q}(\Delta^\circ_{1..k})$ one 
has $\mathbf{\tilde a}^{(n_\cE)}_0 \in \tilde I^{3,d_k}(\Delta^\circ_{1 \ldots k}) $ for $k=1, \ldots ,n_P$. Later on in subsection \ref{sec:sing_class} 
we will also see that $d_i$ labels the type of the singularity at the intersection, i.e.~one has
\beq \label{linkdi-sing}
   \text{singularity on } \Delta^\circ_{1 \ldots k}\ \text{is Type I, II, III, IV} \quad \Longleftrightarrow \quad d_k = (0,1,2,3)\ .   
\eeq
This will become more clear with the classification of singularities that we 
will present in the next section. We will also show that there are restrictions 
on the allowed enhancements and in particular that $d_i \leq d_{i+1}$. 
Using \eqref{locationa0}, together with the fact $W(N_{(i)}^-) = W(N_{(i)})$ in the Sl(2)-orbit theorem of \cite{CKS}, and the general growth result \eqref{general_growth-res} we 
thus find
\beq \label{general_growth_K}
    e^{-K} \sim \big| \big| \mathbf{\tilde a}^{(n_\cE)}_0\big|\big|^2 \sim  c\, \left( y^1 \right)^{d_1} \left( y^2 \right)^{d_2-d_1} \cdots(y^{n_P})^{d_{n_P} - d_{n_P-1}}  \ . 
\eeq
This expression gives the general growth of the K\"ahler potential for any path approaching $P$ 
in the sector \eqref{sector_gen}. Other sectors can be obtained by exchanging the $N_i$ and $y^i$.


\section{Classifying Singularities in Calabi-Yau Moduli Spaces} \label{sec:classification_results}

In this section we summarize some general classification results that highlight the 
power of the mathematical tools introduced in section \ref{math_background}. More precisely, we will discuss a classification 
of Calabi-Yau threefold singularities in subsection \ref{sec:sing_class}, their allowed enhancement patterns 
in subsection \ref{sec:class_enh}, and make some comments on the classification of infinite distance points in subsection~\ref{sec:class_infinitedistanceP}. 
A special emphasis is given to the discussion of the large complex structure and large volume configurations, 
where the presented tools and classifications are of immediate use. 
We stress that the results of this section are relevant in many different contexts that are not related to  
a discussion of the Swampland Distance Conjecture. Therefore, this section can also be read independently 
of the main motivation of this work.

\subsection{A Classification of Calabi-Yau Threefold Singularities} \label{sec:sing_class}

Having summarized the relevant mathematical background we are now 
in the position to present a classification of Calabi-Yau threefold singularities and 
allowed enhancement patterns. While we will mostly discuss the results of \cite{Kerr2017}, 
we will add some additional new insights that are particularly useful for explicit computations. 

The basic idea to classify the arising singularities of $Y_3$ is to classify the allowed 
Deligne splittings $I^{p,q}$. As we described in subsection \ref{sec:charact_sing} these Deligne splittings 
non-trivially depend on the objects $F^p_\Delta$ and $N$ associated to the considered point on $\Delta$. 
The $I^{p,q}$  package this information in an intuitive and useful way. In particular, 
one can introduce to each point of $\Delta$ a \textit{limiting Hodge diamond} containing the  
dimensions of the $I^{p,q}$ given by  
\beq
 \begin{array}{ccccccccc} &&& i^{3,3}&&\\ 
&& i^{3,2}\hspace*{-.2cm}& & \hspace*{-.2cm} i^{2,3} \\ 
& i^{3,1}\hspace*{-.2cm} & & \hspace*{-.2cm} i^{2,2}\hspace*{-.2cm} & & \hspace*{-.2cm} i^{1,3}\\
 i^{3,0} \hspace*{-.2cm} & & i^{2,1} \hspace*{-.2cm} & & \hspace*{-.2cm} i^{1,2} &&\hspace*{-.2cm} i^{0,3}\\
&i^{2,0}\hspace*{-.2cm} & & \hspace*{-.2cm} i^{1,1}\hspace*{-.2cm} & & \hspace*{-.2cm} i^{0,2}\\
&&i^{1,0}\hspace*{-.2cm} & &  \hspace*{-.2cm} i^{0,1} \\
&&&i^{0,0}&& \\  \end{array}\ , 
 \qquad  i^{p, q} = \dim_\bbC I^{p, q}\ . 
\eeq
Since the $I^{p,q}$ represent a finer split of the cohomology near the singularity,  
we can decompose original Hodge numbers for the smooth geometry at the considered point on $\Delta$ 
into the Hodge-Deligne numbers as
\beq \label{hod-to-Dhod}
  h^{p, 3 - p} =  \sum_{q = 0}^3 i^{p, q} \ , \qquad p = 0, \ldots , 3.
\eeq
Furthermore, one can deduce 
several  properties of a limiting Hodge diamond \footnote{A detailed discussion of these 
properties can be found in section 5.2 of \cite{Kerr2017}.}
\begin{align}
   &i^{p, q} = i^{q, p}  = i^{3 - q, 3 - p} \ ,\hspace{-2cm} &&\text{ for all } p, q\ , &   \label{i-symmetries}\\
  &i^{p - 1, q - 1}     \le i^{p, q} ,&&\text{ for } p + q \le 3\ .& 
\end{align}
Given these properties, a first step in classifying singularities is to classify all possible 
Hodge-Deligne diamonds. 

For Calabi-Yau threefolds the classification of limiting Hodge diamonds is greatly simplified 
by the fact that one has $h^{3,0}=1$. Using 
\eqref{hod-to-Dhod} that there are four possible cases $i^{3,d}=1$, $d=0,1,2,3$, which we 
label by Latin numbers following \cite{Kerr2017},    I, II,  III,  IV.
Furthermore, due to the symmetries \eqref{i-symmetries} there are just two independent Hodge-Deligne numbers,
which we pick to be $i^{2,1}$ and $i^{2,2}$. 
In table \ref{HD:enumeration} we will use a more pictorial way to represent 
the limiting Hodge diamonds. For example, the limiting Hodge diamond for $d=2$ is depicted as 

\begin{picture}(100,115)
\put(50,60){$
   \begin{array}{ccccccccc} &&& 0&&\\ 
&&  1\hspace*{-.2cm}& & \hspace*{-.2cm} 1 \\ 
&  0\hspace*{-.2cm} & & \hspace*{-.2cm} i^{2,2}\hspace*{-.2cm} & & \hspace*{-.2cm}  0\\
  0 \hspace*{-.1cm} & &i^{2,1} \hspace*{-.2cm} & & \hspace*{-.2cm} i^{2,1}&&\hspace*{-.1cm} 0\\
&0\hspace*{-.2cm} & & \hspace*{-.2cm}i^{2,2}\hspace*{-.2cm} & & \hspace*{-.2cm} 0\\
&&1 \hspace*{-.2cm} & &  \hspace*{-.2cm}  1\\
&&&0&& \\  \end{array}
$
}
\put(200,20){
\begin{tikzpicture}[scale=0.7,cm={cos(45),sin(45),-sin(45),cos(45),(15,0)}]
  \draw[step = 1, gray, ultra thin] (0, 0) grid (3, 3);
  \draw[fill] (0, 1) circle[radius=0.05];
  \draw[fill] (1, 0) circle[radius=0.05];
  \draw[fill] (1, 2) circle[radius=0.05] node[above]{$c'$};
  \draw[fill] (2, 1) circle[radius=0.05] node[above]{$c'$};
  \draw[fill] (2, 3) circle[radius=0.05];
  \draw[fill] (1, 1) circle[radius=0.05] node[above]{$c$};
  \draw[fill] (2, 2) circle[radius=0.05] node[above]{$c$};
  \draw[fill] (3, 2) circle[radius=0.05];
\end{tikzpicture}}
\put(170,60){$\cong$}
\put(320,70){$c\,=i^{2,2}$}
\put(320,55){$c'=i^{2,1}$}
\end{picture}
\vspace{-.5cm}

\noindent
Furthermore, we will  index the singularity type with $i^{2,2}$ writing 
\beq \label{def-types}
    \text{I}_{i^{2,2}}\, , \quad  \text{II}_{i^{2,2}}\,,\quad  \text{III}_{i^{2,2}}\,,\quad  \text{IV}_{i^{2,2}}\ .
\eeq  
The allowed values for $i^{2,2}$ are obtained form \eqref{hod-to-Dhod} and differ for the different singularity types 
as summarized in the third column of table \ref{HD:enumeration}. In total we thus find $4 h^{2,1}$ possible 
limiting Hodge diamonds depicted in the second column of table \ref{HD:enumeration}.

\newcolumntype{V}{>{\centering\arraybackslash} m{.21\linewidth} }
\newcolumntype{L}{>{\arraybackslash} m{.21\linewidth} }
\begin{table}[h!]
\centering
\begin{tabular}{| c | V | c | L | c |}
\hline
name & Hodge diamond & labels & Young diagram & $N,\eta$ \\
\hline \hline
\rule[-1.5cm]{0cm}{3.4cm} $\mathrm{I}_a$ & 
\begin{tikzpicture}[scale=0.7,cm={cos(45),sin(45),-sin(45),cos(45),(15,0)}]
  \draw[step = 1, gray, ultra thin] (0, 0) grid (3, 3);

  \draw[fill] (0, 3) circle[radius=0.05];
  \draw[fill] (1, 2) circle[radius=0.05] node[above]{$a'$};
  \draw[fill] (2, 1) circle[radius=0.05] node[above]{$a'$};
  \draw[fill] (1, 1) circle[radius=0.05] node[above]{$a$};
  \draw[fill] (2, 2) circle[radius=0.05] node[above]{$a$};
  \draw[fill] (3, 0) circle[radius=0.05];
\end{tikzpicture}
  & \begin{minipage}{.15\textwidth} $a+a'=m$\\ $0 \leq a \leq m$ \\ \end{minipage}
  &  
  \begin{tikzpicture}[scale=0.5]
    \draw (0, 0) rectangle (1, -1);
    \draw (.5, -.5) node {$+$};
    \draw (1, 0) rectangle (2, -1);
    \draw (1.5, -.5) node {$-$};
    \draw (2, -.5) node[right] {$\otimes a$};
    \draw (0, -1) rectangle (1, -2);
    \draw (2, -1.5) node[right] {$\otimes 2a' + 2$};
  \end{tikzpicture}
  & 
  \begin{minipage}{.2\textwidth} rank$(N,N^2,N^3) $\\
                                       $= (a,0,0)$ \\
                                       $\eta N$ has $a$ negative eigenvalues\end{minipage}
\\ \hline
\rule[-1.5cm]{0cm}{3.4cm}$\mathrm{II}_b$  &  
\begin{tikzpicture}[scale=0.7,cm={cos(45),sin(45),-sin(45),cos(45),(15,0)}]
  \draw[step = 1, gray, ultra thin] (0, 0) grid (3, 3);

  \draw[fill] (0, 2) circle[radius=0.05];
  \draw[fill] (1, 3) circle[radius=0.05];
  \draw[fill] (1, 2) circle[radius=0.05] node[above]{$b'$};
  \draw[fill] (1, 1) circle[radius=0.05] node[above]{$b$};
  \draw[fill] (2, 1) circle[radius=0.05] node[above]{$b'$};
  \draw[fill] (2, 2) circle[radius=0.05] node[above]{$b$};
  \draw[fill] (2, 0) circle[radius=0.05];
  \draw[fill] (3, 1) circle[radius=0.05];
\end{tikzpicture}
  & \begin{minipage}{.15\textwidth} $b+b'=m-1$\\ $0 \leq b \leq m-1$ \\ \end{minipage}
  &  
  \begin{tikzpicture}[scale=0.5]
    \draw (0, 0) rectangle (1, -1);
    \draw (.5, -.5) node {$+$};
    \draw (1, 0) rectangle (2, -1);
    \draw (1.5, -.5) node {$-$};
    \draw (2, -.5) node[right] {$\otimes b$};
    \draw (0, -1) rectangle (1, -2);
    \draw (.5, -1.5) node {$-$};
    \draw (1, -1) rectangle (2, -2);
    \draw (1.5, -1.5) node {$+$};
    \draw (2, -1.5) node[right] {$\otimes 2$};
    \draw (0, -2) rectangle (1, -3);
    \draw (2, -2.5) node[right] {$\otimes 2b'$};
  \end{tikzpicture}
  & 
  \begin{minipage}{.2\textwidth} rank$(N,N^2,N^3) $\\
                                       $= (2+b,0,0)$ \\
                                       $\eta N$ has $b$ negative and $2$ positive eigenvalues  \end{minipage}
 \\ \hline
\rule[-1.5cm]{0cm}{3.4cm}$\mathrm{III}_c$ &  
\begin{tikzpicture}[scale=0.7,cm={cos(45),sin(45),-sin(45),cos(45),(15,0)}]
  \draw[step = 1, gray, ultra thin] (0, 0) grid (3, 3);
  \draw[fill] (0, 1) circle[radius=0.05];
  \draw[fill] (1, 0) circle[radius=0.05];
  \draw[fill] (1, 2) circle[radius=0.05] node[above]{$c'$};
  \draw[fill] (2, 1) circle[radius=0.05] node[above]{$c'$};
  \draw[fill] (2, 3) circle[radius=0.05];
  \draw[fill] (1, 1) circle[radius=0.05] node[above]{$c$};
  \draw[fill] (2, 2) circle[radius=0.05] node[above]{$c$};
  \draw[fill] (3, 2) circle[radius=0.05];
\end{tikzpicture}
  & \begin{minipage}{.15\textwidth} $c+c'=m-1$\\ $0 \leq c \leq m-2$ \\ \end{minipage}
  & 
  \begin{tikzpicture}[scale=0.5]
    \draw[step = 1] (0, 0) grid (3, -1);
    \draw (3, -.5) node[right] {$\otimes 2$};
    \draw (0, -1) rectangle (1, -2);
    \draw (.5, -1.5) node {$+$};
    \draw (1, -1) rectangle (2, -2);
    \draw (1.5, -1.5) node {$-$};
    \draw (3, -1.5) node[right] {$\otimes c$};
    \draw (0, -2) rectangle (1, -3);
    \draw (3, -2.5) node[right] {$\otimes 2c' - 2$};
  \end{tikzpicture}
  & 
  \begin{minipage}{.2\textwidth} rank$(N,N^2,N^3) $\\
                                       $= (4+c,2,0)$  \end{minipage}
\\ \hline
\rule[-1.5cm]{0cm}{3.4cm} $\mathrm{IV}_d$   &
\begin{tikzpicture}[scale=0.7,cm={cos(45),sin(45),-sin(45),cos(45),(15,0)}]
  \draw[step = 1, gray, ultra thin] (0, 0) grid (3, 3);

  \draw[fill] (0, 0) circle[radius=0.05];
  \draw[fill] (1, 1) circle[radius=0.05] node[above]{$d$};
  \draw[fill] (1, 2) circle[radius=0.05] node[above]{$d'$};
  \draw[fill] (2, 1) circle[radius=0.05] node[above]{$d'$};
  \draw[fill] (2, 2) circle[radius=0.05] node[above]{$d$};
  \draw[fill] (3, 3) circle[radius=0.05];
\end{tikzpicture}
 & \begin{minipage}{.15\textwidth} $d+d'=m$\\ $1 \leq d \leq m$ \\ \end{minipage}
  &
  \begin{tikzpicture}[scale=0.5]
    \draw[step = 1] (0, 0) grid (4, -1);
    \draw (.5, -.5) node {$-$};
    \draw (1.5, -.5) node {$+$};
    \draw (2.5, -.5) node {$-$};
    \draw (3.5, -.5) node {$+$};
    \draw (4, -.5) node[right] {$\otimes 1$};
    \draw (0, -1) rectangle (1, -2);
    \draw (.5, -1.5) node {$+$};
    \draw (1, -1) rectangle (2, -2);
    \draw (1.5, -1.5) node {$-$};
    \draw (4, -1.5) node[right] {$\otimes d - 1$};
    \draw (0, -2) rectangle (1, -3);
    \draw (4, -2.5) node[right] {$\otimes 2d'$};
  \end{tikzpicture}
  & 
  \begin{minipage}{.2\textwidth} rank$(N,N^2,N^3) $\\
                                       $= (2+d,2,1)$  \end{minipage}
  \\
  \hline
\end{tabular}
\caption{The $4m$ possible limiting Hodge diamonds with Hodge numbers $h^{2,1}=m$. The label next to a dot at a point $(p, q)$ 
represents the value of $i^{p, q}$. A dot at $(p, q)$ without a label represents $i^{p, q} = 1$.} \label{HD:enumeration}
\end{table}

In addition to enumerating the allowed limiting Hodge diamonds one can also characterize the associated nilpotent elements $N$.  
In order to do that one has to classify conjugacy classes of nilpotent elements that are invariant under basis transformations. 
Recall that $N$ is an element of the Lie algebra $\frak{sp}(2h^{2,1} +2,\bbR)$ as discussed after 
\eqref{Npres_S}. The Lie group $\mathrm{Sp}(2h^{2,1} +2,\bbR)$ acts on its Lie algebra $\frak{sp}(2h^{2,1} +2,\bbR)$ via the adjoint action, 
i.e.~$N \mapsto gNg^{-1}$ for $g\in\mathrm{Sp}(2h^{2,1} +2,\bbR)$. Classifying the conjugacy classes of nilpotent elements 
obtained by this equivalence is a well-known problem and it was shown in \cite{Djokovic1982,Collingwood1993}  that it is equivalent to classifying 
signed Young diagrams. While not very involved, we will refrain from presenting the details of this classification here, but 
only list the relevant result in the fourth column of table \ref{HD:enumeration}. The result is that 
to each singularity type I$_a$, II$_b$, III$_a$, IV$_d$ there is a unique associated signed Young diagram which characterizes 
the form of $N$ and $\eta$. This information allows one, for example, to associate a simple normal form of $N$, $\eta$ to the 
singularity type. In order that the reader gets an intuition for such normal forms, we give a possible choice in table \ref{associateNeta}. 
In order to obtain the complete $N$, $\eta$ one needs to use the building blocks of table \ref{associateNeta} and combine them in the canonical 
way to a higher-dimensional matrix. 

\ytableausetup
 {mathmode, boxsize=1.2em}
\newcolumntype{A}{>{\centering\arraybackslash} m{.3\linewidth} }

\begin{table}
\begin{center}
{\small
\begin{tabular}{|c|c|c|}
\hline
    \rule[-.1cm]{0cm}{0.5cm} (signed) Young diagram &  $N$  & $\eta$ \\
   \hline
   \hline
  \begin{ytableau}
    \ \\
    \ 
  \end{ytableau}
  &    \rule[-.5cm]{0cm}{1.2cm} $\left( \begin{array}{cc} 0 & 0 \\ 0 & 0\end{array} \right)$  & $\left( \begin{array}{cc} 0 & -1 \\ 1 & 0\end{array} \right)$\\
   \hline
  \begin{ytableau}
    + & -
  \end{ytableau}
  & \rule[-.5cm]{0cm}{1.2cm}  $\left( \begin{array}{cc} 0 & 0 \\ 1 & 0\end{array} \right)$  & $\left( \begin{array}{cc} 0 & -1 \\ 1 & 0\end{array} \right)$  \\
    \hline
  \begin{ytableau}
    - & +
  \end{ytableau}
  &  \rule[-.5cm]{0cm}{1.2cm}  $\left( \begin{array}{cc} 0 & 0 \\ 1 & 0\end{array} \right)$  & $\left( \begin{array}{cc} 0 & 1 \\ -1 & 0\end{array} \right)$  \\    
       \hline
  \begin{ytableau}
    \ & \ & \ \\
    \ & \ & \
  \end{ytableau}
  &  \rule[-1.3cm]{0cm}{2.8cm}   $\left( \begin{array}{cccccc} 0 & 0 & 0 & 0 & 0 & 0 \\ \ 1 & 0 & 0 & 0 &0 & 0\\0 & \ 1 & 0 &0 & 0 & 0\\ 0 & 0 &0 & 0 & 0 & 0\\ 0 &0 & 0 & -1 & 0 & 0\\ 0 & 0 & 0 & 0 & -1 & 0 \end{array} \right)$
    &  $\left( \begin{array}{cccccc} 0 & 0 & 0 & 0 & 0 &- 1\\ 0 & 0 & 0 & 0 &-1 & 0\\0 & 0 & 0 &- 1 & 0 & 0\\0 & 0 & \ 1 & 0 & 0 & 0\\ 0 & \ 1 & 0 & 0 & 0 & 0\\ \ 1 & 0 & 0 & 0 & 0 & 0 \end{array} \right)$ \\
    \hline
  \begin{ytableau}
    - & + & - & +
  \end{ytableau}
    & \rule[-0.9cm]{0cm}{2.0cm}  $\left( \begin{array}{cccc} 0 & 0 & 0 & 0  \\ \ 1 & 0 & 0 & 0 \\0 & \ 1 & 0 &0 \\ 0 & 0 &-1 & 0 
     \end{array} \right)$
    &  $\left( \begin{array}{cccc}    0 & 0 & 0 &\ 1\\   0 & 0 & \ 1 & 0 \\  0 & - 1 & 0 & 0  \\   - 1 & 0 & 0 & 0 \end{array} \right)$    \\    
    \hline
\end{tabular}
}
\caption{List of all relevant signed Young diagrams and their associated $N$, $\eta$ in some normal form. The complete signed Young diagram and  
$N$, $\eta$ that classify a singularity type are composed out of these building blocks.} \label{associateNeta}
\end{center}
\end{table}

We should stress that in many of the applications and explicit computations the normal forms of table \ref{associateNeta} play 
no role. Rather, it is often useful to have a more basis independent way to determine the singularity type for a given $N$, $\eta$. 
In the last column of table \ref{HD:enumeration} we have included such a distinguishing criterion. To begin with we note that the 
ranks of $N^k$, $k=1,2,3$, often differ for the various singularity types, as one deduces from \eqref{NI=I}, \eqref{I=NP} and the polarization 
condition \eqref{Sj-orthogonality1-I}, \eqref{polarizingSj}. However, there are $(h^{2,1} - 1)$ pairs of I$_a$ and II$_{a - 2}$ that cannot be 
distinguished by only comparing the ranks. In this case one can use again the polarization condition to show that 
these cases differ by the sign of the eigenvalues of $\eta N$. Taking this feature into account indeed the singularity 
type for a given $N$, $\eta$ is uniquely fixed. Clearly, the same conclusion can be reached from using the normal 
forms combining table \ref{HD:enumeration} and \ref{associateNeta}.

\subsection{A Classification of allowed Singularity Enhancements} \label{sec:class_enh}

Having classified the allowed singularity types, we next turn to the discussion of 
allowed singularity enhancements. More precisely, let us assume that on 
the locus $\Delta_1^\circ$ the singularity type is specified by $\mathrm{Type}_a(\Delta_1^\circ)$. We then want 
to answer the question to which singularity types $\mathrm{Type}_b(\Delta_{12'}^\circ)$ this type can enhance further 
when moving to $\Delta^\circ_{12}$, i.e.~we consider 
\beq
   \mathrm{Type}_a(\Delta_1^\circ) \quad \longrightarrow \quad \mathrm{Type}_{\hat a}(\Delta_{12}^\circ)\ ,
\eeq  
where in the following we will denote the allowed enhancements by an arrow. 
It was argued in \cite{Kerr2017} that imposing consistency of the polarization conditions 
\eqref{Sj-orthogonality1-I},  \eqref{polarizingSj} on $\Delta_1^\circ$ and $\Delta_{12}^\circ$ leads to non-trivial 
constraints on possible enhancements. The resulting rules are shown in table \ref{sing_enhancements}, and their derivation is outlined later in this section and in appendix \ref{sec:derpol}. It should be stressed that the enhancement rules are 
actually general and apply to any higher intersection and not only to the case of two 
divisors $\Delta_1$, $\Delta_{2}$ intersecting in $\Delta_{12}^\circ$.

\begin{table}[h!]
\begin{center}
\begin{tabular}{|cl|}
\hline
    \rule[-.16cm]{0cm}{0.6cm} \qquad starting singularity type \qquad\qquad& enhance singularity type \qquad \qquad \\
    \hline \hline 
    \multirow{4}{*}{I$_a$} 
    & \rule[-.2cm]{0cm}{0.8cm} I$_{\hat a}$ for $a\leq \hat a$ \\
    &\rule[-.2cm]{0cm}{0.6cm} II$_{\hat b}$ for $a \leq \hat b $, $a < h^{2,1}$\\
    &\rule[-.2cm]{0cm}{0.6cm} III$_{\hat c}$ for $a \leq \hat c$, $a < h^{2,1}$\\
    &\rule[-.4cm]{0cm}{0.8cm} IV$_{\hat d}$ for $a < \hat d$, $a < h^{2,1}$ \\
    \hline
     \multirow{3}{*}{II$_b$} 
    &\rule[-.2cm]{0cm}{0.8cm} II$_{\hat b}$ for $b\leq \hat b$\\
    &\rule[-.2cm]{0cm}{0.6cm} III$_{\hat c}$ for $2 \leq b \leq \hat c + 2$\\
    &\rule[-.4cm]{0cm}{0.8cm} IV$_{\hat d}$ for $1 \leq b \leq \hat d-1$ \\      
    \hline
     \multirow{2}{*}{III$_c$} 
    &\rule[-.2cm]{0cm}{0.8cm} III$_{\hat c}$ for $c\leq \hat c$ \\
    &\rule[-.4cm]{0cm}{0.8cm} IV$_{\hat d}$ for $c + 2 \leq \hat d$ \\
    \hline 
       IV$_d$ & \rule[-.4cm]{0cm}{1cm} IV$_{\hat d}$ for $d\leq \hat d$  \\
    \hline
\end{tabular}
\begin{picture}(0,0)
\put(-230,67){\begin{tikzpicture}
\draw [->] (0,0) -- (2,.7);
\end{tikzpicture}}
\put(-230,67){\begin{tikzpicture}
\draw [->] (0,0) -- (2,.1);
\end{tikzpicture}}
\put(-230,54){\begin{tikzpicture}
\draw [->] (0,-.2) -- (2,-.66);
\end{tikzpicture}}
\put(-230,40){\begin{tikzpicture}
\draw [->] (0,-.2) -- (2,-1.15);
\end{tikzpicture}}
\put(-230,-5){\begin{tikzpicture}
\draw [->] (0,0) -- (2,.5);
\end{tikzpicture}}
\put(-230,-8){\begin{tikzpicture}
\draw [->] (0,0) -- (2,-.1);
\end{tikzpicture}}
\put(-230,-24){\begin{tikzpicture}
\draw [->] (0,0) -- (2,-.66);
\end{tikzpicture}}
\put(-230,-62){\begin{tikzpicture}
\draw [->] (0,0) -- (2,.25);
\end{tikzpicture}}
\put(-230,-71){\begin{tikzpicture}
\draw [->] (0,0) -- (2,-.32);
\end{tikzpicture}}
\put(-230,-102){\begin{tikzpicture}
\draw [->] (0,0) -- (2,0);
\end{tikzpicture}}
\end{picture}
\caption{List of all allowed enhancements obtained by imposing consistency of the polarization conditions \eqref{Sj-orthogonality1-I},  \eqref{polarizingSj}. These have been shown in \cite{Kerr2017} and the details of their derivation are given in appendix \ref{sec:derpol}.} \label{sing_enhancements}
\end{center}
\end{table}

Using the enhancement rules of table \ref{sing_enhancements} one obtains an instructive picture of the singularity 
structure of a Calabi-Yau threefold $Y_3$ for a given $h^{2,1}$. In figure \ref{Graphs23in} we display the two cases $h^{2,1}=2$ 
and $h^{2,1}=3$. It is interesting to note that, as of now, it is not known whether all allowed enhancements of 
table \ref{sing_enhancements} are actually realized in some Calabi-Yau threefold.

In order to deduce the allowed enhancements one has to use a substantial amount of mathematics. We will limit ourselves 
to some essential facts and refer the reader to appendix \ref{sec:derpol}, where further details on the underlying 
constructions are presented. 
The main focus of this investigation lies on the primitive parts $P^{p,q} \subset I^{p,q}$ that were defined in \eqref{def-Ppq}.  
We note that by using \eqref{def-Ppq} one deduces that $I^{3,j}= P^{3,j}$ and $I^{j,3}= P^{j,3}$ for $j=0,1,2,3$. 
Furthermore, we can apply \eqref{I=NP} to infer that the $I^{p,q}$ split into the $P^{p,q}$ as 
\beq \label{primitive-diamond}
 \begin{array}{ccccccccc} &&& P^{3,3}&&\\ 
&& P^{3,2} & &  P^{2,3} \\ 
& P^{3,1}  & &  P^{2,2} \oplus N P^{3,3} & & P^{1,3}\\
 P^{3,0} \hspace*{.2cm} & & P^{2,1} \oplus N P^{3,2} & &  P^{1,2} \oplus N P^{2,3}&&  \hspace*{.2cm} P^{0,3}\\
&N P^{3,1}  & & \hspace*{-.2cm} N P^{2,2} \oplus N^2 P^{3,3} & &   N P^{1,3}\\
&&N^2 P^{3,2} \hspace*{-.2cm} & &   N^2 P^{2,3} \\
&&&N^3 P^{3,3}&& \\  \end{array}\ .
\eeq 
The primitive subspaces are thus given by 
\begin{align} \label{def-Pi}
&P^{6} =  P^{3,3}\ , \qquad
&&P^{5} = P^{3,2} \oplus  P^{2,3} \ ,\\
&P^{4} = P^{3,1}  \oplus  P^{2,2} \oplus P^{1,3}\ ,
&&P^{3} =  P^{3,0}\oplus P^{2,1} \oplus  P^{1,2} \oplus P^{0,3}\ ,\nn 
\end{align}
where the single superscript on $P^r$ is the weight $r=p+q$ of the contained $P^{p,q}$.
One of the most fundamental results about this construction is that each $P^j$ with $j=3, \ldots ,6$ 
can be shown to define a pure Hodge structure 
of weight $j$. Recall that also the decomposition \eqref{H3-cohom-split} on a smooth 
$Y_3$ provided a pure Hodge structure, which was the starting point of the 
construction of the splittings relevant at the singularities. The 
main idea for looking at enhancements moving from $\Delta_1^\circ$ to an 
intersection $\Delta_{12}^\circ$ is to view $P^j(\Delta_1^\circ)$ as defining the starting 
pure Hodge structures that then splits into even finer mixed Hodge structures. 
Representing the mixed Hodge structures by Deligne splittings, one thus has 
\beq
    P^j(\Delta_1^\circ) \quad \longrightarrow \quad \left[I^{p,q}\right]^j(\Delta_{12}^\circ) \ \text{with}\ {0\leq p+q \leq 2j}\ .
\eeq
One can rearrange the spaces  $\left[I^{p,q}\right]^j(\Delta_{12}^\circ)$
to form the Deligne splitting $I^{p,q}(\Delta_{12}^\circ)$ of the enhanced type. 
To identify the rules when this is possible makes it necessary to use the 
full technology of the Sl(2)-orbit theorem \cite{CKS} as done 
in \cite{Kerr2017} and outlined in appendix \ref{sec:derpol}. 

As a rather simple application of the classification we can evaluate the 
growth of the  K\"ahler potential $e^{-K}$ for the 10 possible enhancements 
of table \ref{sing_enhancements}. Using the general result \eqref{general_growth_K} and 
the link \eqref{linkdi-sing} of $d_i$ to the singularity type one reads of table 
\ref{tab:eKgrowth}. 
\begin{table}[h!]
\centering
\begin{tabular}{|c|c|c|c|}
\hline
\rule[-0.3cm]{0cm}{0.8cm} Type & $d_1$ & $d_2$ & Leading behaviour of $e^{-K}$ \\
\hline\hline
\rule[-0.18cm]{0cm}{0.6cm}$\mathrm{I}_a \rightarrow \mathrm{I}_b $  & 0 & 0 & const. or $e^{-\Im t}$ \\
\hline
\rule[-0.18cm]{0cm}{0.6cm}$\mathrm{II}_a \rightarrow \mathrm{II}_b $  & 1 & 1 & ${\I\;t^1} $\\
\hline
\rule[-0.18cm]{0cm}{0.6cm}$\mathrm{III}_a \rightarrow \mathrm{III}_b$ & 2 & 2 & ${\left(\I\;t^1 \right)^2} $\\
\hline
\rule[-0.18cm]{0cm}{0.6cm}$\mathrm{IV}_a \rightarrow \mathrm{IV}_b$ & 3 & 3 &${\left(\I\;t^1 \right)^{3}} $\\
\hline
\rule[-0.18cm]{0cm}{0.6cm}$\mathrm{I}_a \rightarrow \mathrm{II}_b$ & 0 & 1 & ${\I\;t^2} $\\
\hline
\rule[-0.18cm]{0cm}{0.6cm}$\mathrm{I}_a \rightarrow \mathrm{III}_b$ & 0 & 2 & ${\left(\I\;t^2 \right)}^2 $\\
\hline
\rule[-0.18cm]{0cm}{0.6cm}$\mathrm{I}_a \rightarrow \mathrm{IV}_b$ & 0 & 3 & ${\left(\I\;t^2 \right)^{3}} $\\
\hline
\rule[-0.18cm]{0cm}{0.6cm}$\mathrm{II}_a \rightarrow \mathrm{III}_b$ & 1 & 2 & ${\left(\I\;t^1 \right) \left(\I\;t^2 \right)} $\\
\hline
\rule[-0.18cm]{0cm}{0.6cm}$\mathrm{II}_a \rightarrow \mathrm{IV}_b$ & 1 & 3 & ${\left(\I\;t^1 \right)\left(\I\;t^2 \right)^2} $\\
\hline
\rule[-0.18cm]{0cm}{0.6cm}$\mathrm{III}_a \rightarrow \mathrm{IV}_b$ & 2 & 3 & ${\left(\I\;t^1 \right)^2 \left(\I\;t^2 \right)} $\\
\hline
\end{tabular}
\caption{Leading growth terms of $e^{-K}$ when approaching the singular locus $t^1=t^2=\im\infty$
obtained by using \eqref{linkdi-sing} and \eqref{general_growth_K}.}
\label{tab:eKgrowth}
\end{table}

\subsection{On the Classification of Infinite Distance Points} \label{sec:class_infinitedistanceP}

Having introduced a classification of singularities and singularity enhancements 
arising in general Calabi-Yau threefold geometries, we next turn to the 
discussion of infinite distance points. To define such points let us pick 
a point $P$ in the complex structure moduli space $\cM_{\rm cs}$ including $\Delta$. 
$P$ is said to be an \textit{infinite distance point}, if the length measured 
with the Weil-Petersson metric $g_{\rm WP}$ of \textit{every} path to this point is infinite.  
Accordingly, we would call $P$ a \textit{finite distance point} if there exists at least one 
path to this point that has finite length in the metric $g_{\rm WP}$. In the following we will discuss the classification 
of finite and infinite distance points using the classification of singularities presented in 
subsection \ref{sec:sing_class}.

To begin with, we note that any two points $P,\ Q$ that are not on the discriminant locus 
$\Delta$ are connected by a path of finite distance in the Weil-Petersson metric. 
This implies that in order to have an infinite distance point,
at least one of the points has to lie on $\Delta$ and we chose this to be $P$. One 
then has to distinguish two situations: (1) $P$ lies on a divisor $\Delta_k$ away 
from any intersection locus, (2) $P$ lies on an intersection locus $\Delta_{k k'}$ 
of two (or more) divisors $\Delta_k$ and $\Delta_{k'}$. In the following we will discuss the two cases in 
turn. 

Considering a point $P$ on a divisor $\Delta_k$ that does not lie on any intersection with 
other divisors corresponds to considering a one-parameter degeneration of the Calabi-Yau manifold $Y_3$.
In this case one can prove a simple criterion when such a point is at infinite distance. 
More precisely, it was shown in \cite{Wang1997} that a point on $\Delta_k^\circ$ is at finite distance if and 
only if $N F^3(\Delta_k^\circ) = 0$. Using the nilpotent orbit \eqref{nilp-orbit} this is nothing else then the condition
\beq
    P \in \Delta_k^\circ \ \text{at finite distance} \quad \Longleftrightarrow \quad N \mathbf{a}_0 = 0\ . 
\eeq
It is not difficult to translate this condition to the statement that 
the singularity on $\Delta_k^\circ$ is Type I. Thus one concludes
\bea \label{one-modulus}
   \text{$P \in \Delta^\circ_k$ is finite distance point} \quad &\Longleftrightarrow& \quad \text{Type I}\ ,\\
   \text{$P \in \Delta^\circ_k$ is infinite distance point} \quad &\Longleftrightarrow& \quad \text{Type II}\,,\ \text{Type III}\,,\ \text{Type IV}\ . \nn
\eea  
This shows that the classification of singularities is sufficiently fine to separate infinite and finite distance points $\Delta^\circ_k$. In fact, 
it clearly contains more information, since the index on the type, as introduced in \eqref{def-types}, is not relevant for this distinction. 

Let us now turn to the more involved case that the considered point lies on an intersection locus $\Delta^\circ_{kk'}$. 
This implies that one is not considering a one-parameter degeneration and path-dependence becomes 
a very important issue. It is currently not known an equivalent condition to \eqref{one-modulus} is true. 
The directions that are not difficult to prove are 
\bea \label{two-moduli}
   \text{$P \in \Delta^\circ_{kk'}$ is finite distance point} \quad &\Longleftarrow& \quad \text{Type I}\ ,\\
   \text{$P \in \Delta^\circ_{kk'}$ is infinite distance point} \quad &\Longrightarrow& \quad \text{Type II}\,,\ \text{Type III}\,,\ \text{Type IV}\ . \nn
\eea 
To see this we note that in order to show that Type I implies that the point is finite distance, it suffices to find a single 
path that is at finite distance. This path can be easily chosen such that the question reduces to a one-parameter degeneration 
with nilpotent operator $N_k+N_{k'}$ and one can use \eqref{one-modulus}. Clearly, the statement in the 
second line in \eqref{two-moduli} is just equivalent to the statement in the first line. 
Note that \eqref{two-moduli}, and its obvious higher-dimensional generalizations, can 
also be stated as \cite{Wang2015}
\beq \label{Patinfitedistance}
      \text{$P$ is infinite distance point} \quad \Longrightarrow \quad \exists\, N_i \mathbf{a}_0 \neq 0 \ ,
\eeq
where $\mathbf{a}_0$ is associated via \eqref{nilp-orbit} to the point on $\Delta_{i_1 \ldots i_l}$.
Attempting to prove a one-to-one correspondence as in \eqref{one-modulus}
requires to carefully deal with a path dependence.\footnote{It was conjectured 
in \cite{Wang2015} that a statement equivalent to \eqref{one-modulus} is true.} 
We believe that this is a very important problem that, however, goes 
beyond the scope of the current paper.

\subsection{The Large Complex Structure and Large Volume Point} \label{sec:lcs_point}

A prime example of an infinite distance point in complex structure moduli space  
is the so-called \textit{large complex structure point}. To begin 
with, we first have to more formally define such points. General definitions 
have been discussed in \cite{Morrison1993}. However, with the classification of singularities 
presented in subsection \ref{sec:sing_class}, we can give a very elegant general definition. 
We call a point a large complex structure point if it is a type IV$_{h^{2,1}(Y_3)}$ point on $\Delta$
that arises from the intersection of $h^{2,1}(Y_3)$ divisors $\Delta_I$, $I=1, \ldots ,h^{2,1}(Y_3)$ 
each being of type II, III, or IV. By this we mean that a generic point, i.e.~a point on $\Delta_I^\circ$, on these $\Delta_I$
has these types. While we did not show the equivalence of this definition with the ones in \cite{Morrison1993}, we will 
see that it is in perfect match with the expectations from mirror symmetry. 

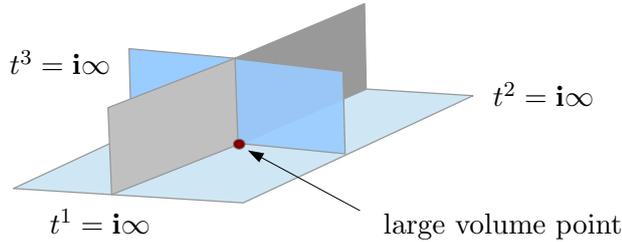
\begin{figure}[h!]
\vspace*{.5cm}
\begin{center} 
\includegraphics[width=7.5cm]{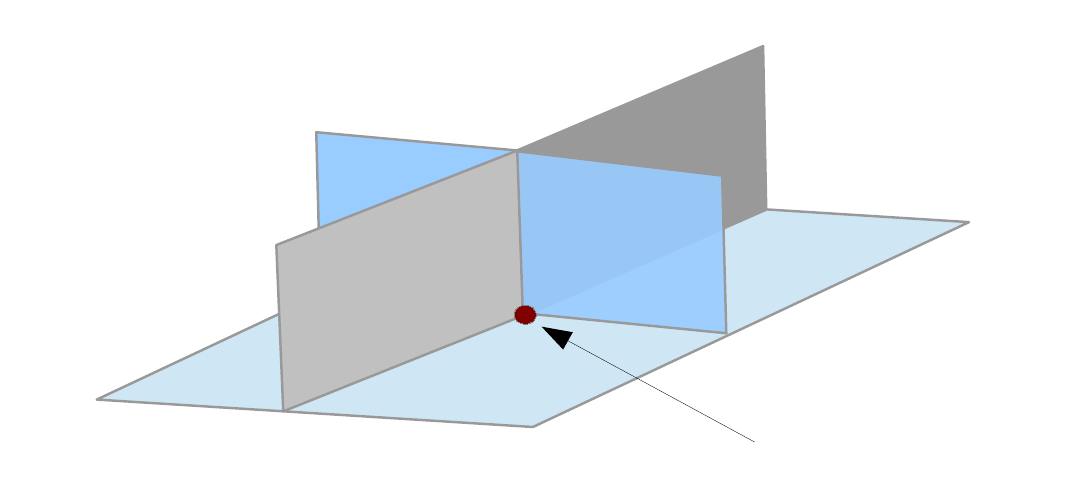} 
\vspace*{-.2cm}
\end{center}
\begin{picture}(0,0)
\put(316,68){$t^2 = \im \infty$}
\put(150,20){$t^1=\im\infty$}
\put(135,80){$t^3=\im\infty$}
\put(275,20){large volume point}
\end{picture}
\caption{The large volume point arises on the discriminant locus of the K\"ahler moduli space
 at the intersection of  $h^{1,1}(\tilde Y_3)=3$ divisors $t^I = \im \infty$.} \label{largevolumepoint}
\end{figure}

The large complex structure points are of crucial importance in the first 
mirror symmetry proposals \cite{Candelas1991a}. More precisely, 
mirror symmetry states that the large complex structure point is mapped to large 
volume point by identifying complex structure and complexified K\"ahler structure deformations
in Type IIA and Type IIB compactifications. One thus has a mirror Calabi-Yau threefold 
geometry $\tilde Y_3$ associated to $Y_3$. On this mirror one defines the complexified 
K\"ahler coordinates $t^I$, $I=1,\ldots, h^{1,1}(\tilde Y_3)$ by 
\beq
    B_2 + \im J = t^I \omega_I \ ,
\eeq
where $B_2$ is the NS-NS two-form field and $J$ is the K\"ahler form.  The large volume point 
is given by 
\beq
    t^I  \ \rightarrow \ \im \infty \ , \quad I=1,\ldots, h^{1,1}(\tilde Y_3)\ . 
\eeq
In other words the large volume point arises from the intersection of $h^{1,1}(\tilde Y_3)$ 
divisors in the K\"ahler moduli space that are individually given by $t^I = \im \infty$. 
We depict this in figure \ref{largevolumepoint}. 
Clearly, in order 
to consider the complete mirror moduli space to $\cM_{\rm cs}$ we have to investigate the 
allowed values of $t^I$. These are encoded by the K\"ahler cone, which we will briefly introduce 
next. 

In order to define the K\"ahler cone it is easiest to first introduce the dual Mori cone. 
The Mori cone is spanned by equivalence classes of the irreducible, proper curves on
$\tilde Y_3$, i.e.~one can form positive linear combinations $\sum_i a_i [C^i]$, $a_i>0$ of homology classes $[C^i]$ of such 
curves. The dual cone is obtained by 
\beq
     J \in H^{1,1}(Y_3):\quad \int_{C} J \geq 0 \ ,
\eeq
for all curves $C$ in the (closure of the) Mori cone. Hence, when picking a K\"ahler form inside the K\"ahler cone 
one ensures that all proper curves have positive volume. For the following discussion it is important 
to point out that the K\"ahler cone is in general not simplicial. Roughly speaking this implies that 
we cannot represent the cone by picking $h^{1,1}(Y_3)$ generating two-forms $\omega_I$ and 
consider the linear combination $a^I \omega_I$, $a^I \geq 0$. 
In order to connect to the discussion of the previous subsections, we 
will now make a crucial simplifying assumption. More precisely, we will 
consider only situations that admit a \textit{simplicial K\"ahler cone}. 
While many of our formulas are valid generally, this assumption will 
help us to interpret our results more easily. 

Our starting point will be the local form of the mirror periods at the large volume point.
These can be computed in various different ways, for example, by evaluating the central charges for a 
set of D0-, D2-, D3-, D4-branes by using the $\Gamma$-class (see, e.g.~refs.\cite{Cota2018, Gerhardus2016, Bloch:2016izu, Iritani2011, Katzarkov2008}). 
For these branes one can introduce an appropriate K-theory basis
\begin{equation} \label{basis_cO}
  (\cO_{\tilde Y_3}, \cO_{D_I}, \mathcal{C}^J, \cO_p),
\end{equation}
where $D_J$ are $h^{1,1}(\tilde Y_3)$ divisors, $p$ are points and the K-theory basis $\mathcal{C}^J := \iota_!\cO_{C^J}\big(K^{1/2}_{C^J}\big)$ are for $h^{1,1}(\tilde Y_3)$ curves $C^I$ (see \cite{Gerhardus2016}, section 2.3 for their precise definition).

We require that the curves and divisors are dual, i.e.~that
$C^J \cdot D_I =\delta_I^J$, and that the Poincar\'e dual two-forms $\omega_I$ to $D_I$ span the K\"ahler cone.
Furthermore, we define 
\beq
    K_{IJK}  = \int_{\tilde Y_3} \omega_I  \wedge \omega_J \wedge \omega_K\ , \qquad 
    b_I     = \frac{1}{24} \int_{\tilde Y_3} \omega_I \wedge c_2(\tilde Y_3)\ ,
\eeq
where $K_{IJK}$ are the triple intersection numbers and $c_2(\tilde Y_3)$ is the second 
Chern class of $\tilde Y_3$. Using these abbreviations one finds the mirror period vector 
\begin{equation}
  \Pi^{\Omega}(t^I) =
  \begin{pmatrix}
    1\\
    t^I\\
    \frac{1}{2}K_{IJK} t^J t^K + \frac{1}{2}K_{IJJ} t^J - b_I + \cO(e^{2\pi \im t})\\
    \frac{1}{6}K_{IJK} t^I t^J t^K - (\frac{1}{6} K_{III} + b_I) t^I + \frac{\im \zeta(3) \chi(\tilde{Y}_3)}{8\pi^3} + \cO(e^{2\pi \im t})
  \end{pmatrix},
\end{equation}
where $\chi(\tilde Y_3) = \int_{\tilde Y_3} c_3(\tilde Y_3)$ is the Euler number of $\tilde Y_3$. 

Having determined the local form of the periods near the large volume point, 
we use them to compute the monodromy matrix $T_A$.
Note that by \eqref{monodromy-trafo} the action of $T_A$ is induced by sending $t^A \mapsto t^A - 1$, when taking $z^A = e^{2\pi \im t^A}$.
Explicitly we find the $(2h^{1,1}+2) \times (2h^{1,1}+2)$-matrix 
\beq
  T_A  = \left(
  \begin{array}{cccc}
              1            & 0                              & 0            & 0\\
   -\delta_{AI} & \delta_{IJ}                    & 0            & 0\\
    0            & -K_{AIJ}                       & \delta_{IJ}  & 0\\
     0            & \frac{1}{2}(K_{AAJ} + K_{AJJ}) & -\delta_{AJ} & 1
  \end{array} \right),
\eeq
where the upper left corner corresponds to the element $\cO_{\tilde Y_3}-\cO_{\tilde Y_3}$ in the basis \eqref{basis_cO}.
It is interesting to point out that due to the basis choice \eqref{basis_cO} 
the $T_A$ only depends on the intersection numbers with no $b_I$ appearing. 
Given these monodromies one checks that they are unipotent and we can determine the log-monodromies $N_A$ 
by simply evaluating $N_A = \log T_A$ following their definition \eqref{def-Nk}. 
We thus find  
\beq
  N_A  = \left(
  \begin{array}{cccc}
    0                   & 0                  & 0            & 0\\
    -\delta_{AI}        & 0                  & 0            & 0\\
    -\frac{1}{2}K_{AAI} & -K_{AIJ}           & 0            & 0\\
    \frac{1}{6}K_{AAA}  & \frac{1}{2}K_{AJJ} & -\delta_{AJ} & 0
  \end{array}\right)\ . \label{lcsNa} 
\eeq
This rather simple expression determines all large complex volume log-monodromies 
about single divisors in the discriminant locus of the K\"ahler moduli space 
specified by $t^A = \im \infty$.  
As discussed above in \eqref{postive-sum}, log-monodromies around intersecting 
divisors are determined by positive linear combinations of these $N_A$. 
For example, the log-monodromies relevant for the discriminant locus given by 
$t^A= \im \infty$, $t^{A'}= \im \infty$ are given by $a N_A + b N_{A'}$
with $a,b>0$. 

In order to also classify the corresponding singularity types using table \ref{HD:enumeration}, 
we still need to determine the polarisation $\eta$. This can be done by evaluating 
the negative of the Mukai pairing \cite{Cota2018, Bloch:2016izu, Iritani2011, Katzarkov2008}.
On the K-theory space the Mukai pairing of branes $\xi$ and $\xi'$ is defined by
\begin{equation}
  \langle \xi, \xi' \rangle = \int_{\tilde{Y}_3} \mathrm{ch}(\xi^\vee) \mathrm{ch}(\xi') \mathrm{Td}(\tilde{Y}_3)\, ,
\end{equation}
where $-^\vee$ is the dual operation, $\mathrm{ch}(-)$ is the Chern character and $\mathrm{Td}(-)$ is the Todd class. In the basis \eqref{basis_cO} one finds
\begin{equation}
  \eta = \left(
  \begin{array}{cccc}
    0                         & -\frac{1}{6}K_{JJJ} - 2b_J     & 0                & -1\\
    \frac{1}{6}K_{III} + 2b_I & \frac{1}{2}(K_{IIJ} - K_{IJJ}) & \delta_{IJ}      &  0\\
    0                         & -\delta_{IJ}                   & 0                &  0\\
    1                         & 0                              & 0                &  0
  \end{array}\right)\ ,
  \label{eta-lcs}
\end{equation}
and it always satisfies $\det \eta = 1$. The inverse of $\eta$ is also computed
\begin{equation}
  \eta^{-1} = \left(
  \begin{array}{cccc}
    0 & 0           & 0                              & 1\\
    0 & 0           & -\delta_{IJ}                   & 0\\
    0 & \delta_{IJ} & \frac{1}{2}(K_{IIJ} - K_{IJJ}) & -\frac{1}{6}K_{III} - 2b_I \\
    1 & 0           &  \frac{1}{6}K_{JJJ} + 2b_J     & 0
  \end{array}\right)\ .
\end{equation}
These expressions now depends both on the intersection numbers, as well as the 
second Chern class. As a side remark, let us note that the complete set of $N_A$'s 
together with $\eta$ and the Hodge numbers $h^{2,1}(\tilde Y_3)$, $h^{1,1}(\tilde Y_3)$ contain the relevant information for Wall's classification 
theorem of homotopy types of complex compact Calabi-Yau threefolds \cite{Wall1966}. It is interesting 
to combine this fact with the following classification 
of singularities.

Given the explicit forms \eqref{lcsNa} and \eqref{eta-lcs} of $N_A,\eta$ it is now 
straightforward to determine the singularity type using 
the last column table \ref{HD:enumeration}. 
Due to the lower-triangular form of $N_A$ its powers $N_A^2$ and $N_A^3$ are 
easily computed. We immediately see that $N_A^3$ is only non-zero if $K_{AAA}$ is 
non-vanishing. This is thus precisely the condition for a type IV$_d$ singularity. Similarly, 
if and only if $K_{AAI}$ is non-vanishing for one or more $I$ we find that $N^2_A$ is non-vanishing.
Hence, the $N_A$ is of type III$_c$ if $K_{AAA}=0$ and $K_{AAI}$ non-vanishing for some $I\neq A$.  
The precise type III$_c$ and IV$_d$ are now determined by evaluating the rank of the matrix 
$K_{AIJ}$ with the result listed in table \ref{special_N_class}. 

It remains to discuss the cases I$_a$ and II$_b$ that occur if all $K_{AAI}=0$. 
As we have discussed in 
subsection \ref{sec:sing_class} they can, in general, only be distinguished if we also consider $\eta$. 
In fact, we can compute $\eta N_A$ and determine its number of positive and negative eigenvalues. 
Explicitly, we find that 
\begin{equation}
  \eta N_A = \left(
  \begin{array}{cccc}
    2b_A                       & -\frac{1}{2}K_{AJJ}     &  \delta_{AJ}              & 0\\
   \frac{1}{2} ( K_{IAA} - K_{IIA} )& - K_{AIJ} &0 &  0\\
    \delta_{AI}                     &0                 & 0                &  0\\
    0                         & 0                              & 0                &  0
  \end{array}\right)\ ,
  \label{Neta-lcs}
\end{equation}
where we  need to impose $K_{AAI}=0$ for all $I$. It can be now easily seen that this matrix has positive eigenvalues. In fact, evaluating 
$V^\mathrm{T} \eta N_A V = 2$ for $V = (1,0, \ldots ,0, (1-b_A) \delta_{AI}, 0)^\mathrm{T}$ we find a positive direction. Hence, the case I$_a$ is actually never realized for the $N_A$, $\eta$
given in \eqref{lcsNa}, \eqref{eta-lcs}. We thus conclude that we can distinguish also the precise type II$_b$ by evaluating the 
rank of the matrix $K_{AIJ}$ as listed in table \ref{special_N_class}.

\begin{table}[!h]
\centering
\begin{tabular}{| c | c | c |c| }
\hline
name & $\text{rank}(K_{AAA})$ & $\text{rank}(K_{AAI})$ & $\text{rank}(K_{AIJ})$ \\ 
\hline \hline
II$_b$ & 0 & 0 & $b $  \\
III$_c$ &  0 &1 & $c+2$ \\
IV$_d$ & 1 &  1 & $d  $  \\
  \hline
\end{tabular}
\caption{This table list the conditions on $N_A,\,\eta$ given in \eqref{lcsNa} and \eqref{eta-lcs} 
that ensure a certain singularity type on the discriminant divisor $t^A =\im \infty$ for a single coordinate. 
Note that $\text{rank}(K_{AAA})$ and $\text{rank}(K_{AAJ})$ are either $0$, $1$ depending 
on whether these quantities are trivially zero or non-zero.} \label{special_N_class}
\end{table}

To conclude this section, let us note that the large volume point 
$t^A=\im\infty$ for all $A=1,\ldots, h^{1,1}(\tilde{Y}_3)$ has precisely the properties mirror dual to a large complex structure point 
defined at the beginning of this subsection. To see this, let us first show that it is a point of type IV$_{h^{1,1}}$. 
In order to do that we have to analyse the sum of all $N_A$ with positive coefficients. A convenient choice 
is to pick the K\"ahler coordinates $v^A = \I \, t^A$, which are positive in a simplicial K\"ahler cone. Hence we 
consider 
\beq
   N = \sum_A v^A N_A = \left(
  \begin{array}{cccc}
    0                   & 0                  & 0            & 0\\
    -v^I        & 0                  & 0            & 0\\
    -\frac{1}{2} v^A K_{AAI} & -v^A K_{AIJ}           & 0            & 0\\
    \frac{1}{6}v^A K_{AAA}  & \frac{1}{2}K_{AJJ} & -v^J & 0
  \end{array}\right)\ .
\eeq 
If we now compute $N^3$, we simply find a matrix which only has a single entry proportional to  
the volume $\frac{1}{6}\, K_{IJK} v^I v^J v^K$. Hence, the rank of $N^3$ is $1$ and we conclude from the last 
column in table \ref{special_N_class} that the singularity is type IV$_d$. 
To determine $d$ we need to evaluate the rank of $N$ itself. However, the contraction $v^A K_{A IJ}$ is crucial in defining the metric 
on K\"ahler moduli space and is full rank \cite{Candelas1991}. 
So indeed, we find that the singularity $t^A = \im \infty$ is of type IV$_{h^{1,1}(\tilde Y_3)}$. Furthermore, all the intersecting 
divisors have type II, III, or type IV as discussed above.


\section{Charge orbits and the Swampland Distance Conjecture} \label{sec:infinite_towers}

In this section we analyse the Swampland Distance Conjecture (SDC) using the powerful 
geometric tools about the complex structure moduli space introduced so far. To begin 
with, let us first recall the statement of the SDC adapted to our setting. 
It implies that when approaching any infinite distance point $P$ along any path $\gamma$
one should encounter a universal behaviour of infinitely many states of the theory sufficiently 
close to $P$. More precisely, picking a point $Q'$ in a sufficiently small 
neighbourhood of the infinite distance point $P$, and then moving along the geodesic towards $P$ onto a point $P'$, the SDC asserts that one should be able to identify an infinite tower of states with masses $M_m$, $m=1, \ldots ,\infty$, behaving as 
\beq \label{Mn-decrease}
    M_{m}\left(P'\right) \approx M_m\left(Q'\right) \, e^{- \gamma d\left(Q',P'\right)}\ ,
\eeq
where $M_{m}\left(P'\right)$ and $M_m\left(Q'\right) $ are the masses of the states at $P'$ and $Q'$, respectively. Here $d(Q',P')$ is the distance along the geodesic in the Weil-Petersson metric determined from the 
K\"ahler potential \eqref{Kpot_cs} and $\gamma$ is some positive constant. In other words, the SDC not only asserts that there is an infinite 
tower of states becoming massless at $P$, but also that this has to happen exactly 
in an exponentially suppressed way \eqref{Mn-decrease}.

The goal of this section is to identify such a candidate set of states. As in \cite{Grimm:2018ohb},
we propose that these states arise from BPS D3-branes wrapped on certain 
three-cycles in the Calabi-Yau space $Y_3$. In the case of one-modulus degenerations studied in \cite{Grimm:2018ohb} arguments were presented, by using walls of marginal stability, that the proposed tower is actually populated by BPS states. In this work, we will focus solely on identifying the tower of states, and will not be able to show that they are indeed populated by BPS states. We leave such an analysis for future work, and for now will assume that the identified tower of states is indeed populated by BPS states. 

Asserting that the constructed tower indeed consists of BPS states with charges $\mathbf{Q}$, 
we can use the central charge $Z(\mathbf{Q})$ to compute their mass $M=|Z(\mathbf{Q})|$.
The explicit form of $Z(\mathbf{Q})$ is given by \footnote{Note that we have 
exchanged $\mathbf{\Pi}$ and $\mathbf{Q}$ in $S$ in order to absorb the minus sign in \eqref{def-S}.}
\beq \label{def-Z}
     Z(\mathbf{Q}) = e^{\frac{K}{2}} \int_{Y_3} H \wedge \Omega = e^{\frac{K}{2}}\ S(\mathbf{\Pi},\mathbf{Q})\ ,
\eeq
where $H$ is the three-form with coefficients $\mathbf{Q}$ in the integral basis $\gamma_I$, the $\Omega$ is the (3,0)-form
introduced in \eqref{Omega-exp} with periods $\mathbf{\Pi}$, and $K$ is the K\"ahler potential 
given in \eqref{Kpot_cs}.

We construct the infinite set of states relevant for the SDC by introducing, what we 
call a \textit{charge orbit}. In the one-parameter case this is the same as the monodromy orbit of \cite{Grimm:2018ohb}. It will be obtained by acting on a seed charge vector $\mathbf{q}_0$ with the monodromy matrices relevant in a local 
patch around the infinite distance point $P$. Due to the multi-parameter nature of our analysis, we will change notation with respect to reference \cite{Grimm:2018ohb} and denote the infinite charge orbit by 
\beq
   \mathbf{Q}\left(\mathbf{q}_0| m_1, \ldots , m_n\right) \;,
\eeq
where $m_1, \ldots , m_n$ is a set of integers labelling the considered states, as we discuss below. The charge orbit 
will be infinite, if there are infinitely many allowed values for $m_1, \ldots , m_n$.

\subsection{Single parameter charge orbits} 
\label{sec:spco}

\begin{figure}[h!]
\vspace*{.5cm}
\begin{center} 
\includegraphics[width=6cm]{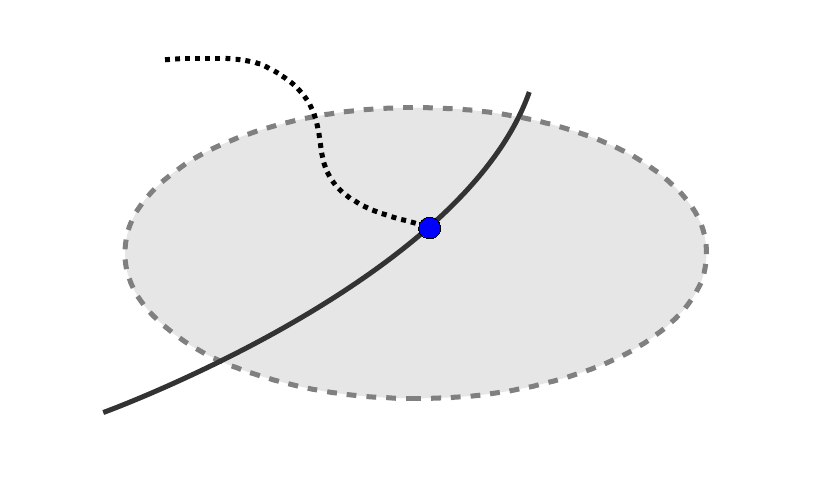} 
\vspace*{-1cm}
\end{center}
\begin{picture}(0,0)
\put(180,30){$\cE$}
\put(290,40){$\longrightarrow$ charge orbit $\mathbf{Q}$}
\put(247,82){$ \Delta_{1}$}
\put(235,42){$ P$}
\end{picture}
\caption{Associating a charge orbit to a point $P \in \Delta_1$ in local patch $\cE$ in moduli space. In this single parameter 
degeneration no intersection locus of $\Delta_1$ within $\Delta$ is in $\cE$.} \label{patch_single}
\end{figure}
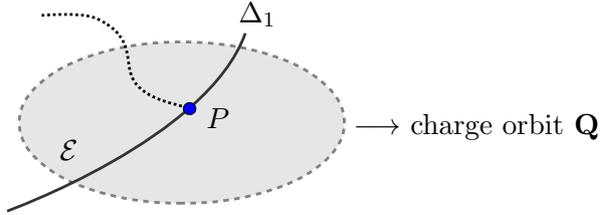

To give a comprehensive introduction of the charge orbit, we will first discuss a single parameter degeneration $t^1 \rightarrow \im \infty$, where we consider only the divisor $\Delta_1 \subset \Delta$ disregarding any 
further intersections. In other words we consider a local patch $\cE$ intersecting $\Delta_1$, but not containing any 
other component of the discriminant locus (see figure \ref{patch_single}).
Such one-parameter degenerations have been discussed at length in \cite{Grimm:2018ohb}. We will introduce a slightly modified description in the following which will then match more seamlessly unto the multi-parameter analysis. 

In analogy to a one-parameter nilpotent orbit \eqref{nilp-orbit} and a one-parameter Sl(2)-orbit \eqref{Sl(2)-orbit}, 
we define the charge orbit as 
\beq \label{charge-orbit_1}
   \mathbf{Q}(\mathbf{q}_0|m_1) \equiv \text{exp}\left[m_1 N_1 \right] \mathbf{q}_0\ ,
\eeq
where $m_1$ is an integer. 
Note that since the monodromy matrix 
$T_1 =\text{exp}\left[ N_1\right] $ the $Q$ are simply the charges 
obtained by acting with the monodromy matrix $T_1^{m_1}$. Since we consider an infinite 
distance point $P$ the results of subsection \ref{sec:class_infinitedistanceP} imply that 
$N_1$ is non-trivial and thus $T_1$ is of infinite order, i.e.~there exists no $m$ such that $T_1^{m} = T_1$. 
In order that the orbit is actually infinite, we further have to demand that 
\beq \label{Nq_not_zero}
   N_1  \mathbf{q}_0 \neq 0 \ . 
\eeq
Hence the definition of an infinite charge orbit agrees with the one in \cite{Grimm:2018ohb}. 

Let us next consider the second crucial part of the distance conjecture, namely that the infinite tower of states becomes exponentially light towards the infinite distance point. 
As mentioned above, we will assume that the considered states are BPS D3-branes, such that 
their masses are measured by $|Z|$, with the central charge $Z$ given in \eqref{def-Z}.
Near the point $P$ we can use the one-variable nilpotent orbit $\mathbf{\Pi}_{\rm nil} =  \text{exp} \left[t^1 N_1\right]  \mathbf{a}_0$, to approximate the behaviour of the central charge
\beq \label{Zasy}
    Z_{\rm asy}(\mathbf{Q}) =  e^{\frac{K}{2}}\ S(\mathbf{\Pi}_{\rm nil},\mathbf{Q})\ .
\eeq
Note that using the results of subsection \ref{sec:period_approx_nil} the asymptotic central charge 
$Z_{\rm asy}(\mathbf{Q})$ differs from $Z(\mathbf{Q})$ by terms proportional to the 
exponential $e^{2\pi \im t^1}$, which are strongly suppressed in the limit $\I \,t^1 \rightarrow \infty$.
Inserting \eqref{charge-orbit_1} into \eqref{Zasy}, we realize that $|Z(\mathbf{Q})| \approx |Z_{\rm asy}(\mathbf{Q})|
\rightarrow 0$ is equivalent to demanding
\beq \label{Zasy-zero}
    |Z_{\rm asy}(\mathbf{q}_0)| \rightarrow 0\ .
\eeq
This can be deduced by moving the exponential $e^{m_1 N_1}$ onto $\mathbf{\Pi}_{\rm nil}$ and absorbing 
it by a shift $\Re{t^1} \rightarrow \Re{t^1} - 1$. Hence, in order to find an 
\textit{infinite massless charge orbit} $\mathbf{Q}$ we  
have to demand that the seed charge $\mathbf{q}_0$ satisfies \eqref{Nq_not_zero} and \eqref{Zasy-zero}.

Let us now construct the seed $\mathbf{q}_0$ for an infinite massless charge orbit $\mathbf{Q}$. We first 
note that there is a particular set of charges that is massless which in \cite{Grimm:2018ohb} were termed to be of type II. They are obtained as elements of the space 
\beq \label{def-TypeII_one}
    \cM_{\rm II}(\mathbf{\Pi}_{\rm nil}) = \Big\{\mathbf{q}^\cI \gamma_\cI \in H^{3}(Y_3,\mathbb{Z}):\quad S( \mathbf{q} ,N_1^k \mathbf{a}_0)=0 \, , \ \forall\, k\Big\}\ ,
\eeq
where we have considered vectors $\mathbf{q}$ over the integers $\mathbb{Z}$. Note that this space depends on the data $(N_1,  \mathbf{a}_0)$ defining the 
nilpotent orbit. 
Stated differently, these are precisely the states that are orthogonal to the nilpotent orbit $\mathbf{\Pi}_{\rm nil}$. 
Their asymptotic central charge \eqref{Zasy} vanishes trivially, which implies that the full central charge 
$Z$ vanishes by exponentially suppressed terms $e^{2\pi \im t^1}$. 

BPS states which become massless as $\I \,t^1 \rightarrow \infty$, but which are not of type II, are called type I states. In \cite{Grimm:2018ohb} arguments were presented for why, given a one-parameter degeneration, the populated BPS states are of type I, and therefore the tower of states of the distance conjecture should be composed of an infinite number of type I states.

It was also shown in \cite{Grimm:2018ohb} that the mass of type I states decreases exponentially fast for one-parameter variations approaching infinite distance. This can be easily seen since the states become massless as a power law in $\I \,t^1$, while the leading behaviour of the K\"ahler potential (\ref{vt-growth}) is logarithmic in $\I \,t^1$. This matches the behaviour predicted by the distance conjecture.

Let us now determine a the set of states that become massless at $P$. To begin with we 
give a sufficient condition for a charge $\mathbf{q}$ 
to become massless at $P$. In order to do that we note that the central charge 
$Z(\mathbf{q})$ can also be written with the help of the Hodge inner product 
$S(C \mathbf{a},\bar{\mathbf{b}}) = \int_{Y_3} a \wedge * \bar{b}$, which is the inner product 
associated to the Hodge norm \eqref{hodgenorm}. Using the fact that $C \mathbf{\Pi} = - \im\, \mathbf{\Pi}$ 
together with \eqref{Omega_norm} we find that $Z(\mathbf{q})$ can be written as 
\beq \label{Z<q}
      | Z(\mathbf{q}) |= \frac{| S(C \mathbf{\Pi},\mathbf{q} )|}{ || \mathbf{\Pi} || } \leq ||\mathbf{q}||\ ,
\eeq
where we have used the Cauchy-Schwarz inequality $| S(C \mathbf{v},\mathbf{\bar u}) | \leq || \mathbf{v}||\, ||\mathbf{u}||$. 
We thus conclude that if the norm $||\mathbf{q}||$ goes to zero at the singularity, the charge $\mathbf{q}$ yields 
a massless state. 
Now we can use the growth theorem \eqref{location_1}, \eqref{vt-growth} to infer that 
\beq \label{q0massless}
  ||\mathbf{q}|| \rightarrow 0 \quad \Longleftrightarrow \quad \mathbf{q} \in W_i^{\mathbb{Q}} \ \text{for}\ i \leq 2\ ,
\eeq
which identifies vector spaces that contain massless states. It is important to stress that the condition \eqref{q0massless}
is a sufficient, but not necessary condition that a charge $\mathbf{q}$ is massless.

Finally, we relate the result \eqref{q0massless} to the classification of singularities discussed in subsection \ref{sec:sing_class}. 
 We use the fact that $W_j^\bbC = \bigoplus_{p+q\leq j} I^{p,q}$ and apply the classification of Hodge diamonds for the singularity 
Types I, II, III, and IV given in table \ref{HD:enumeration}. Using \eqref{primitive-diamond} and \eqref{def-Pi} we realize that
\begin{align}
  & \text{Type I}: && W_2^{\mathbb{C}}  \subset \cM_{\rm II}\ ,&\quad& W_1^{\mathbb{C}} = 0 \ , &\quad&  W_0^{\mathbb{C}}  = 0 \ ,   &\nn \\
  & \text{Type II}: &&  W_2^{\mathbb{C}} = N_1 P^4 \ ,&\quad& W_1^{\mathbb{C}} = 0 \ , &\quad & W_0^{\mathbb{C}}  = 0 \ ,    & \\
   &   \text{Type III}:  && W_2^{\mathbb{C}}= N_1 P^4 \oplus N^2_1 P^5 \ , &\quad&  W_1^{\mathbb{C}}  = N^2_1 P^5 \ ,&\quad& W_0^{\mathbb{C}}  = 0 \ , \nn \\
     &   \text{Type IV}:  &&  W_2^{\mathbb{C}}= N_1 P^4 \oplus N^2_1 P^6 \oplus N^3_1 P^6\ , &\quad&  W_1^{\mathbb{C}} =N_1^3 P^{6} \ ,&\quad& W_0^{\mathbb{C}} = N_1^3 P^{6}    \ . \nn &
\end{align}
We stress that only for the Type IV singularities all spaces $W_2^\bbC$, $W_1^\bbC$ and $W_0^\bbC$ are always non-zero
due to the existence of the non-trivial vectors $N^j \mathbf{a}_0$, $j\leq 3$.  
Finally, combining this with the requirement that $N \mathbf{q}_0 \neq 0$ as well as the fact that 
$N_1 W_i \subset W_{i-2}$ we find that only Type IV singularities straightforwardly admit an 
infinite massless charge orbit $\mathbf{Q}$. 

Let us have a closer look at the $\mathbf{q}_0$ in the case of a Type IV singularity. From the above discussion 
we require $\mathbf{q}_0 \in W_2^\mathbb{Q}$. Furthermore, we note that 
$W_2^\bbC =  I^{1,1} \oplus I^{0,0} =N_1 P^{2,2} \oplus N^2_1 P^{3,3}$
and stress that 
\beq
    S(N_1 P^{2,2}, N_1^k \mathbf{a}_0) = 0\ , 
\eeq
for all $k$, since $ \mathbf{a}_0$ spans $P^{3,3}$. The latter condition shows that $N_1 P^{2,2}$ is a type II state. 
Since we require the orbit to be composed of type I states, we can therefore determine that ${\bf q}_0$ must have a non-trivial component in $N_1^2 P^{3,3}$, so ${\bf q}_0 \notin N_1 P^{2,2}$. In fact, we propose a particular element of the $\mathbb{R}$-split $P^{3,3}$, which can be written as
\be
{\bf q}_0 \sim_{\mathbb{Z}} N_1^2 \mathbf{\tilde a}_0^{(1)} \;.
\label{q01pa}
\ee
Here we have introduced new notation $\sim_{\mathbb{Z}}$ which is rather involved but has a precise definition as follows. 

Consider an element ${\bf a}$ in $W^{\mathbb{C}}_l$, where $l$ is the smallest possible index. If it is possible to add to ${\bf a}$ some other elements in $W^{\mathbb{C}}_l \cap \mathrm{Ker}\,N$ such that one obtains an element in $W^{\mathbb{Q}}_l$, then $ \sim_{\mathbb{Z}} {\bf a} $ is defined as the associated element of $W^{\mathbb{Q}}_l$. If it is not possible, then $ \sim_{\mathbb{Z}} {\bf a} $ is defined to vanish. 

In utilising $\sim_{\mathbb{Z}}$ in (\ref{q01pa}), we will {\it assume} that defined this way ${\bf q}_0$ is non-vanishing. This is true in any example we have studied, but we have no proof that it is always the case. Note that for the particular case of the one-parameter example (\ref{q01pa}), acting with $N_1$ on ${\bf q}_0$ will only receive a contribution from the piece $N_1^2 \mathbf{\tilde a}_0$, but the other components may be necessary in general for quantisation purposes. Note also that we have utilised $\mathbf{\tilde a}_0$, rather than $\mathbf{a}_0$, as introduced in subsection \ref{sec:sl2-orbit}. Finally, it is important to emphasise that in general $\mathbf{\tilde a}_0$ may depend on the coordinates along the singular locus $\mathbf{\tilde a}_0\left(\xi\right)$, and so the combination of elements involved in defining $\sim_{\mathbb{Z}}$ can vary with $\xi$. 

This conclusion seems to imply that the SDC cannot be shown using this construction for the cases Type II and Type III. We know from the discussion of subsection \ref{sec:class_infinitedistanceP} that points 
on these loci are at infinite distance. In examples with $h^{2,1} =1$ the classification of table  \ref{HD:enumeration}
shows that Type III can never be realized. However, Type II singularities do occur in explicit examples and have been 
discussed in more detail in \cite{Grimm:2018ohb}. These constitute interesting cases that require further investigation. 
For higher-dimensional moduli spaces, we 
will now show that the above construction can be generalized yielding a remarkable way to satisfy the SDC 
if intersection loci of divisor $\Delta_i$ appear.

\subsection{Defining the general charge orbit} \label{general_chargeorbit}

Having discussed the one-parameter degenerations, we next propose a general 
form of the charge orbit $\mathbf{Q}(\mathbf{q}_0|m_1, \ldots ,m_n)$ labelling the states relevant for the 
SDC close to an infinite distance point $P$. We stress that this requires that $\mathbf{Q}$ labels
\textit{infinitely many} states that become \textit{massless} at the point $P$. Hence we have to carefully define 
an appropriate orbit that ensures these properties. 
We first give the general expression and 
then show that it has the desired features.

To begin with, let us stress that the definition of $\mathbf{Q}$ is, at first, not  global on $\cM_{\rm cs}$. Rather 
we have to adjust the orbit according to the location of $P$ in the discriminant $\Delta$. 
Nevertheless, the definition of $\mathbf{Q}$ also is not only depending on the location of $P$, but rather takes into account two additional features: 
\vspace*{-.2cm}
\begin{itemize}
\item[(1)] the intersecting patterns and singularity enhancements 
of the $\Delta_i$ in some sufficiently small neighbourhood $\cE$  containing $P$, \vspace*{-.1cm}
\item[(2)] the sector $\cR$ of the path that is traversed when approaching the point $P$. \vspace*{-.1cm}
\end{itemize}
While the first condition will be used in showing when $\mathbf{Q}$ labels infinitely many states, the second 
condition is crucial to ensure that they become massless. 
It will be an important task to carefully spell out these two properties of $\mathbf{Q}$ in the following. 
The reason that these features occur stems from our construction of 
$\mathbf{Q}$ using the Sl(2)-orbit theorem introduced in subsection \ref{sec:sl2-orbit}
and the growth theorem discussed in subsection \ref{sec:growth_theorems}. 

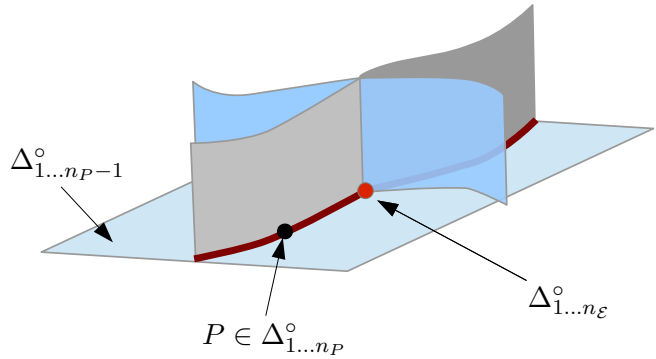
\begin{figure}[h!]
\begin{center} 
\includegraphics[width=9.5cm]{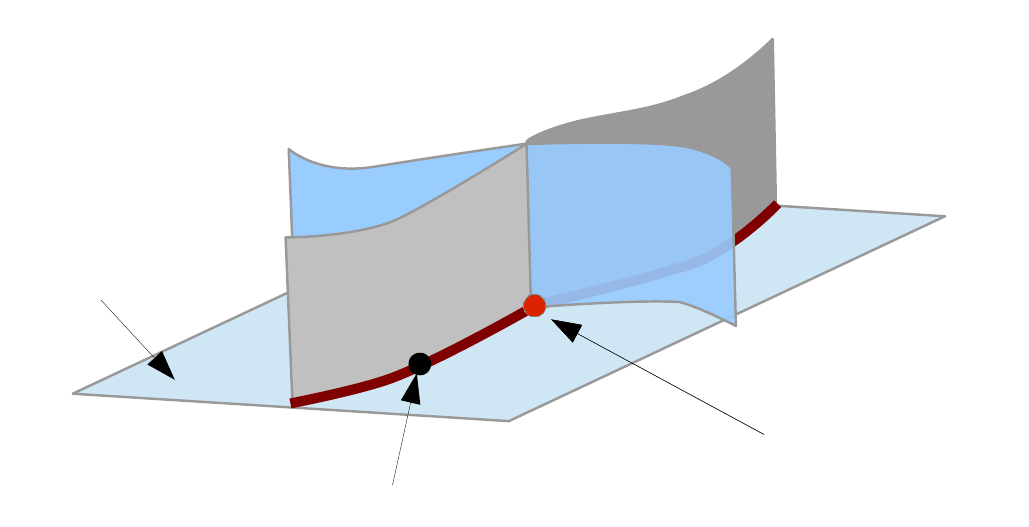} 
\end{center}
\begin{picture}(0,0)
\put(170,26){$P \in \Delta_{1 \ldots n_P}^{\circ}$}
\put(292,40){$\Delta_{1 \ldots n_\cE}^{\circ}$}
\put(98,92){$ \Delta_{1 \ldots n_{P}-1}^{\circ}$}
\end{picture}
\vspace*{-.5cm}
\caption{Illustration of the general setup showing a patch $\cE$ around a point $P$ which lies on the intersection of $n_P$ singular divisors, but away from any further intersections $P \in \Delta_{1 \ldots n_P}^{\circ}$. Within the patch, there is also a further enhancement due to an intersection of additional divisors at $\Delta_{1 \ldots n_{\cE}}^{\circ}$.} \label{patchgen}
\end{figure}

To display our proposal for the charge orbit, it is convenient to recall some more notation
from subsections \ref{sec:period_approx_nil} and \ref{sec:sl2-orbit}.
We consider a patch $\cE$ around the point $P \in \Delta$ 
which might contain any type of higher intersections of divisors $\Delta_i$.
This patch is defined by requiring that the nilpotent orbit \eqref{nilp-orbit} provides a 
good approximation in $\cE$ to the full periods. In other words, we 
can drop the exponential corrections in $\cE$ as discussed in detail in subsection \ref{sec:period_approx_nil}. 
Let us denote the divisors intersecting in the patch $\cE$ by $\Delta_i$ with $i=1,\ldots, n_{\cE}$. 
As usual we denote the monodromy logarithms associated to $\Delta_i$ by $N_i$. 
Furthermore, we will consider a point $P$ on the intersection of the first $n_P$ divisors 
$\Delta_k$, i.e.~
\beq
  P \in \Delta_{1 \ldots n_P}^\circ\ , 
\eeq 
where we recall that $\,^\circ$ indicates that we consider points away from any further intersection
as introduced in subsection \ref{sec:charact_sing}. In order to use the growth theorem 
for the norm of $\mathbf{Q}$ when approaching $P$ we introduce the sectors $\cR_{r_1 \ldots  r_{n_P}}$
as before. They are defined by first setting
\beq \label{path1}
   \cR_{1 \ldots n_P} \equiv \left\{t^i: \quad \frac{\I\, t^{1} }{\I \, t^2 } > \lambda \, ,\ldots , \ 
\frac{\I\, t^{n_P-1}}{\I \, t^{n_P} } > \lambda  \,  ,\  \I \, t^{n_P} > \lambda\, ,
\quad \R\, t^i < \delta\, \right\} ,
\eeq
for some fixed $\lambda,\delta>0$. The other orderings of the indices on $\cR_{1 \ldots n_P}$ 
are defined by simple permutations of the indices in all of \eqref{path1}. 
In this work we will only consider paths that traverse a single sector $\cR_{r_1 \ldots  r_{n_P}}$. Completely arbitrary paths cannot be analysed so easily and might require to patch together sectors of the form \eqref{path1}.
It should, however, be stressed that this is a very mild path dependence. We do not 
expect that our conclusions change for more general paths.
The setup is illustrated in figure \ref{patchgen}.

Let us now turn to the proposal for the charge orbit $\mathbf{Q}$. Given a path towards $P$ that 
traverses a single sector $\cR_{r_1 \ldots r_{n_P}}$ we fix an ordering of $n_P$ matrices $N_i$ as 
$(N_{r_1}, \ldots ,N_{r_{n_P}})$. 
By a simple relabelling we can pick this ordering to be $(N_1, \ldots ,N_{n_P})$ without loss of generality. 
The ordering of the remaining $N_i$, $i=n_P, \ldots ,n_\cE$ does not need to be fixed as of now. For convenience 
we will pick the simplest ordering such that we have in total $(N_1, \ldots ,N_{n_\cE})$.
In analogy the Sl(2)-orbit \eqref{Sl(2)-orbit} we now define the 
charge orbit as 
\beq \label{charge-orbit_gen}
   \mathbf{Q}(\mathbf{q}_0 | m_1, \ldots ,m_\cE) \equiv \text{exp}\Big(\sum_{i=1}^{n_{\cE}} m_i N^-_i \Big) \mathbf{q}_0\ ,
\eeq
with integers $m_i$. The first non-trivial part of the construction 
is the use of the matrices $N_i^-$ in \eqref{charge-orbit_gen}. These are part of the commuting 
$\slt$s discussed in subsection \ref{sec:sl2-orbit} and are non-trivially constructed from the $N_i$ given
 in a particular ordering. Clearly, we pick the ordering introduced before, which was partly dictated by the considered path 
 towards $P$. 
Note that the construction of $N_i^-$ depends on all other $N_j$ with $j\geq i$. If one considers situations with 
$n_P < n_\cE$ this implies that they contain information about the other divisors 
intersection in $\cE$ even though $P$ can be away from them. Also note that for a one-parameter case one 
trivially has $N_1 = N_1^-$, such that \eqref{charge-orbit_gen} is a natural generalization of \eqref{charge-orbit_1}.

In order to fully specify the charge orbit \eqref{charge-orbit_gen} it is crucial to determine the 
properties of the intersections in $\cE$ such that a seed 
charge $\mathbf{q}_0$ exists that ensures that $\mathbf{Q}(\mathbf{q}_0 | m_1, \ldots ,m_\cE)$  
yields an infinite set of charges that become massless when approaching $P$. 
Let us thus consider a general enhancement chain within $\cE$ of the form
\beq \label{enhance_chain}
  \mathsf{Type\ A}_1 \rightarrow   \ldots  \rightarrow \underbrace{\boxed{\mathsf{Type\ A}_{n_P}}}_{\text{location of }P} \rightarrow  \ldots   \rightarrow \mathsf{Type\ A}_{n_\cE}\ ,
\eeq
where we list the singularity types on the intersection loci $\Delta_{1}^\circ$, $\Delta_{12}^\circ$, \ldots ,$\Delta_{1 \ldots n_P}^\circ$, \ldots , $\Delta_{1 \ldots n_\cE}^\circ$ and indicated by a box singularity of the locus $\Delta_{1 \ldots n_P}^\circ$ containing $P$. Note that we have fixed an ordering of the first $n_P$ elements $N_{i}$ according to the considered path. 

Let us now summarize the results that we will show in this section. 

\noindent
\textbf{Existence and construction of a charge orbit.} We find an \textit{infinite} charge orbit $\mathbf{Q}$
that becomes \textit{massless} at the location of a point $P \in \Delta^\circ_{1 \ldots n_P}$ if one of the two 
conditions are satisfied: 
\vspace*{-.5cm}
\begin{itemize} 
\item[(R1)]  $P$ is on a locus $\Delta^\circ_{1 \ldots n_P}$ carrying a  Type IV singularity. In other words, if 
$\mathsf{Type\ A}_{n_P} = \text{IV}$ in the enhancement chain \eqref{enhance_chain}. 
\item[(R2)]  $P$ is on a locus $\Delta^\circ_{1 \ldots n_P}$ carrying a  Type II or Type III singularity and
    there exists a higher intersection, $\Delta^\circ_{1 \ldots n_P+1}, \ldots ,\Delta^\circ_{1 \ldots n_\cE}$, 
on which the singularity type increases. In other words, we have $\mathsf{Type\ A}_{n_P} = \text{II or III} $ and the 
enhancement chain \eqref{enhance_chain} contains either one of the enhancements   II $\rightarrow$   III,   
II $\rightarrow$  IV or   III $\rightarrow$   IV after the singularity type at $P$. \vspace*{-.2cm}
\end{itemize}
Importantly, as indicated at the beginning of this section, these results are true for \textit{any path} approaching
$P$ that stays within the growth sector \eqref{path1}.
We will generally show these statements employing the full power of the mathematical machinery introduced 
in section \ref{math_background} and section \ref{sec:classification_results}. Furthermore, we will explicitly construct the 
seed charge $\mathbf{q}_{0}$ for all of the enhancement chains allowed by (R1) and (R2).
Given a chain \eqref{enhance_chain} satisfying (R1) and (R2), we  show the existence of a seed charge $\mathbf{q}_{0}$ with 
the following simple features:
\bea \label{q0_properties}
    \boxed{\mathsf{Type\ A}_{n_P}}\neq \text{IV}\ : &\quad & \left\{ \begin{array}{ll} N^-_{(i)} \mathbf{q}_0 = 0 \hspace*{.4cm} & \text{for all}\ i\ \text{with}\ 1\leq i \leq n_P\ ,\\
    N^-_{(j)} \mathbf{q}_0 \neq 0  & \text{for some}\ j \ \text{with} \ n_P < j \leq n_\cE\ , \end{array}\right. 
    \\
      \boxed{\mathsf{Type\ A}_{n_P}}= \text{IV}\ :  &\quad & \left\{ \begin{array}{ll} N^-_{(i)} \mathbf{q}_0 = 0 \ \ & \text{for all}\ i\ \text{with}\ 1\leq i < n_P\ ,\\
        N^-_{(n_P)} \mathbf{q}_0 \neq 0, & \big(N^-_{(n_P)}\big)^2 \mathbf{q}_0 = 0.
        \end{array}\right.  \nn
\eea
We will show that together with the fact that $P$ is on an infinite distance locus, this ensures 
that $\mathbf{q}_0$ is massless along any path within the growth sector \eqref{path1}.

To systematically establish these claims we first discuss  in subsection \ref{sec:general_masscomments} 
some general facts about the mass of the states associated to 
$\mathbf{Q}$ and $\mathbf{q}_0$ when approaching $P$. We then turn to our main tool and 
discuss in detail in subsection \ref{sec:cantwodiv} configurations which consists of two intersecting 
divisors in $\cE$, i.e.~we will study the general $n_{\cE}=2$ configuration. 
We will not only see that (R1) and (R2) are true in this case, but also 
describe how a given $\mathbf{q}_0$ can be tracked through an enhancement. 
Concretely we will consider two types of enhancement chains
\bea
   n_P = 1 &:& \quad \boxed{\mathsf{Type\ A}}\rightarrow  \mathsf{Type\ B}\ ,\\ 
   n_P = 2 &:& \quad \mathsf{Type\ A}\rightarrow  \boxed{\mathsf{Type\ B}}\ ,
\eea
where, as above, the box indicates the location of the point $P$. In this simpler situation 
we will easier to construct the relevant seed charges $\mathbf{q}_0$ and explain how in the cases stated above 
induce an infinite, massless orbit when approaching $P$. 
The general case of having an arbitrary enhancement chain \eqref{enhance_chain} will be 
subsequently studied in subsection \ref{multi-N-analysis}.

Note that while this covers many possible singularities and singularity enhancements in the Calabi-Yau moduli space, there
are a number of enhancements that do not lead to a simple charge orbit that is both infinite and massless for any path in a 
sector. For example, we will see that if the chain \eqref{enhance_chain} ends on an enhancement II $\rightarrow$ III with $P$ being at the Type III locus, a natural candidate orbit with the desired features exists only if one excludes certain paths in the sector. More generally, we find that all chains \eqref{enhance_chain} of the form 
\beq
\mathsf{Type\ A}_1  \rightarrow  \ldots   \rightarrow \mathsf{Type\ A}_{n_\cE-1}\rightarrow \boxed{\text{Type II or Type III}}\ ,
\eeq
do not lead to a natural infinite and massless orbit that is path-independent within a sector by using the methods presented in this work. We will discuss possible 
extensions to tackle these cases in more detail 
in subsection \ref{discussion_properties+extensions}.

\subsection{Masslessness of the charge orbit} \label{sec:general_masscomments}

Let us first discuss the conditions on the charge orbit $\mathbf{Q}$ defined in \eqref{charge-orbit_gen}
such that it consists of states that become light at $P$
and can serve as the states of SDC. 
To do that we have to determine the behaviour of the central charge $|Z(\mathbf{Q})|$ when approaching the point $P$. In other words we have to ensure that  
\beq \label{M-to-zero}
    M(\mathbf{Q}) = |Z(\mathbf{Q})| \ \longrightarrow \ 0 \ . 
\eeq
To identify sufficient conditions for \eqref{M-to-zero} we use the general growth theorem \eqref{general_growth-res} for the Hodge norm $|| \mathbf{Q}||$. 
In order to do that we note that the central charge 
$Z(\mathbf{Q})$ can also be written with the help of the Hodge inner product $S(C \mathbf{a},\bar{\mathbf{b}})$ associated to the 
Hodge norm \eqref{hodgenorm}.
Using the fact that $C \mathbf{\Pi} = - \im \mathbf{\Pi}$ 
together with \eqref{Omega_norm} we find that $|Z(\bQ)|$ can be written as 
\beq \label{Z<Q_gen}
      | Z(\mathbf{ Q}) |= \frac{| S(C \mathbf{\Pi}, \mathbf{Q} )|}{ || \mathbf{\Pi} || } \leq ||\mathbf{ Q}||\ . 
\eeq
where we have used the Cauchy-Schwarz inequality $| S(C \mathbf{v},\mathbf{\bar u}) | \leq || \mathbf{v}||\, ||\mathbf{u}||$. 
We thus conclude that if the norm $||\mathbf{Q}||$ goes to zero at the singularity, the charge orbit $\mathbf{Q}$ yields 
massless states. 

The general discussion of subsection \ref{sec:growth_theorems} provides us with a powerful tool to 
determine the behaviour of $||\mathbf{ Q}||$ near the point $P$. More precisely, we 
introduced the multi-variable growth theorem, which allows us to evaluate the asymptotic 
behaviour of $ \mathbf{ Q}$ from its location in 
\beq
   W_{l_1}\big(N_{(1)} \big) \cap W_{l_2}\big(N_{(2)} \big)  \cap  \ldots  \cap W_{l_{n_P}}\big(N_{(n_P)}\big) \;,
\eeq 
with the $N_{(i)}$ introduced in \eqref{lbr}. Given our definition of $\mathbf{Q}$, the 
restriction to a growth sector and ordering as discussed in subsection \ref{general_chargeorbit}, we would rather 
like to work with the $N_{(i)}^-$ constructed from the commuting $\slt$s containing $N_i^-$. Here 
another fact from the Sl(2)-orbit theorem of \cite{CKS} can be applied, which states that 
\beq \label{W(N)=W(N-)}
   W_{l} \big(N_{(i)} \big) = W_l\big(N_{(i)}^-\big) \ . 
\eeq
Hence, we can apply the results of subsection \eqref{sec:growth_theorems} by simply replacing $N_{(i)} \rightarrow N_{(i)}^-$
when staying in the ordering of the $N_i$ used to determine $N_i^-$. 
 
The next step is to establish that the growth of $||\mathbf{Q}||$ is identical to this of $||\mathbf{q}_0||$. In order 
to do that we have to show that the location of $\mathbf{Q}$ and $\mathbf{q}_0$ in the spaces 
\beq
   W_{l_1}\big(N_{(1)}^-\big) \cap W_{l_2}\big(N_{(2)}^-\big)  \cap  \ldots  \cap W_{l_{n_P}}\big(N_{(n_P)}^-\big) \;,
\eeq 
agree, where we recall the notation $N^-_{(n)} = \sum_{i=1}^n N_i^- $. 
Now the existence of $n_\cE$
\textit{commuting} $\slt$-triples \eqref{Sl2-triples} containing the $N^-_i$ becomes relevant. 
In fact, each of these triples contain the operators $Y_i$ that gives the location of a vector $\mathbf{v}$
in $W_{l}\big(N_{(j)}^-\big)$. Using \eqref{Yk-action} and \eqref{Fp-Wi_split} one has 
\beq
   Y_{(j)} \mathbf{v} = l_j \mathbf{v} \quad \Rightarrow \quad  \mathbf{v} \in W_{l_j+3}(N^-_{(j)})\ ,
\eeq 
where $Y_{(j)} = Y_1+ \ldots +Y_j$ as in \eqref{Yk-action}. Crucially, the location of $\mathbf{q}_0$ and 
$N^-_{j} \mathbf{q}_0$ agree, which implies 
that if $\mathbf{q}_0$ is massless also $\text{exp}(m_{n_P+1} N_{n_P+1}^-+  \ldots  + m_{n_\cE} N_{n_\cE}^-) \mathbf{q}_0$ is massless. Concerning the growth and the masslessness thus only the terms $\text{exp}(m_1 N_1^-+  \ldots  + m_{n_P} N_{n_P}^-)$
are relevant. However, due to the exponential the location of the highest $l_i$-components of $\mathbf{Q}$ and
$\mathbf{q}_0$ agree. In fact, it was already shown in \cite{CKS} that the growth does not change upon multiplying 
by this exponential term. We hence conclude that with respect to the leading growth one has 
\beq
  || \mathbf{Q}||  \sim || \mathbf{q}_0|| \ ,
\eeq
and hence $\mathbf{Q}$ is massless for all values of $m_1, \ldots ,m_\cE$ 
as long as $\mathbf{q}_0$ is massless. 

Let us finally give a \textit{sufficient} condition for having $|| \mathbf{q}_0|| \rightarrow 0$ along any path 
in the considered growth sector.  
Using the general growth theorem \eqref{general_growth-res} with \eqref{general_growth-loc}, it is not hard show that 
$||\mathbf{q}_0||\rightarrow 0$ is true if one has 
\bea \label{massless_q0}
   \mathbf{q}_0 & \in &  W_{l_1}\big(N_{(1)}^-\big) \cap W_{l_2}\big(N_{(2)}^-\big)  \cap  \ldots  \cap W_{l_{n_P}}\big(N_{(n_P)}^-\big)\ , \\
    &&\text{with}\ \ l_{n_P} < 3, \quad l_{1} , \ldots , l_{n_P - 1}\leq 3\ . \nn
\eea 
This condition uses that if $l_i\leq 3$, $i=1, \ldots ,n_P - 1$ then we 
can estimate $\I\, t^{i+1} / \I\, t^{i} < \lambda^{-1}$ in \eqref{path1} and hence find that $|| \mathbf{q}_0||$ vanishes for any path.
Let us stress that this statement of masslessness can only be obtained 
on the sector $\cR_{1 \ldots n_P}$ defined in \eqref{path1}, due to the path dependence in the growth theorem. 

While we have discussed in detail the masslessness of the orbit at infinite distance, the distance conjecture further states that the states in the orbit should become massless exponentially fast in the geodesic proper distance. This is much more difficult to prove generally for multi-parameter settings since one must calculate geodesics. We leave a detailed analysis of this for future work, but will give some evidence that it is natural to expect that the exponential behaviour is universal. First we note that the masses of the BPS states are still power-law in the  $\I\, t^{i}$, as was the case for the one-parameter case. Therefore, if the geodesic proper distance grows only logarithmically in the $ \I\, t^{i} $ the states will becomes massless exponentially fast. 

To see evidence for the logarithmic behaviour in the multi-parameter cases we can approximate the behaviour of the field space metric through the leading behaviour of the K\"ahler potential. The growth theorem applied to the K\"ahler potential 
 implies as shown in \eqref{general_growth_K} that the asymptotic leading behaviour within a given growth sector takes the form
\be
K_{\rm asy} = - \sum_i  r_i \, \text{log} \left(  \I\, t^{i} \right) \;,
\ee
where the $r_i = d_i - d_{i-1}$, with $d_0=0$, are positive integers no larger than 3. The K\"ahler metric derived from this asymptotic K\"ahler potential, which we emphasise may not necessarily be the leading behaviour of the metric, takes the form
\be
g_{i\bar{\jmath}} \sim \mathrm{diag\;}\left(\frac{r_i}{ \left(\I\, t^{i} \right)^2} \right) \;.
\ee
The proper distance $d_{\gamma}(P,Q)$ along a path $\gamma$ in field space with affine parameter $s$ then take the form
\be
d_{\gamma}(P,Q) = \int_{\gamma} \sqrt{g_{i\bar{\jmath}}\frac{dt^i}{ds}\frac{d\bar{t}^{\bar{\jmath}}}{ds} } ds \;.
\ee
If we restrict to a path with fixed $\R\;t^i$ we can write this as
\be
d_{\gamma}(P,Q) = \int_{\gamma} \left[\sum_i r_i \left(\frac{d \log \I\;t^i}{ds}\right)^2\right]^{\frac12} ds \;.
\ee
For sufficiently simple paths this manifestly grows logarithmically. The distance $d(P,Q)$ along a geodesic 
path is relevant for the exponential behaviour \eqref{Mn-decrease} of the SDC and we expect that it shares the logarithmic behaviour in 
the asymptotic regime. 

\subsection{The two-divisor analysis} \label{sec:cantwodiv}

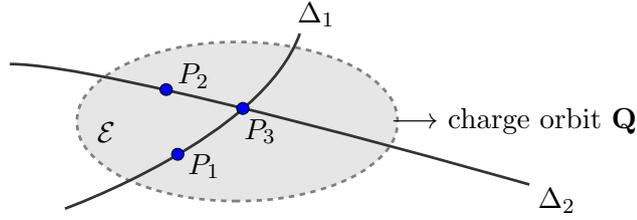
\begin{figure}[h!]
\vspace*{.5cm}
\begin{center} 
\includegraphics[width=7.5cm]{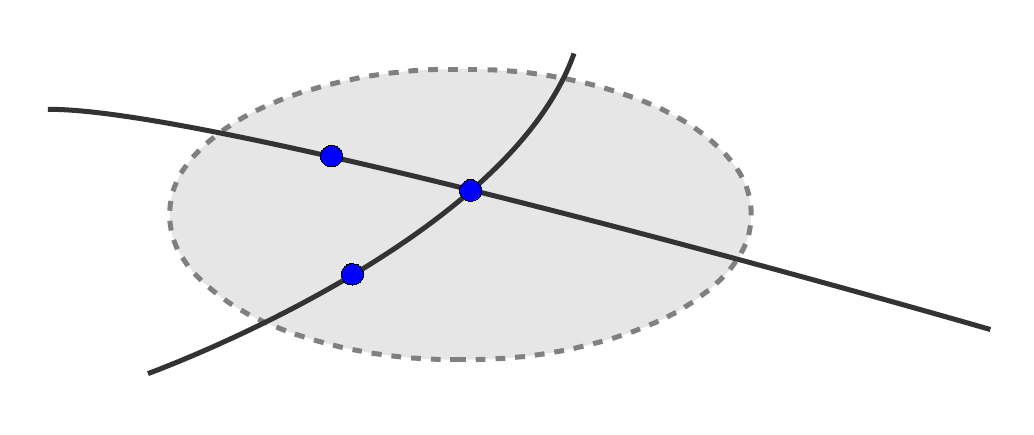} 
\vspace*{-1cm}
\end{center}
\begin{picture}(0,0)
\put(194,23){$P_1$}
\put(190,57){$P_2$}
\put(215,36){$P_3$}
\put(160,35){$\cE$}
\put(270,40){$\longrightarrow$ charge orbit $\mathbf{Q}$}
\put(235,80){$ \Delta_{1}$}
\put(325,10){$ \Delta_{2}$}
\end{picture}
\caption{The canonical case of two singular divisors intersecting on a local patch $\cE$ in moduli space. 
The considered infinite distance points $P$ can be located either on $\Delta_1^\circ$, $\Delta_2^\circ$ or $\Delta_{12}^\circ$ 
as exemplified by $P_1$, $P_2$, and $P_3$.} \label{patch2}
\end{figure}

Most of our general arguments about enhancement chains and charge orbits will 
be built on the case of just two singularity loci intersecting. 
To study this canonical situation, we will consider a patch $\cE$ in which two discriminant divisors $\Delta_1$, $\Delta_2$ with associated monodromy 
logarithms $N_1$, $N_2$ intersect. This is depicted 
in figure \ref{patch2}. The point $P$ under consideration now can be at different locations in this configuration. 
We can have either $P \in \Delta_1^\circ$, $P \in \Delta_2^\circ$, or $P \in \Delta_{12}^\circ = (\Delta_1 \cap \Delta_2)^\circ$.

The restriction of \eqref{charge-orbit_gen} to the two-dimensional case $n_{\cE}=2$ is given by 
\beq \label{charge-orbit_2}
   \mathbf{Q}(\mathbf{q}_0|m_1,m_2) \equiv \text{exp}\Big(m_1 N^-_1 + m_2 N^-_2 \Big) \mathbf{q}_0\ .
\eeq
Recall that the definition of $N_1^-$, $N_2^-$ requires to fix an ordering.
We thus distinguish three cases 
\bea \label{cases(1)(2)}
   (1)\quad P \in \Delta_1^\circ: &\quad& \text{ordering} \ (N_1, N_2) \ \rightarrow\ (N_1^-=N_1, N^-_2) \ ,\\
    (2)\quad P \in \Delta_2^\circ: &\quad& \text{ordering} \ (N_2, N_1) \ \rightarrow\ (N_1^-=N_2, N^-_2) \ ,\nn
\eea
and the sector-dependent case 
\beq \label{case(3)}
     (3)\quad P \in \Delta_{12}^\circ:  \ \left\{\begin{array}{c} (N_1, N_2) \, \rightarrow\, (N_1^-=N_1, N^-_2) \qquad \text{path}\  \left\{ \ \frac{\I\, t^1 }{\I \, t^2 },\ \I \, t^2 > \lambda \ \right\}\, ,\\[.2cm]
     (N_2, N_1) \, \rightarrow\, (N_1^-=N_2, N^-_2)  \qquad \text{path}\  \left\{ \ \frac{\I\, t^2 }{\I \, t^1 },\ \I \, t^1 > \lambda \ \right\}\,.
                                                                                   \end{array}\right. 
\eeq 
Note that the construction of $N_2^-$ is a rather non-trivial task, as outlined in the appendices. 
Our aim is to identify the possible enhancements for which a ${\bf q}_0$ exists such 
that the charge orbit is massless and infinite.

We can choose, with generality, to focus on the ordering (1) above and correspondingly 
focus only on the upper growth sector in \eqref{case(3)}. We go through each enhancement 
chain $\mathsf{Type\ A} \to \mathsf{Type\ B}$ and track candidate charges $\bqz$ through
the enhancement. In particular, we will check the conditions 
(R1) and (R2) and identify the $\bqz$ that induces an infinite massless orbit. 
Moreover, we will show how our construction does not necessarily yield an infinite orbit that
is massless on any path within a sector if (R1) and (R2) are violated. 
We show that the enhancement of type $\boxed{\rI_a}\rightarrow \mathsf{Type\ B}$ 
do not admit infinite orbits by examining the example $\rI_a \to \rIV_d$. 
The path dependence will be discussed for the example $\text{II}_b\rightarrow \boxed{\text{III}_c}$. 
Finally, we will also examine chains with no type enhancement by discussing the 
example $\rII_b \to \rII_c$.

For every enhancement $\mathsf{Type\ A} \to \mathsf{Type\ B}$, we denote the Sl(2)-splitting of the limiting mixed Hodge structure of $\mathsf{Type\ A}$ and $\mathsf{Type\ B}$ by $(F_{(1)}, W^{(1)})$ and $(F_{(2)}, W^{(2)})$, respectively. Then we have a pair of commuting $\slt$-operators $(N^-_1, N^+_1, Y_1)$ and $(N^-_2, N^+_2, Y_2)$. The Deligne splitting of $(F_{(i)}, W^{(i)})$ is denoted by 
\beq 
H^3\left(Y_3,\mathbb{C} \right) = \bigoplus_{p,q} I^{p, q}_{(i)}, \qquad \quad I_{(i)}^{p,q} = \bigoplus_{k \geq 0} \, \big(N^-_{(i)}\big)^k \, P^{p+k,q+k}(N^-_{(i)})\ ,
\eeq 
where we have also displayed the decomposition \eqref{I=NP} into primitive parts.
The bracket notation matches that introduced in (\ref{lbr}) and (\ref{dsl}), so for example $W^{(2)}_l\equiv W_l\left(N_1^-+N_2^-\right)$.

\subsubsection{The enhancement $\rI_a \to \rIV_d$}

Let us first discuss the enhancement chains $\boxed{\rI_a} \rightarrow \rIV_d$ and  $\rI_a \rightarrow \boxed{\rIV_d}\,$,
i.e.~where we consider $P$ at either on a $\rI_a$ locus or a $\rIV_d$ locus. This will also allow us to 
introduce the strategy on how we relate the Hodge-Deligne diamonds along enhancements.

Focusing first on $\boxed{\rI_a} \rightarrow \rIV_d$, we recall that the conditions \eqref{two-moduli}
imply that a divisor of type $\rI_a$ is at finite distance. Hence we do not necessarily 
expect any infinite tower of massless states as we approach the type $\rI_a$ divisor in our formalism. 
We will check that we can indeed not identify an infinite charge orbit associated to this locus. 

We first spell out the decomposition into primitive parts \eqref{primitive-diamond} associated with the mixed Hodge structure $\left(F_{(1)}, W^{(1)}\right)$ of type $\rI_a$
\begin{equation} \label{eqn:Lefschetz-decomposition-Ia}
  H^3\left(Y_3,\mathbb{C} \right) = \textcolor{red}{P^3(N_1^-)} \oplus \left[ \textcolor{green}{P^4(N_1^-)} \oplus \textcolor{blue}{N_1^- P^4(N_1^-)}  \right]\ ,
\end{equation}
where the $P^i(N_1^-)$ are the primitive spaces defined in \eqref{def-Pi}. Note the we have used 
different colours for later expositional convenience. We depict the decomposition into primitive parts also in the Hodge-Deligne diamond in figure \ref{fig:Lefschetz-decomposition-Ia}.
\begin{figure}[!h]
\begin{center}
  \begin{tikzpicture}[cm={cos(45),sin(45),-sin(45),cos(45),(15,0)}]
    \draw[step = 1, gray, ultra thin] (0, 0) grid (3, 3);

    \draw[fill, red] (0, 3) circle[radius=0.03];
    \draw[fill, red] (1, 2) circle[radius=0.03] node[above=2pt] {$a'$};
    \draw[fill, green] (2, 2) circle[radius=0.03] node[above=2pt] {$a$};
    \draw[->] (1.9, 1.9) -- (1.1, 1.1);
    \draw[fill, blue] (1, 1) circle[radius=0.03] node[below=2pt] {$a$};
    \draw[fill, red] (2, 1) circle[radius=0.03] node[above=2pt] {$a'$};
    \draw[fill, red] (3, 0) circle[radius=0.03];

    \draw[dashed, gray, ultra thin, opacity=0] (-0.5, 3.5) -- (3.5, -0.5);
    \draw[dashed, gray, ultra thin, opacity=0] (0.5, 3.5) -- (3.5, 0.5);
  \end{tikzpicture}
\end{center}
\caption{The Hodge-Deligne diamond of type $\rI_a$ with its decomposition into primitive parts \eqref{eqn:Lefschetz-decomposition-Ia}. The action of $N_1^-$ are labelled by arrows, and we use colours to highlight the primitive subspaces $P^3(N_1^-)$, $P^4(N_1^-)$ and their images under the action of $N_1^-$. Since the two $\slt$-triples are commuting, the primitive subspace $P^3(N_1^-)$, $P^4(N_1^-)$ and their images under $N_1^-$ are preserved by $N_2^-$.} \label{fig:Lefschetz-decomposition-Ia}
\end{figure}

As discussed in section \ref{sec:class_enh}, the $P^k(N^-_1)$ carry a pure Hodge structure of weight $k$ on $\Delta_1^\circ$, while at $\Delta^{\circ}_{12}$ these degenerate into mixed Hodge structures. Specifically, we have a pure Hodge structure of weight $3$ with Hodge number $(0, a', a', 0)$ on $P^3(N_1^-)$, and a pure Hodge structure of weight $4$ with Hodge number $(0, 0, a, 0, 0)$ on $P^4(N_1^-)$. Then the second $\slt$-triple $(N^-_2, N^+_2, Y_2)$ induces polarised mixed Hodge structures polarised by $N_2^-$ coming from variation of Hodge structures on $P^3(N_1^-)$ and $P^4(N_1^-)$. We show the Deligne splitting of these two mixed Hodge structures and their images under the action of $N_1^-$ in the figure \ref{fig:Ia-IVd-with-colour}.\footnote{Note that we do not depict the full grid for the higher weight Hodge structures, see for example appendix \ref{sec:derpol} for this.}  The sum \eqref{eqn:Lefschetz-decomposition-Ia} of the mixed Hodge structures then gives a mixed Hodge structure, $I^{p,q}_{(2)}$, of type $\mathrm{IV}_d$ with $d = r + a$ where $r \ge 1$ is an integer.

\begin{figure}[!h]
\begin{center}
  \begin{tabular}{c c c}
    \begin{tikzpicture}[cm={cos(45),sin(45),-sin(45),cos(45),(15,0)}]
      \draw[step = 1, gray, ultra thin] (0, 0) grid (3, 3);
  
      \draw[fill, red] (3, 3) circle[radius=0.03];
      \draw (3, 3) circle[radius=0.1] node[left] {$\mathbf{\tilde{a}}_0^{(2)}$};
      \draw[->, red] (2.9, 2.9) -- (2.1, 2.1);
      \draw[fill, red] (2, 2) circle[radius=0.03] node[right=2pt] {$r$};
      \draw[->, red] (1.9, 1.9) -- (1.1, 1.1);
      \draw[fill, red] (2, 1) circle[radius=0.03] node[right=2pt] {$a' - r$};
      \draw[fill, red] (1, 2) circle[radius=0.03] node[left=2pt] {$a' - r$};
      \draw[fill, red] (1, 1) circle[radius=0.03] node[right=2pt] {$r$};
      \draw[->, red] (0.9, 0.9) -- (0.1, 0.1);
      \draw (1, 1) circle[radius=0.1] node[left] {$\bqz$};
      \draw[fill, red] (0, 0) circle[radius=0.03];
    \end{tikzpicture} &
    \begin{tikzpicture}[cm={cos(45),sin(45),-sin(45),cos(45),(15,0)}]
      \draw[step = 1, gray, ultra thin] (0, 0) grid (3, 3);
  
      \draw[fill, green] (2, 2) circle[radius=0.03] node[above=2pt] {$a$};
    \end{tikzpicture} &
    \begin{tikzpicture}[cm={cos(45),sin(45),-sin(45),cos(45),(15,0)}]
      \draw[step = 1, gray, ultra thin] (0, 0) grid (3, 3);
  
      \draw[fill, blue] (1, 1) circle[radius=0.03] node[below=2pt] {$a$};
    \end{tikzpicture}\\
    \textcolor{red}{$P^3(N_1^-)$} & \textcolor{green}{$P^4(N_1^-)$} & \textcolor{blue}{$N_1^- P^4(N_1^-)$}
  \end{tabular}
\end{center}
\caption{The left picture shows a mixed Hodge structure, determined by some integer $r \ge 1$, on $P^3(N_1^-)$. The middle picture shows a mixed Hodge structure on $P^4(N_1^-)$. The right picture shows the image of the middle picture under the action of $N_1^-$. In these diamonds, the coloured arrows label the action of $N_2^-$. The colourings are in agreement with equation \eqref{eqn:Lefschetz-decomposition-Ia} and figure \ref{fig:Lefschetz-decomposition-Ia}. The sum of these three Hodge-Deligne diamonds is the diamond of $\left(F_{(2)}, W^{(2)}\right)$, associated to the mixed Hodge structure $I^{p,q}_{(2)}$, of type $\rIV_d$. The circles around the dots in the first diamond indicate the location of $\bqz$ and $\mathbf{\tilde{a}}_0^{(2)}$.} \label{fig:Ia-IVd-with-colour}
\end{figure}

We can now identify an element  ${\bf q}_0$ in $I^{p,q}_{(2)}$ that looks similar to the one occurring in the one-parameter case (\ref{q01pa}). The relevant ${\bf q}_0$ is shown in figure \ref{fig:Ia-IVd-with-colour}, and it can be written as
\be
{\bf q}_0 \sim_{\mathbb{Z}} \sumNsqr{2} \mathbf{\tilde{a}}_0^{(2)} \;.
\label{q02pa14}
\ee
We can see also from figure \ref{fig:Ia-IVd-with-colour} that ${\bf q}_0$ is not in the kernel of $N_2^-$, and so the charge orbit (\ref{charge-orbit_2}) is indeed infinite and given by
\begin{equation} \label{infiniteness_14}
  \bQ(\bqz | m_1, m_2) = \bqz + m_2 N_2^- \bqz, \textrm{ for } m_1, m_2 \in \bbZ.
\end{equation}

Next we would like to check if this infinite orbit is indeed massless on $\Delta^{\circ}_1$. This can of course be checked by using condition \eqref{massless_q0}, but in this section of two-divisor analysis we will also spell out the growths of Hodge norm explicitly to familiarise the reader with the formalism. To do this we follow a similar procedure to the one-parameter case in section \ref{sec:spco}. We first determine the location of $\bqz$, i.e.~$\bqz \in W_{l_1}\left(N_1^-\right) \cap W_{l_2}\big(N_{(2)}^-\big)$. The grades $l_1$ and $l_2$ can be read off from figures \ref{fig:Lefschetz-decomposition-Ia} and \ref{fig:Ia-IVd-with-colour} as the height of the position of $\bqz$. This then readily gives 
\beq \label{q0loc_14}
   \bqz \in W_{3}\left(N_1^-\right) \cap W_{2}\big(N_{(2)}^-\big)\ .
\eeq 
Since approaching a point $P \in \Delta_1^\circ$  
 requires to send $\mathrm{Im}\,t^1 \rightarrow \infty$ while keeping $\mathrm{Im}\, t^2$ finite 
 we use the growth theorem \eqref{vt-growth} to read off that
\be
||\bqz|| \sim c\,(\Im t^1)^0 \;,
\ee
which implies that $||\bqz||$ does not tend to $0$ at $P$. Hence the charge orbit $\bQ(\bqz | m_1, m_2)$ is not necessarily massless. In terms of the condition \eqref{massless_q0}, we see that the grade relevant to the type $\rI_a$ divisor is $l_1 = 3$ and it obviously does not satisfy \eqref{massless_q0}. 
 
Let us now turn to the enhancement $\rI_a \to \boxed{\rIV_d}\,$, i.e.~to the case that $P$ is located at $\Delta^{\circ}_{12}$.
We now have to utilize the multi-parameter growth theorem as outlined in section \ref{sec:growth_theorems}.
Using the location \eqref{q0loc_14} in the two-parameter growth \eqref{(1)vt-growth_2} we 
find 
\beq \label{eqn:growth-I-IV}
  ||\bqz|| \sim c\, \frac{1}{\Im t^2} \;.
\eeq 
From this growth we can easily see that the $\bqz$ defined in \eqref{q02pa14} indeed generates a massless charge orbit, 
which is infinite due to \eqref{infiniteness_14}. In order to discuss the path dependence of this result, we first recall that we have fixed the upper sector in \eqref{case(3)}. It is now obvious form \eqref{eqn:growth-I-IV} that $\mathbf{q}_0$ is massless along
any path in this sector approaching $P$ at $t^1 = t^2 = \im \infty$. This confirms that (R1) applies in this case.

\subsubsection{The enhancement $\rII_b \to \rIV_d$}

The other enhancement cases where the type increases can be analysed in the same way. The case we discuss 
next is the enhancement $\rII_b \to \rIV_d$, again considering the two possible locations for $P$.

We first consider placing the $P$ on the type $\rII_b$ divisor, i.e.~$\boxed{\rII_b} \to \rIV_d$.
The decomposition into primitive parts of the type $\rII_b$ mixed Hodge structure $\left(F_{(1)}, W^{(1)}\right)$ is
\begin{equation} \label{eqn:Lefschetz-decomposition-IIb}
  H^3\left(Y_3,\mathbb{C} \right) = \textcolor{red}{P^3(N^-_1)} \oplus \left[ \textcolor{green}{P^4(N^-_1)} \oplus \textcolor{blue}{N^-_1 P^4(N^-_1)}  \right] \;.
\end{equation}
We depict this decomposition in the Hodge-Deligne diamond of $\rII_b$ in figure \ref{fig:Lefschetz-decomposition-IIb}.
\begin{figure}[!h]
\begin{center}
  \begin{tikzpicture}[cm={cos(45),sin(45),-sin(45),cos(45),(15,0)}]
    \draw[step = 1, gray, ultra thin] (0, 0) grid (3, 3);

    \draw[fill, green] (1, 3) circle[radius=0.03];
    \draw[->] (0.9, 2.9) -- (0.1, 2.1);
    \draw[fill, green] (2, 2) circle[radius=0.03] node[above=2pt] {$b$};
    \draw[->] (1.9, 1.9) -- (1.1, 1.1);
    \draw[fill, green] (3, 1) circle[radius=0.03];
    \draw[->] (2.9, 0.9) -- (2.1, 0.1);
    \draw[fill, red] (2, 1) circle[radius=0.03] node[above=2pt] {$b'$};
    \draw[fill, red] (1, 2) circle[radius=0.03] node[above=2pt] {$b'$};
    \draw[fill, blue] (0, 2) circle[radius=0.03];
    \draw[fill, blue] (1, 1) circle[radius=0.03] node[below=2pt] {$b$};
    \draw[fill, blue] (2, 0) circle[radius=0.03];

  \end{tikzpicture}
\end{center}
\caption{The Hodge-Deligne diamond of type $\rII_b$ with its decomposition into primitive parts \eqref{eqn:Lefschetz-decomposition-IIb}. The action of $N_1^-$ are labelled by arrows, and we use colours to highlight the primitive subspaces $P^3(N_1^-)$, $P^4(N_1^-)$ and their images under the action of $N_1^-$. Since the two $\slt$-triples are commuting, the primitive subspaces $P^3(N_1^-)$, $P^4(N_1^-)$ and their images under $N_1^-$ are preserved by $N_2^-$.} \label{fig:Lefschetz-decomposition-IIb}
\end{figure}
The enhancement $\rII_b \to \rIV_d$ is equivalent to a decomposition of the Hodge diamond of $\rIV_d$ as shown in figure \ref{fig:IIb-IVd-with-colour}.
\begin{figure}[!h] 
\begin{center}
  \begin{tabular}{c c c}
    \begin{tikzpicture}[cm={cos(45),sin(45),-sin(45),cos(45),(15,0)}]
      \draw[step = 1, gray, ultra thin] (0, 0) grid (3, 3);
  
      \draw[fill, red] (1, 2) circle[radius=0.03] node[left=2pt] {$b' - r$};
      \draw[fill, red] (2, 2) circle[radius=0.03] node[above=2pt] {$r$};
      \draw[->, red] (1.9, 1.9) -- (1.1, 1.1);
      \draw[fill, red] (1, 1) circle[radius=0.03] node[below=2pt] {$r$};
      \draw[fill, red] (2, 1) circle[radius=0.03] node[right=2pt] {$b' - r$};
    \end{tikzpicture} &
    \begin{tikzpicture}[cm={cos(45),sin(45),-sin(45),cos(45),(15,0)}]
      \draw[step = 1, gray, ultra thin] (0, 0) grid (3, 3);
  
      \draw[fill, green] (3, 3) circle[radius=0.03];
      \draw (3, 3) circle[radius=0.1] node[left=2pt] {$\mathbf{\tilde{a}}_0^{(2)}$};
      \draw[->, green] (2.9, 2.9) -- (2.1, 2.1);
      \draw[fill, green] (2, 2) circle[radius=0.03] node[left=2pt] {$b$};
      \draw[->, green] (1.9, 1.9) -- (1.1, 1.1);
      \draw[fill, green] (1, 1) circle[radius=0.03];
    \end{tikzpicture} &
    \begin{tikzpicture}[cm={cos(45),sin(45),-sin(45),cos(45),(15,0)}]
      \draw[step = 1, gray, ultra thin] (0, 0) grid (3, 3);
  
      \draw[fill, blue] (2, 2) circle[radius=0.03];
      \draw (2, 2) circle[radius=0.1] node[left=2pt] {$\bqz$};
      \draw[->, blue] (1.9, 1.9) -- (1.1, 1.1);
      \draw[fill, blue] (1, 1) circle[radius=0.03] node[right=2pt] {$b$};
      \draw[->, blue] (0.9, 0.9) -- (0.1, 0.1);
      \draw[fill, blue] (0, 0) circle[radius=0.03];
    \end{tikzpicture}\\
    $\textcolor{red}{P^3(N_1^-)}$ & $\textcolor{green}{P^4(N_1^-)}$ & $\textcolor{blue}{N_1^- P^4(N_1^-)}$
  \end{tabular}
\end{center}
\caption{The left picture shows a mixed Hodge structure on $P^3(N_1^-)$, the middle picture a mixed Hodge structure on $P^4(N_1^-)$, and the right picture shows the image of the middle picture under the action of $N_1^-$. In these diamonds, the arrows label the action of $N_2^-$. The colourings are in agreement with equation \eqref{eqn:Lefschetz-decomposition-IIb} and figure \ref{fig:Lefschetz-decomposition-IIb}. The sum of these three Hodge-Deligne diamonds is the diamond of $\left(F_{(2)}, W^{(2)}\right)$ of type $\rIV_d$. Again, $\bqz$ and $\mathbf{\tilde{a}}_0^{(2)}$ are denoted explicitly.
} \label{fig:IIb-IVd-with-colour}
\end{figure}
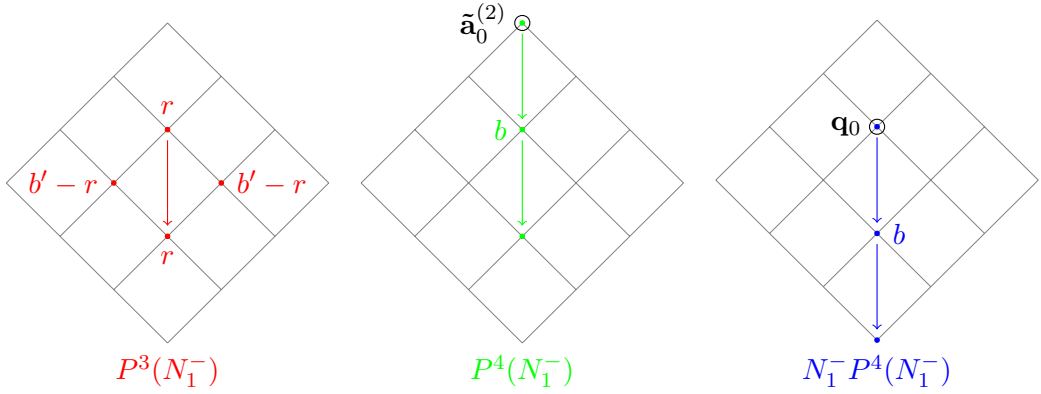

We can now identify the element in $I^{p,q}_{(2)}$ which gives ${\bf q}_0$ as
\be
\bqz  \sim_{\mathbb{Z}} \sumN{1} \mathbf{\tilde a}_0^{(2)} \;.
\label{q02pa24}
\ee
Again from figure \ref{fig:IIb-IVd-with-colour} we see that $\bqz$ is not in the kernel of $N_2^-$ and so we have an infinite orbit
\begin{equation}
  \bQ(\bqz | m_1, m_2 ) = \bqz + m_2 N_2^- \bqz + \frac{1}{2} m_2^2 \big( N_2^- \big)^2 \bqz, \textrm{ for } m_1, m_2 \in \bbZ \;.
\end{equation}
The location of $\bqz$ is determined as well from figure \ref{fig:IIb-IVd-with-colour} to be
\beq
   \bqz \in W_2\left(N_1^-\right) \cap W_4\big(N_{(2)}^-\big)\ .
 \eeq
Considering a path towards $P$ in the Type II locus $\Delta^{\circ}_1$ amounts to keeping $t^2$ finite
and sending $t^1 = \im \infty$. The growth theorem \eqref{vt-growth} thus implies $\lVert \bqz \rVert \sim c\, \frac{1}{\Im t^1}$.
In accord with the condition \eqref{massless_q0} we thus find that $\bqz$ is massless at $P$.
 We therefore deduce that the full infinite charge orbit is massless on $\Delta^{\circ}_1$. 
 This case belongs to the condition (R2) in section \ref{general_chargeorbit} and exemplifies one 
 of the key results of our work.
 
Having identified the orbit we can return to the point discussed in section \ref{sec:spco}, that the orbit should not only contain an infinite number of type II states. This can be easily checked to be the case. In particular, the orbit contains an infinite number of elements with non-vanishing components in $P^{0,2}\left(\Delta^{\circ}_1\right)$, which have non-trivial contraction with ${\bf \tilde{a}}_0^{(1)}$. The fact that the orbit is still infinite, even after a quotient by type II charges as proposed in \cite{Grimm:2018ohb}, will hold for all the cases where we identify such an orbit.

We can also change the position of $P$, considering $\rII_b \to \boxed{\rIV_d}$ instead. Following a similar analysis as above, we find that the choice of seed charge
\begin{equation}
  \bqz \sim_{\bbZ} \sumN{1} \sumN{2} \mathbf{\tilde{a}}_0^{(2)} \;,
\end{equation}
yields an infinite massless charge orbit $\bQ(\bqz | m_1, m_2)$ at $\Delta^{\circ}_{12}$. It is useful to notice that we have used the $N_{(2)}^-$ which is at the type $\rIV_d$ divisor, and the $N_{(1)}^-$ which is at the type $\rII_b$ divisor just before the enhancement. This fact is crucial in defining the corresponding general version of the charge orbit in table \ref{tab:Chains_with_q0}. We also remark that such a $\bqz$ always exists, because the enhancement condition \eqref{sing_enhancements} for $\rII_b \to \rIV_d$ requires that $b \ge 1$. This case belongs to the condition (R1) in section \ref{general_chargeorbit}.

\subsubsection{The enhancement $\rIII_c \to \rIV_d$}
Turning to the case $\rIII_c \to \rIV_d$
we follow the same procedure as the previous two cases, first considering $\boxed{\rIII_c} \to \rIV_d$. The decomposition into primitive parts of the type $\rIII_c$ mixed Hodge structure $\left(F_{(1)}, W^{(1)}\right)$ is
\begin{equation} \label{eqn:Lefschetz-decomposition-IIIc}
  H^3\left(Y_3,\mathbb{C} \right) = \textcolor{red}{P^3(N_1^-)} \oplus \left[ \textcolor{green}{P^4(N_1^-)} \oplus \textcolor{olive}{N_1^-P^4(N_1^-)} \right] \oplus \left[ \textcolor{blue}{P^5(N_1^-)} \oplus \textcolor{cyan}{N_1^- P^5(N_1^-)} \oplus \textcolor{teal}{\left(N_1^-\right)^2 P^5(N_1^-)} \right].
\end{equation}
We depict this decomposition in the Hodge-Deligne diamond of $\rIII_c$ in figure \ref{fig:Lefschetz-decomposition-IIIc}.
\begin{figure}[!h]
\begin{center}
  \begin{tikzpicture}[cm={cos(45),sin(45),-sin(45),cos(45),(15,0)}]
    \draw[step = 1, gray, ultra thin] (0, 0) grid (3, 3);
    
    \draw[fill, red] (2.1, 0.9) circle[radius=0.03] node[right=2pt] {$c' - 1$};
    \draw[fill, red] (0.9, 2.1) circle[radius=0.03] node[left=2pt] {$c' - 1$};
    \draw[fill, olive] (1, 1) circle[radius=0.03] node[below=2pt] {$c$};
    \draw[->] (1.9, 1.9) -- (1.1, 1.1);
    \draw[fill, green] (2, 2) circle[radius=0.03] node[above=2pt] {$c$};
    \draw[fill, blue] (2, 3) circle[radius=0.03];
    \draw[->] (1.9, 2.9) -- (1.1, 2.1);
    \draw[fill, blue] (3, 2) circle[radius=0.03];
    \draw[->] (2.9, 1.9) -- (2.1, 1.1);
    \draw[fill, cyan] (2, 1) circle[radius=0.03];
    \draw[->] (0.9, 1.9) -- (0.1, 1.1);
    \draw[fill, cyan] (1, 2) circle[radius=0.03];
    \draw[->] (1.9, 0.9) -- (1.1, 0.1);
    \draw[fill, teal] (1, 0) circle[radius=0.03];
    \draw[fill, teal] (0, 1) circle[radius=0.03];
    \end{tikzpicture}
\end{center}
\caption{The Hodge-Deligne diamond of type $\rIII_c$ with its decomposition into primitive parts \eqref{eqn:Lefschetz-decomposition-IIIc}. The action of $N_1^-$ are labelled by arrows, and we use colours to highlight the primitive subspaces $P^3(N_1^-)$, $P^4(N_1^-)$, $P^5(N_1^-)$ and their images under the action of $N_1^-$. Since the two $\slt$-triples are commuting, the primitive subspaces $P^3(N_1^-)$, $P^4(N_1^-)$, $P^5(N_1^-)$ and their images under $N_1^-$ are preserved by $N_2^-$.} \label{fig:Lefschetz-decomposition-IIIc}
\end{figure}
The enhancement $\rIII_c \to \rIV_d$ is equivalent to a decomposition of the Hodge diamond of $\rIV_d$ as shown in figure \ref{fig:IIIc-IVd-with-colour}.
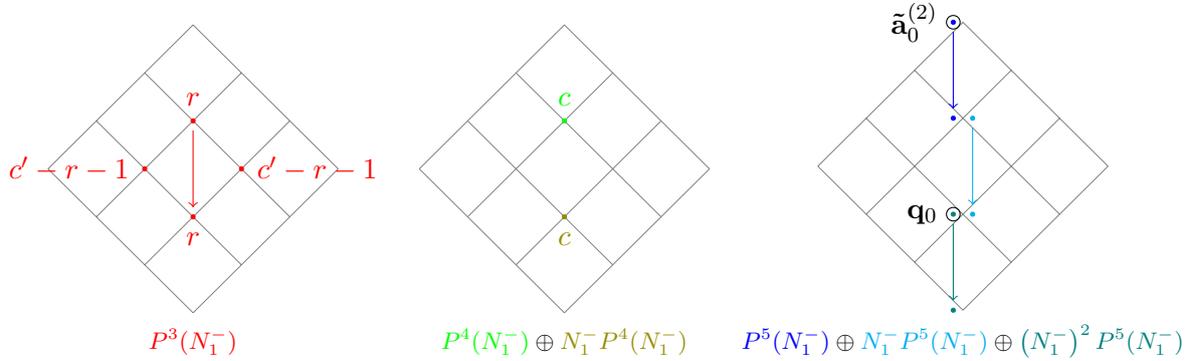
\begin{figure}[!h]
\begin{center}
  \begin{tabular}{c c c}
    \begin{tikzpicture}[scale=0.9, cm={cos(45),sin(45),-sin(45),cos(45),(15,0)}]
      \draw[step = 1, gray, ultra thin] (0, 0) grid (3, 3);
  
      \draw[fill, red] (1, 2) circle[radius=0.03] node[left=2pt] {$c' - r - 1$};
      \draw[fill, red] (2, 2) circle[radius=0.03] node[above=2pt] {$r$};
      \draw[->, red] (1.9, 1.9) -- (1.1, 1.1);
      \draw[fill, red] (1, 1) circle[radius=0.03] node[below=2pt] {$r$};
      \draw[fill, red] (2, 1) circle[radius=0.03] node[right=2pt] {$c' - r - 1$};
    \end{tikzpicture} & 
    \begin{tikzpicture}[scale=0.9, cm={cos(45),sin(45),-sin(45),cos(45),(15,0)}]
      \draw[step = 1, gray, ultra thin] (0, 0) grid (3, 3);
  
      \draw[fill, green] (2, 2) circle[radius=0.03] node[above=2pt] {$c$};
      \draw[fill, olive] (1, 1) circle[radius=0.03] node[below=2pt] {$c$};
    \end{tikzpicture} &
    \begin{tikzpicture}[scale=0.9, cm={cos(45),sin(45),-sin(45),cos(45),(15,0)}]
      \draw[step = 1, gray, ultra thin] (0, 0) grid (3, 3);
    
      \draw[fill, blue] (2.9, 3.1) circle[radius=0.03];
      \draw (2.9, 3.1) circle[radius=0.1] node[left=2pt] {$\mathbf{\tilde a}_0^{(2)}$};
      \draw[->, blue] (2.8, 3) -- (2.0, 2.2);
      \draw[fill, blue] (1.9, 2.1) circle[radius=0.03];
      \draw[fill, cyan] (2.1, 1.9) circle[radius=0.03];
      \draw[->, cyan] (2, 1.8) -- (1.2, 1);
      \draw[fill, cyan] (1.1, 0.9) circle[radius=0.03];
      \draw[fill, teal] (0.9, 1.1) circle[radius=0.03];
      \draw (0.9, 1.1) circle[radius=0.1] node[left=2pt] {$\bqz$};
      \draw[->, teal] (0.8, 1) -- (0, 0.2);
      \draw[fill, teal] (-0.1, 0.1) circle[radius=0.03];
    \end{tikzpicture}\\
    \footnotesize$\textcolor{red}{P^3(N^-_1)}$ & \footnotesize$\textcolor{green}{P^4(N^-_1)} \oplus \textcolor{olive}{N^-_1P^4(N^-_1)}$ & \footnotesize$\textcolor{blue}{P^5(N^-_1)} \oplus \textcolor{cyan}{N^-_1 P^5(N^-_1)} \oplus \textcolor{teal}{\left(N^-_1\right)^2 P^5(N^-_1)}$
  \end{tabular}
\end{center}
\caption{Pictures showing the mixed Hodge structures induced on $P^3(N_1^-)$, $P^4(N_1^-)$ and $P^5(N_1^-)$, together with their images under the action of $N_1^-$ and $(N_1^-)^2$.  In these diamonds, the coloured arrows label the action of $N_2^-$. The colourings are in agreement with \eqref{eqn:Lefschetz-decomposition-IIIc} and figure \ref{fig:Lefschetz-decomposition-IIIc}. The sum of these three Hodge-Deligne diamonds is the diamond of $(F_{(2)}, W^{(2)})$ of type $\rIV_d$. As before, $\bqz$ and $\mathbf{\tilde{a}}_0^{(2)}$ are denoted explicitly.} \label{fig:IIIc-IVd-with-colour}
\end{figure}

In this case we have that $\bqz$ is given by
\be
\bqz  \sim_{\mathbb{Z}} \sumNsqr{1} \mathbf{\tilde a}_0^{(2)} \;.
\label{q02pa34}
\ee 
and we see that there is an infinite orbit
\begin{equation}
  \bQ(\bqz | m_1, m_2) = \bqz + m_2 N_2^- \bqz, \textrm{ for } m_1, m_2 \in \bbZ \;.
\end{equation}
The location $\bqz \in W_1(N_1^-) \cap W_2(N_{(2)}^-)$ implies by using \eqref{vt-growth} the asymptotics 
$\lVert \bqz \rVert  \sim c\, \left( \Im t^1 \right)^{-2}$ in the limit $t^1\rightarrow \im \infty$. 
Therefore, again for $P \in \Delta^{\circ}_1$ we have an infinite massless charge orbit. 
This case belongs to the condition (R2) in section \ref{general_chargeorbit}.

We can also explore the candidate $\bqz$ for the enhancement $\rIII_c \to \boxed{\rIV_d}$ and we find the same seed $\bqz$ as in \eqref{q02pa34}. The orbit stays massless approaching $\Delta_{12}$ along any path in the considered growth sector. It is useful to notice that in defining the seed charge $\bqz$ around $\rIV_d$, we are using the $\sumN{1}$ which is at the type $\rIII_c$ divisor just before the enhancement. This fact is crucial in defining the corresponding general charge orbit in table \ref{tab:Chains_with_q0}. This case belongs to condition (R1) in section \ref{general_chargeorbit}.

\subsubsection{The enhancement $\rII_b \to \rIII_c$} \label{sec:II-III}

Let us next consider $\rII_b \to \rIII_c$ and first focus on $\boxed{\rII_b} \to \rIII_c$. Following the same procedure as the previous cases, we refer to equation \eqref{eqn:Lefschetz-decomposition-IIb} and figure \ref{fig:Lefschetz-decomposition-IIb} for the decomposition into primitive parts of the type $\rII_b$ mixed Hodge structure $\left(F_{(1)}, W^{(1)}\right)$. Then the enhancement $\rII_b \to \rIII_c$ is equivalent to a decomposition of the Hodge diamond of $\rIII_c$ as shown in figure \ref{fig:IIb-IIIc-with-colour}.
\begin{figure}[!h]
\begin{center}
\begin{tabular}{c c c}
  \begin{tikzpicture}[cm={cos(45),sin(45),-sin(45),cos(45),(15,0)}]
    \draw[step = 1, gray, ultra thin] (0, 0) grid (3, 3);

    \draw[fill, red] (1, 2) circle[radius=0.03] node[left=2pt] {$b' - r$};
    \draw[fill, red] (2, 2) circle[radius=0.03] node[above=2pt] {$r$};
    \draw[->, red] (1.9, 1.9) -- (1.1, 1.1);
    \draw[fill, red] (1, 1) circle[radius=0.03] node[below=2pt] {$r$};
    \draw[fill, red] (2, 1) circle[radius=0.03] node[right=2pt] {$b' - r$};
  \end{tikzpicture} &
  \begin{tikzpicture}[cm={cos(45),sin(45),-sin(45),cos(45),(15,0)}]
    \draw[step = 1, gray, ultra thin] (0, 0) grid (3, 3);

    \draw[fill, green] (3, 2) circle[radius=0.03];
    \draw (3, 2) circle[radius=0.1] node[right=2pt] {$\mathbf{\tilde{a}}_0^{(2)}$};
    \draw[->, green] (2.9, 1.9) -- (2.1, 1.1);
    \draw[fill, green] (2, 1) circle[radius=0.03];
    \draw[fill, green] (2, 3) circle[radius=0.03];
    \draw[->, green] (1.9, 2.9) -- (1.1, 2.1);
    \draw[fill, green] (1, 2) circle[radius=0.03];
    \draw[fill, green] (2, 2) circle[radius=0.03] node[above=2pt] {$b - 2$};
  \end{tikzpicture} &
  \begin{tikzpicture}[cm={cos(45),sin(45),-sin(45),cos(45),(15,0)}]
    \draw[step = 1, gray, ultra thin] (0, 0) grid (3, 3);

    \draw[fill, blue] (2, 1) circle[radius=0.03];
    \draw (2, 1) circle[radius=0.1] node[right=2pt] {$\bqz$};
    \draw[->, blue] (1.9, 0.9) -- (1.1, 0.1);
    \draw[fill, blue] (1, 0) circle[radius=0.03];
    \draw[fill, blue] (1, 2) circle[radius=0.03];
    \draw[->, blue] (0.9, 1.9) -- (0.1, 1.1);
    \draw[fill, blue] (0, 1) circle[radius=0.03];
    \draw[fill, blue] (1, 1) circle[radius=0.03] node[below=2pt] {$b - 2$};
  \end{tikzpicture}\\
  $\textcolor{red}{P^3(N_1^-)}$ & $\textcolor{green}{P^4(N_1^-)}$ & $\textcolor{blue}{N_1^- P^4(N_1^-)}$
\end{tabular}
\end{center}
\caption{The left picture shows a mixed Hodge structure, determined by some non-negative integer $r$, on $P^3(N_1^-)$. The middle picture shows a mixed Hodge structure on $P^4(N_1^-)$. The right picture shows the image of the middle picture under the action of $N_1^-$. In these diamonds, the coloured arrows label the action of $N_2^-$. The colourings are in agreement with equation \eqref{eqn:Lefschetz-decomposition-IIb} and figure \ref{fig:Lefschetz-decomposition-IIb}. The sum of these three Hodge-Deligne diamonds is the diamond of $\left(F_{(2)}, W^{(2)}\right)$, associated to the mixed Hodge structure $I^{p,q}_{(2)}$, of type $\rIII_c$. The circle around the dot in the last diamond indicates the location of $\bqz$, and the circle around the dot in the middle diamond indicates the location of  $\mathbf{\tilde{a}}_0^{(2)}$.} \label{fig:IIb-IIIc-with-colour}
\end{figure}

In this case, the $\bqz$ is chosen to be
\be
{\bf q}_0 \sim_{\mathbb{Z}} \sumN{1} \mathbf{\tilde a}_0^{(2)} \;.
\label{q02pa23}
\ee
and we see that the orbit
\begin{equation} \label{infinte_2}
  \bQ(\bqz | m_1, m_2 ) = \bqz + m_2 N_2^- \bqz, \textrm{ for } m_1, m_2 \in \bbZ
\end{equation}
is indeed infinite.

The location of $\bqz$ is determined to be 
\beq  \label{q0-location-23_}
   \bqz \in W_2\left(N_1^-\right) \cap W_3\big(N_{(2)}^-\big) .
\eeq 
This implies that for $P$ on $\Delta^{\circ}_1$, i.e.~when taking the limit $t^1 \rightarrow \im \infty$, 
we find by using \eqref{vt-growth} that $||\bqz|| \sim c \, (\mathrm{Im}\;t^1)^{-1}$. Together with 
\eqref{infinte_2} we have an infinite massless charge orbit. This case belongs to the condition (R2) in section 
\ref{general_chargeorbit}.

We now turn to the situation $\rII_b \to \boxed{\rIII_c}\,$. As we will show, in this case, 
the masslessness of the charge orbit around the type $\rIII_c$ divisor will depend on the path along which we approach it. 
For concreteness our choice of $\bqz$ is still \eqref{q02pa23}, but it is important to note that one cannot find 
any other $\mathbf{q}_0$ that generates an infinite orbit and is path-independently massless. 
The fate of the orbit as we approach the point $P$ on the type $\rIII_c$ divisor is different from the previous cases. In 
fact, using the growth theorem \eqref{(1)vt-growth_2} for the $\mathbf{q}_0$-locations \eqref{q0-location-23_} one finds
\beq
||\bqz|| \sim c\, \frac{\mathrm{Im}\;t^2}{\mathrm{Im}\;t^1} \;
\label{bqzt1t2cii}
\eeq
in the upper growth region in \eqref{case(3)}. We thus conclude that the charge orbit remains massless at $P$ if we approach it with a path satisfying 
\be
\mathrm{Massless\;Path\;}: \; \mathrm{Im}\;t^2\rightarrow \infty ,\; \mathrm{Im}\;t^1 \rightarrow \infty \;,\;\; \mathrm{such\;that} \;\frac{\mathrm{Im}\;t^2}{\mathrm{Im}\;t^1} \rightarrow 0 \;.
\ee
The only other possible path, compatible with the considered growth sector, is 
\be
\mathrm{Massive\;Path\;}: \; \mathrm{Im}\;t^2\rightarrow \infty ,\; \mathrm{Im}\;t^1 \rightarrow \infty \;,\;\; \mathrm{such\;that} \;\frac{\mathrm{Im}\;t^2}{\mathrm{Im}\;t^1} \rightarrow \lambda > 0  \;.
\ee
In other words, we cannot claim that the considered $\bqz$ is actually massless independent of the path. Therefore, this case was excluded from the conditions (R1), (R2) specifying our general construction. 
Clearly, in this case also the location \eqref{q0-location-23_} of $\bqz$ does not satisfy the condition \eqref{massless_q0}. This case belongs to the situation described at the end of section \ref{general_chargeorbit}.

\subsubsection{A case without type enhancement $\rII_b \to \rII_c$}

\begin{figure}[!h]
\begin{center}
\begin{tabular}{c c c}
  \begin{tikzpicture}[cm={cos(45),sin(45),-sin(45),cos(45),(15,0)}]
    \draw[step = 1, gray, ultra thin] (0, 0) grid (3, 3);

    \draw[fill, red] (1, 2) circle[radius=0.03] node[left=2pt] {$b' - r$};
    \draw[fill, red] (2, 2) circle[radius=0.03] node[above=2pt] {$r$};
    \draw[->, red] (1.9, 1.9) -- (1.1, 1.1);
    \draw[fill, red] (1, 1) circle[radius=0.03] node[below=2pt] {$r$};
    \draw[fill, red] (2, 1) circle[radius=0.03] node[right=2pt] {$b' - r$};
  \end{tikzpicture} &
  \begin{tikzpicture}[cm={cos(45),sin(45),-sin(45),cos(45),(15,0)}]
    \draw[step = 1, gray, ultra thin] (0, 0) grid (3, 3);

    \draw[fill, green] (3, 1) circle[radius=0.03];
    \draw (3, 1) circle[radius=0.1] node[above=2pt] {$\mathbf{\tilde{a}}_0^{(2)}$};
    \draw[fill, green] (2, 2) circle[radius=0.03] node[above=2pt] {$b$};
    \draw[fill, green] (1, 3) circle[radius=0.03];
  \end{tikzpicture} &
  \begin{tikzpicture}[cm={cos(45),sin(45),-sin(45),cos(45),(15,0)}]
    \draw[step = 1, gray, ultra thin] (0, 0) grid (3, 3);

    \draw[fill, blue] (2, 0) circle[radius=0.03];
    \draw[fill, blue] (1, 1) circle[radius=0.03] node[below=2pt] {$b$};
    \draw[fill, blue] (0, 2) circle[radius=0.03];
  \end{tikzpicture}\\
  $\textcolor{red}{P^3(N_1^-)}$ & $\textcolor{green}{P^4(N_1^-)}$ & $\textcolor{blue}{N_1^- P^4(N_1^-)}$
\end{tabular}
\end{center}
\caption{The left picture shows a mixed Hodge structure on $P^3(N_1^-)$ of weight $3$ with Hodge numbers $(0, b', b', 0)$. The middle picture shows a mixed Hodge structure on $P^4(N_1^-)$ of weight $4$ with Hodge numbers $(0, 1, b, 1, 0)$. The right picture shows the image of the middle picture under the action of $N_1^-$. In these diamonds, the coloured arrows label the action of $N_2^-$. The colourings are in agreement with equation \eqref{eqn:Lefschetz-decomposition-IIb} and figure \ref{fig:Lefschetz-decomposition-IIb}. The sum of these three Hodge-Deligne diamonds is the diamond of $(F_{(2)}, W^{(2)})$ of type $\rII_c$. The circle around the dot in the middle diamond indicates the location of $\mathbf{\tilde{a}}_0^{(2)}$.} \label{fig:IIb-IIc-with-colour}
\end{figure}
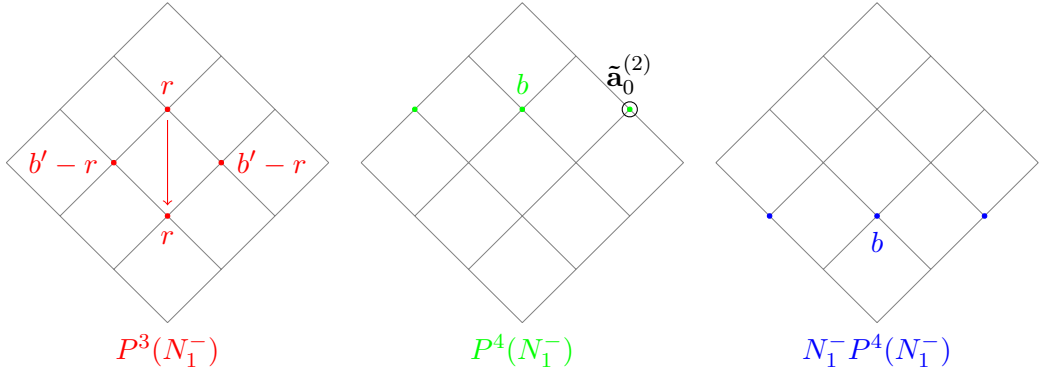

To end our two-divisor analysis let us explore a case where no type enhancement is present. 
As usual, the decomposition into primitive parts of the type $\rII_b$ mixed Hodge structure 
$\left(F_{(1)}, W^{(1)}\right)$ is given by equation \eqref{eqn:Lefschetz-decomposition-IIb}, and it is depicted in figure \ref{fig:Lefschetz-decomposition-IIb}. Then $\rII_b \to \rII_c$ is equivalent to a decomposition of the Hodge-Deligne diamond of type $\rII_c$ shown in figure \ref{fig:IIb-IIc-with-colour}.

If we try to find a $\bqz$ such that the generated orbit is infinite and massless at either $\rII_b$ or $\rII_c$ following the methods in previous cases, then we realise that such a $\bqz$ does not exist. In particular if $b = c$, meaning that there is no enhancement at all, then the second $\slt$-triple is trivial
\begin{equation}
  (N_2^-, N_2^+, Y_2) = (0, 0, 0).
\end{equation}

Nevertheless, this case is relevant in the discussion of multi-divisor enhancements in the following section. To exemplify this 
we consider the following simple case of a $3$-term enhancement chain
\begin{equation}
  \rII_a \to \boxed{\rII_b} \to \rIII_c \label{II-II-III}\, .
\end{equation}
In this chain we have already considered in \eqref{q02pa23} a $\bqz$ from the last step of type enhancement. The next step is to estimate the Hodge norm of $\bqz$ and this requires the location of $\bqz$ in every monodromy weight filtration of the mixed Hodge structures of type $\rII_a, \rII_b$ and $\rIII_c$. According to the analysis in subsection \ref{sec:II-III}, we have $\bqz$ staying in $W_2(\sumN{2})$ and $W_3(\sumN{3})$. Then the analysis in this section tells us that generally $\bqz \in W_3(\sumN{1})$. It could still be possible that we have $\bqz \in W_2(\sumN{1})$. However, either of these two possible locations satisfies the massless condition \eqref{massless_q0} and hence implies that the charge orbit generated by $\bqz$ is massless at $P$ located at the type $\rII_b$ singular locus. Analogously we can also analyse the other cases without type enhancements $\rI_a \to \rI_b$, $\rIII_a \to \rIII_b$ and $\rIV_a \to \rIV_b$. The results of this analysis are similar to the 
$ \rII_a \to  \rII_b$ case and will be used to justify the analysis in the next section. In particular, it will allow us to introduce the notation \eqref{W-notation}, which indicates that all enhancements of non-changing type will not influence our constructions.

This completes our  two-divisor analysis. We will now use these results to perform the general multi-divisor analysis.

\subsection{The general multi-divisor analysis} \label{multi-N-analysis}

In the previous subsection we have shown when in the case of two intersecting divisors it is possible to identify an infinite massless charge orbit depending on the type of singularity of the divisors and at the intersection as well as the location of $P$. 
In this subsection we will generalise the analysis to multiple intersecting divisors. We will first give all possible 
enhancement chains and then stepwise apply the two-divisor result by treating the two intersecting divisors which 
themselves are loci of intersection of an arbitrary number of divisors. This is the general setup described in subsection \ref{general_chargeorbit}. By explicitly constructing $\bqz$ we will thus be able to show the conditions (R1), (R2)
for it to generate an infinite massless charge orbit when approaching $P$. 

\subsubsection{Masslessness of the general charge orbit}

In this subsection we show that one can construct for each enhancement 
chain \eqref{enhance_chain} an appropriate seed charge $\bqz$
that defines a massless state when approaching $P$ along any 
path in a fixed growth sector \eqref{path1}. Crucially, as 
stated already in subsection \ref{general_chargeorbit}, such a $\bqz$ only exists if the 
singularity type at the location of $P$ is either II, III, IV. These are also 
the singularities that occur if we demand $P$ to be at infinite distance. 

We thus have to consider the three following general enhancement chains
\begin{align}
  & \text{I}_{a_1}  \shortrightarrow  \shdots \shortrightarrow \text{I}_{a_k} \shortrightarrow   
   \text{II}_{b_1}  \shortrightarrow  \shdots \shortrightarrow  \boxed{\text{II}_{b_m}} \shortrightarrow \shdots \label{TypeIIchain} \\
   & \text{I}_{a_1}  \shortrightarrow  \shdots \shortrightarrow \text{I}_{a_k} \shortrightarrow   
   \text{II}_{b_1}  \shortrightarrow  \shdots \shortrightarrow  \text{II}_{b_m} \shortrightarrow 
    \text{III}_{c_1}  \shortrightarrow  \shdots \shortrightarrow  \boxed{\text{III}_{c_n}}\shortrightarrow \shdots \label{TypeIIIchain}\\
    & \text{I}_{a_1}  \shortrightarrow  \shdots \shortrightarrow \text{I}_{a_k} \shortrightarrow   
   \text{II}_{b_1}  \shortrightarrow  \shdots \shortrightarrow  \text{II}_{b_m} \shortrightarrow 
    \text{III}_{c_1}  \shortrightarrow  \shdots \shortrightarrow  \text{III}_{c_n}\shortrightarrow
     \text{IV}_{d_1}  \shortrightarrow  \shdots \shortrightarrow  \boxed{\text{IV}_{d_r}}\shortrightarrow \shdots\ , \label{TypeIVchain}
\end{align}
where the box indicates the singularity at the location of $P$. Note that $k$, $m$, $n$, and $r$ 
are integers and we allow for chains that do not admit all types. For example, in \eqref{TypeIIchain}, \eqref{TypeIIIchain}, and \eqref{TypeIVchain}
we can have $k=0$, i.e.~start the enhancement at type II. Furthermore, let us stress that we 
have only displayed the enhancement chains until the singularity at $P$. This part will be relevant 
in studying the masslessness of the associated $\mathbf{q}_0$ as we will see below. In order 
to show that the full orbit $\mathbf{Q}$ is infinite, the enhancements after the singularity at $P$ 
become relevant.  We will discuss these parts in subsection \ref{sec:infinteness}. 

It will turn out to be sufficient 
to only focus on the type I, II, III, IV without having information about the index required 
in the complete classification of subsection \ref{sec:sing_class}. Since we also want to simplify the 
expressions,  we thus introduce 
the shorthand notation
\bea
  \text{I}   & \equiv &    \text{I}_{a_1}  \rightarrow  \ldots \rightarrow \text{I}_{a_k} \ , \nn \\
  \text{II}   & \equiv &    \text{II}_{b_1}  \rightarrow  \ldots \rightarrow \text{II}_{b_m} \ , \label{Type-notation}\\
    \text{III}   & \equiv &    \text{III}_{c_1}  \rightarrow  \ldots \rightarrow \text{III}_{c_n} \ ,\nn \\
      \text{IV}   & \equiv &    \text{IV}_{d_1}  \rightarrow  \ldots \rightarrow \text{IV}_{d_p} \ .\nn
\eea
Now it is straightforward to display all appearing enhancement chains that can occur before 
the singularity at $P$. We list them in the first column of table \ref{tab:Chains_with_q0}.

\begin{table}[!h!]
\centering
{\small \begin{tabular}{|r|r|c|}
\hline
 \multicolumn{1}{|c|}{\rule[-0.3cm]{0cm}{0.8cm} Chain}                           & \multicolumn{1}{c|}{$\quad \mathbf{q}_0$   }               & location of $\bqz$\\
\hline\hline
\rule[-0.27cm]{0cm}{0.8cm}$\oneCh{\rII}$                    & $\sumN{n_P} \taE$               & $W_2\big(N^-_{(\text{II})} \big)$\\
\hline
\rule[-0.27cm]{0cm}{0.8cm}$\twoCh{\rI}{\rII}$               & $\sumN{n_P} \taE$               & $W_3\big(N^-_{(\text{I})} \big) \cap W_2\big(N^-_{(\text{II})} \big)$\\
\hline\hline
\rule[-0.27cm]{0cm}{0.8cm}$\oneCh{\rIII}$                   & $\sumNsqr{n_P} \taE$            & $W_1\big(N^-_{(\text{III})} \big)$\\
\hline
\rule[-0.27cm]{0cm}{0.8cm}$\twoCh{\rI}{\rIII}$              & $\sumNsqr{n_P} \taE$            & $W_3\big(N^-_{(\text{I})} \big) \cap W_1\big(N^-_{(\text{III})} \big)$\\
\hline
\rule[-0.27cm]{0cm}{0.8cm}$\twoCh{\rII}{\rIII}$             & $\sumNsqr{n_P} \taE$            & $W_2\big(N^-_{(\text{II})} \big) \cap W_1\big(N^-_{(\text{III})} \big)$\\
\hline
\rule[-0.27cm]{0cm}{0.8cm}$\threeCh{\rI}{\rII}{\rIII}$      & $\sumNsqr{n_P} \taE$            & $W_3\big(N^-_{(\text{I})} \big) \cap W_2\big(N^-_{(\text{II})} \big)  \cap W_1\big(N^-_{(\text{III})} \big)$\\
\hline\hline
\rule[-0.27cm]{0cm}{0.8cm}$\oneCh{\rIV}$                    & $\sumNsqr{n_P} \taE$            &  $W_2\big(N^-_{(\text{IV})} \big)$\\
\hline
\rule[-0.27cm]{0cm}{0.8cm}$\twoCh{\rI}{\rIV}$               & $\sumNsqr{n_P} \taE$            &  $W_3\big(N^-_{(\text{I})} \big)  \cap W_2\big(N^-_{(\text{IV})} \big)$\\
\hline
\rule[-0.27cm]{0cm}{0.8cm}$\twoCh{\rII}{\rIV}$              & $\sumN{n_P - r}\sumN{n_P} \taE$ & $W_2\big(N^-_{(\text{II})} \big)   \cap W_2\big(N^-_{(\text{IV})} \big)$\\
\hline
\rule[-0.27cm]{0cm}{0.8cm}$\twoCh{\rIII}{\rIV}$             & $\sumNsqr{n_P - r} \taE$        & $W_1\big(N^-_{(\text{III})} \big) \cap W_2\big(N^-_{(\text{IV})} \big)$\\
\hline
\rule[-0.27cm]{0cm}{0.8cm}$\threeCh{\rI}{\rII}{\rIV}$       & $\sumN{n_P - r}\sumN{n_P} \taE$ & $W_3\big(N^-_{(\text{I})} \big) \cap W_2\big(N^-_{(\text{II})} \big) \cap W_2\big(N^-_{(\text{IV})} \big)$\\
\hline
\rule[-0.27cm]{0cm}{0.8cm}$\threeCh{\rI}{\rIII}{\rIV}$      & $\sumNsqr{n_P} \taE$            & $W_3\big(N^-_{(\text{I})} \big) \cap W_1\big(N^-_{(\text{III})} \big) \cap W_2\big(N^-_{(\text{IV})} \big)$\\
\hline
\rule[-0.27cm]{0cm}{0.8cm}$\threeCh{\rII}{\rIII}{\rIV}$     & $\sumNsqr{n_P - r} \taE$        & $\big(N^-_{(\text{II})} \big) \cap W_1\big(N^-_{(\text{III})} \big) \cap W_2\big(N^-_{(\text{IV})} \big)$\\
\hline
\rule[-0.27cm]{0cm}{0.8cm}$\fourCh{\rI}{\rII}{\rIII}{\rIV}$ & $\sumNsqr{n_P - r} \taE$        & $W_3\big(N^-_{(\text{I})} \big) \cap W_2\big(N^-_{(\text{II})} \big)  \cap W_1\big(N^-_{(\text{III})} \big) \cap W_2\big(N^-_{(\text{IV})} \big)$\\
\hline
\end{tabular}}
\caption{The table contains all possible enhancement chains that can arise before the singularity at $P$. We use the 
notation \eqref{Type-notation} in the first column. The  $\bqz$ associated to each chain is listed in the second column. Note 
that $N_{(n_P-r)}^-$ is the element associated to the last type III singularity in the locus, with $r$ as in \eqref{TypeIVchain}. 
The third column lists the location of $\bqz$ using the notation introduced in \eqref{W-notation}.} \label{tab:Chains_with_q0}
\end{table}

The second column of table \ref{tab:Chains_with_q0} lists the seed charge $\mathbf{q}_0$
that we propose for the corresponding chain. This charge has been constructed
such that it has a universal location in the spaces $W(N^-_{(k)})$ relevant in the 
growth theorem of subsection \ref{sec:growth_theorems}. In fact, tracking $\bqz$ through the 
various enhancements as in subsection \ref{sec:cantwodiv} we 
find for the three general chains \eqref{TypeIIchain}-\eqref{TypeIVchain} the locations
\begin{align}
  & P \in \text{Type II locus}: &\quad&\bqz \in W_3\big(N^-_{(\text{I})} \big)  \cap
   W_2\big(N^-_{(\text{II})} \big)  &
    \label{TypeIIchainW} \\
   & P \in \text{Type III locus}: &\quad&\bqz \in W_3\big(N^-_{(\text{I})} \big)  \cap
   W_2\big(N^-_{(\text{II})} \big)  \cap W_1\big(N^-_{(\text{III})} \big) &  \label{TypeIIIchainW}\\
    & P \in \text{Type IV locus}: &\quad&\bqz \in W_3\big(N^-_{(\text{I})} \big)  \cap
   W_2\big(N^-_{(\text{II})} \big)  \cap W_1\big(N^-_{(\text{III})} \big) \cap W_2\big(N^-_{(\text{IV})} \big)  \ , & \label{TypeIVchainW}
\end{align}
where we have introduced the shorthand notation
\footnote{Note the unusual pattern in $W_2\big(\sumN{\rII}\big)$, which contains an additional parameter $k'$ with $k + 1 \le k' \le k + m$. The crucial information here is that if ${\sf Type\ A}_{n_P} = \rII$, then we must have $\bqz \in W_2\big(\sumN{n_P}\big)$ to ensure masslessness. Before the end of the type $\rII$ chain, the location of an $\bqz$ could be pushed up by $1$ to $W_3$ but this will not affect the masslessness. In $W\big(\sumN{\rI}\big), W\big(\sumN{\rIII}\big)$ and $W\big(\sumN{\rIV}\big)$, such a phenomenon is not present. This justifies the shorthand notation \eqref{W-notation}.}
\begin{align}
  W_3\big(N^-_{(\text{I})} \big)   & \equiv W_3\big(N^-_{(1)} \big) \cap \shdots \cap W_3\big(N^-_{(k)} \big) \, , \nn\\
  W_2\big(N^-_{(\text{II})} \big)  & \equiv W_3\big(N^-_{(k+1)} \big)  \cap  \shdots \cap W_3\big(N^-_{(k')} \big) \cap W_2\big(N^-_{(k'+1)} \big) \cap \shdots \cap W_2\big(N^-_{(k+m)} \big) \, , \label{W-notation}\\
  W_1\big(N^-_{(\text{III})} \big) & \equiv W_1\big(N^-_{(k+m+1)} \big) \cap \shdots \cap W_1\big(N^-_{(k+m+n)} \big) \, , \nn \\
  W_2\big(N^-_{(\text{IV})} \big)  & \equiv W_2\big(N^-_{(k+m+n+1)} \big) \cap \shdots \cap W_2\big(N^-_{(k+m+n+r)} \big) \, . \nn
\end{align}
Note that the $W_l(N^-_{(i)})$ in each line  \eqref{TypeIIchainW}-\eqref{TypeIVchainW} is always $W_l\big(N^-_{(n_P)} \big)$ corresponding 
to the location of $P$. The shorthand notation \eqref{W-notation} is also used in the last column of table \ref{tab:Chains_with_q0}
giving the location of the listed $\mathbf{q}_0$.

We now collected all the information to show that $\mathbf{q}_0$ becomes massless along any path approaching 
$P$ within a growth sector. This is straightforward since we have already established the general result \eqref{massless_q0},
which gives a sufficient condition for this behaviour. It is easy to check using the last column of table \ref{tab:Chains_with_q0}
that \eqref{massless_q0} is satisfied.

\subsubsection{Infiniteness of the general charge orbit} \label{sec:infinteness}

Having shown the masslessness of the charge orbit $\bQ(\bqz | m_1, \ldots, m_{n_\mathcal{E}})$, we will in this subsection check its infiniteness. The procedure is similar to the one used in subsection \ref{sec:cantwodiv}. Let us first repeat the definition of the charge orbit \eqref{charge-orbit_gen} and expand the exponential
\begin{align}
  \bQ(\bqz | m_1, \ldots, m_{n_\mathcal{E}}) & = \exp\Big(\sum_{i = 1}^{n_\cE} m_i N_i^- \Big) \bqz \nonumber\\
                                             & = \bqz + \sum_{i = 1}^{n_\cE} m_i N_i^- \bqz + \ldots,
\end{align}
where each $m_i$ is an integer, and the $\ldots$ indicate terms that are at least quadratic in $m_i$.

Furthermore, we notice that if there exists an $N_{(k)}^-$ with $k$ taking any value $k=1, \ldots ,n_\cE$
which does not annihilate $\bqz$ then the orbit is infinite.
 To see this, we set $m_1 = \cdots = m_{k - 1} = m_k = m$ and $m_{k + 1} = \cdots = m_{n_\cE} = 0$. 
 The orbit reduces to
\begin{equation}
  \bQ(\bqz | m, \ldots, m, 0, \ldots, 0) = \bqz + m\, N_{(k)}^- \bqz + \frac{1}{2} m^2\, \big(N_{(k)}^- \big)^2 \bqz + \frac{1}{6} m^3 \,\big(N_{(k)}^- \big)^3 \bqz,
\end{equation}
where we have used $\big(N_{(k)}^- \big)^4 \bqz = 0$. If the orbit $\bQ(\bqz | m, \ldots, m , 0, \ldots, 0)$ is not infinite, then there is an $m' \ne m$ such that $\bQ(\bqz | m', \ldots, m', 0, \ldots, 0) =  \bQ(\bqz | m, \ldots, m, 0, \ldots, 0)$, hence $N_{(k)}^- \bqz = 0$.
This contradiction implies that the orbit $ \bQ(\bqz | m_1, \ldots, m_{n_\mathcal{E}})$ is infinite, provided the existence of an $N_{(k)}^-$ that does not annihilate $\bqz$.

Let us now show that such an $N_{(k)}^-$ exists for the enhancement chains \eqref{enhance_chain} satisfying the conditions (R1) or (R2) of subsection \ref{general_chargeorbit}. The simpler condition to show is (R1), which considers enhancement 
chains for in which $P$ is at a Type IV locus. In this case the $N_{(k)}^-$  not annihilating the seed charge $\bqz$
is simply $N_{(k)}^- = N_{(n_P)}^-$. This immediately follows from the fact that in the Type IV case one has 
$\big(N_{(n_P)}^-\big)^3 \mathbf{\tilde a}_0^{(n_\cE)} \neq 0$, which implies that the relevant $\mathbf{q}_0$s given 
in table \ref{tab:Chains_with_q0} satisfy $ N_{(n_P)}^-  \mathbf{q}_0 \neq 0$.  

Turning to condition (R2), we recall that it states that for every enhancement 
chain with $P$ at a Type II or Type III locus at least one further enhancement has to occur after the location of 
$P$. Considering this 
enhancement to occur from the $(n_P + j - 1)$-term to $(n_P + j)$-term the 
general expressions of the relevant chains are
\begin{eqnarray} \label{eqn:example_chain_enhancement_from_II_to_III}
  \cdots \to \rII_{b_1} \to \cdots \to  & \underset{\textrm{at }n_P}{\boxed{\rII_b}} \to \cdots \to \rII_{b_m} \to \underset{\textrm{at }(n_P + j)}{\rIII_{c_1}} \to \cdots,\nonumber\\
  \cdots \to \rII_{b_1} \to \cdots \to  & \underset{\textrm{at }n_P}{\boxed{\rII_b}} \to \cdots \to \rII_{b_m} \to \underset{\textrm{at }(n_P + j)}{\rIV_{d_1}} \to \cdots,\\
  \cdots \to \rIII_{c_1} \to \cdots \to & \underset{\textrm{at }n_P}{\boxed{\rIII_c}} \to \cdots \to \rIII_{c_n} \to \underset{\textrm{at }(n_P + j)}{\rIV_{d_1}} \to \cdots\ .\nonumber
\end{eqnarray}
We claim that in these cases the $N_{(k)}^-$ not annihilating $\mathbf{q}_0$ is given by $N_{(k)}^- = N_{(n_P + j)}^-$. 
Indeed, since the type of the singularity increases, also the highest power of $N_{(i)}^-$ 
not annihilating $\mathbf{\tilde a}_0^{(n_\cE)}$ increases. Using the relevant definitions of $\mathbf{q}_0$ of table \ref{tab:Chains_with_q0} this implies that $N_{(n_P + j)}^-$ does not annihilate $\mathbf{q}_0$. 
In conclusion we have found for chains satisfying (R1) and (R2) relevant 
$N_{(k)}^-$ that do not annihilate the seed charge $\bqz$ 
and thus have shown the infiniteness of the charge orbit $\bQ(\bqz | m_1, \ldots, m_{n_\mathcal{E}})$.

\subsection{A two parameter example: mirror of $\mathbb{P}^{(1,1,1,6,9)}[18]$} \label{sec:charge_orbits_in_example}

The discussions so far have been general, but rather abstract. In this section we show how to explicitly realise our approach to identifying the orbit. We consider the degree-$18$ Calabi-Yau hypersurface
 inside the weighted projective space $\bbP^{(1, 1, 1, 6, 9)}$. This hypersurface is denoted by 
 $\tilde Y_3 = \bbP^{(1, 1, 1, 6, 9)}[18]$ and has $h^{1,1}(\tilde Y_3)=2$. The Calabi-Yau hypersurface 
of which we will consider the complex structure moduli space 
is the mirror $Y_3$ of $\tilde Y_3$. Note that the geometry and the periods of the  pair $(\tilde{Y}_3, Y_3)$ 
have been studied in detail in \cite{Candelas:1994hw} as one of the early applications of mirror symmetry. 
 
 We will consider a patch 
 $\cE$ containing the large complex structure point of $Y_3$, which by mirror symmetry corresponds to 
 the large volume point of $\tilde Y_3$. Hence we can use the formulas of subsection \ref{sec:lcs_point} 
 to derive the monodromy logarithms $N_1,N_2$ and determine the corresponding singularity type. 
 The Calabi-Yau threefold $\tilde{Y}_3$ sits inside the toric ambient space with toric data
\begin{equation}
  \begin{array}{| l|r r r r r|r r|} 
  \hline
   \rule[-.2cm]{0cm}{.7cm} &   &   &    &    &    & l^{(1)} & l^{(2)}\\
   \hline
    K                      & 1 & 0 &  0 &  0 &  0 &      -6 &       0\\
    D_0                    & 1 & 0 &  0 & -1 & -1 &       1 &      -3\\
    D_1                    & 1 & 1 &  0 & -1 & -1 &       0 &       1\\
    D_2                    & 1 & 0 &  1 & -1 & -1 &       0 &       1\\
    D_3                    & 1 &-1 & -1 & -1 & -1 &       0 &       1\\
    D'                     & 1 & 0 &  0 &  2 & -1 &       2 &       0\\
    D''                    & 1 & 0 &  0 & -1 &  1 &       3 &       0\\
    \hline
  \end{array}
  \end{equation}
where the first column labels the toric divisors. Restricting all divisors to the hypersurface $\tilde Y_3$ in this ambient space, 
the generators of the K\"ahler cone are chosen to be
\begin{equation}
  J_1 = D_0 + 3D_1\, ,\qquad J_2 = D_1\, .
\end{equation}
The intersection numbers $K_{ijk} = J_i \cdot J_j \cdot J_k$ in this bases are determined to be \footnote{While not 
relevant later on, we note that the second Chern class for this example yields $b_1 = \frac{1}{24} c_2 \cdot J_1= \frac{17}{4}$, $b_2= \frac{1}{24} c_2 \cdot J_2 = \frac{3}{2}$.}
\begin{equation} \label{intersections_example2}
  K_{111} = 9\,,\qquad K_{112} = 3\,,\qquad K_{122} = 1\,,\qquad K_{222} = 0\, .
\end{equation}
Inserting \eqref{intersections_example2} into 
the general expression \eqref{lcsNa} we derive 
\begin{equation}
  N_1 =
  \begin{pmatrix}
    0            & 0           & 0           & 0  & 0 & 0\\
    -1           & 0           & 0           & 0  & 0 & 0\\
    0            & 0           & 0           & 0  & 0 & 0\\
    -\frac{9}{2} & -9          & -3          & 0  & 0 & 0\\
    -\frac{3}{2} & -3          & -1          & 0  & 0 & 0\\
    \frac{3}{2}  & \frac{9}{2} & \frac{1}{2} & -1 & 0 & 0
  \end{pmatrix},\qquad
  N_2 =
  \begin{pmatrix}
    0            & 0           & 0  & 0 & 0  & 0\\
    0            & 0           & 0  & 0 & 0  & 0\\
    -1           & 0           & 0  & 0 & 0  & 0\\
    -\frac{1}{2} & -3          & -1 & 0 & 0  & 0\\
    0            & -1          & 0  & 0 & 0  & 0\\
    0            & \frac{3}{2} & 0  & 0 & -1 & 0
  \end{pmatrix}.
\end{equation}
Furthermore, using  table \ref{special_N_class} we immediately determine the singularity types 
\begin{align}
   &\Delta_1 = \big\{ t^1 = \im\infty \big\}\, :   &&N_1 & &  \text{Type IV}_1\ , &\nn \\
   &\Delta_2 = \big\{ t^2 = \im\infty \big\}\ :   &&N_2 & &  \text{Type III}_0\ ,&\\
   &\Delta_{12} = \big\{ t^1 = \im\infty,t^2 = \im\infty \big\}\ : & &N_1 + N_2&&   \text{Type IV}_2\ , &\nn
\end{align}
where we note that $\Delta_{12}$ is nothing else then the large complex structure or large volume point and hence 
has the maximal enhancement IV$_{h^{2,1}}$.
 
In order to construct the charge orbits, we next have to explicitly construct the vector $\mathbf{\tilde a}_0^{(2)}$, 
i.e.~the limiting vector at $\Delta_{12}$, and the two nilnegative elements $N_1^-$, $N_2^-$ in the commuting $\slt$-pair associated to the 
enhancements III$_0 \ \rightarrow\ $IV$_2$ and  IV$_1 \ \rightarrow\ $IV$_2$. The corresponding derivation 
is lengthy, but follows the steps outlined in subsection \ref{sec:sl2-orbit}.  The details of this computation are presented in the appendices \ref{app:Sl2-splitting}, \ref{app:constructSl2s} and \ref{sec:exa}. Firstly, one uses the large complex structure periods rotated to an $\mathbb{R}$-split 
representation to derive 
\begin{equation} \label{tildea0_(2)}
\small  \mathbf{\tilde a}_0^{(2)} =
  \left(      1   ,
    0    ,
      0   ,
      -\frac{17}{4}   ,
      -\frac{3}{2}  ,
    0     
  \right)^\mathrm{T}.
\end{equation}
The commuting $\slt$-pair for the enhancement IV$_1 \ \rightarrow\ $IV$_2$ are then shown to be
\beq
 \small N_1^- =  \begin{pmatrix}
  0            & 0           & 0           & 0  & 0 & 0\\
  -1           & 0           & 0           & 0  & 0 & 0\\
  0            & 0           & 0           & 0  & 0 & 0\\
  -\frac{9}{2} & -9          & -3          & 0  & 0 & 0\\
  -\frac{3}{2} & -3          & -1          & 0  & 0 & 0\\
  \frac{3}{2}  & \frac{9}{2} & \frac{1}{2} & -1 & 0 & 0
\end{pmatrix},\qquad N_2^- =
    \left(
      \begin{array}{cccccc}
        0 & 0 & 0           & 0 & 0 & 0\\
        0 & 0 & 0           & 0 & 0 & 0\\
        0 & 0 & 0           & 0 & 0 & 0\\
        0 & 0 & 0           & 0 & 0 & 0\\
        0 & 0 & \frac{1}{3} & 0 & 0 & 0\\
        0 & 0 & 0           & 0 & 0 & 0
      \end{array}
    \right).
\eeq
In contrast, for the enhancement III$_0 \ \rightarrow\ $IV$_2$ we find the $\slt$-pair 
\begin{equation}
 \small N_1^- =   \begin{pmatrix}
  0            & 0           & 0  & 0 & 0  & 0\\
  0            & 0           & 0  & 0 & 0  & 0\\
  -1           & 0           & 0  & 0 & 0  & 0\\
  -\frac{1}{2} & -3          & -1 & 0 & 0  & 0\\
  0            & -1          & 0  & 0 & 0  & 0\\
  0            & \frac{3}{2} & 0  & 0 & -1 & 0
  \end{pmatrix},\quad N_2^- = 
    \left(
    \begin{array}{cccccc}
      0             & 0            & 0            & 0  & 0           & 0\\
      -1            & 0            & 0            & 0  & 0           & 0\\
      \frac{3}{2}   & 0            & 0            & 0  & 0           & 0\\
      -\frac{15}{4} & -\frac{9}{4} & -\frac{3}{2} & 0  & 0           & 0\\
      -\frac{3}{2}  & -\frac{3}{2} & -1           & 0  & 0           & 0\\
      \frac{3}{2}   & \frac{9}{4}  & \frac{1}{2}  & -1 & \frac{3}{2} & 0
    \end{array}
    \right).
\end{equation}
With these results we immediately compute the infinite charge orbits for this patch in moduli space. Using 
\eqref{charge-orbit_2} for the cases \eqref{cases(1)(2)}, \eqref{case(3)} and inserting $\mathbf{q}_0$ proposed 
in table \ref{tab:Chains_with_q0} we find
\begin{align} \label{ex_chargeoribits}
(1)\ & P \in \Delta^\circ_1 : & \ &\, \ \quad \mathbf{Q} =  \left(0, 0, 0, 9, 3,   - 9 m_1\right)^\mathrm{T} \, ,& \nn \\
(2)\ & P \in \Delta^\circ_2: &  \ &\, \ \quad \mathbf{Q} =   \left(0,0,0,1,0,-m_2 \right)^\mathrm{T}\, , &\\
(3)\ &P \in \Delta^\circ_{12}:& \ &\left\{ \begin{array}{ll}\mathbf{Q} =  \left({0, 0, 0, 9, 3, - 9 m_1} \right)^\mathrm{T}  &\text{for }  
\left\{ \ \frac{\I\, t^1 }{\I \, t^2 },\ \I \, t^2 > \lambda \ \right\} \, ,\\[.3cm] 
\mathbf{Q} =   \left(0,0,0,1,0,-m_2 \right)^\mathrm{T}  &\text{for }  \left\{ \ \frac{\I\, t^2 }{\I \, t^1 },\ \I \, t^1 > \lambda \ \right\}\, .
\end{array}\right.& \nn
\end{align}
Let us stress that by our general arguments all three orbits are infinite and 
massless at the location of the $P$ under consideration. The infiniteness is immediate 
due to the dependence on $m_1,m_2$, while the masslessness can alternatively be explicitly 
checked by a tedious but straightforward computation using the results of appendix \ref{sec:exa}. 
It is also nice to see that the charges are actually quantized. This is non-trivial, since $\mathbf{\tilde a}_0^{(2)}$ 
as well as $N_1^-$, $N_2^-$ contain rational entries. 

We close this section by discussing how the general properties and ideas we have outlines are realised in the charge orbits \eqref{ex_chargeoribits}. 
Firstly, we recall that $\Delta_2^\circ$ is a Type III locus and hence the one parameter arguments of subsection \ref{sec:spco}
and reference \cite{Grimm:2018ohb} would suggest that there is no infinite orbit. Indeed the orbit $\mathbf{Q}$ is independent 
of $m_1$ and hence not generated by the $N_1^- = N_2$ associated to $\Delta_2^\circ$. 
However, we see in \eqref{ex_chargeoribits}
that this orbit is `inherited' from the enhancement locus, i.e.~induced by the second monodromy logarithm $N_2^-$ not 
directly associated to $\Delta_2^\circ$. Secondly, we stress that the expression for $\mathbf{Q}$ in case (3) is indeed 
path dependent. If one approaches $\Delta_{12}^\circ$ via a path almost touching $\Delta_1^\circ$, we find that the orbit 
agrees with the one of case (1). This is not surprising, since this is the infinite orbit of the Type IV$_1$ singularity along 
$\Delta_1^\circ$ which is transferred to $\Delta_{12}^\circ$. Moving towards $\Delta_{12}^\circ$ along
 a path almost touching $\Delta_2^\circ$ we find a completely different charge orbit depending on $m_2$.

\subsection{Discussion on properties of the charge orbit} \label{discussion_properties+extensions}

To summarise, in this section we have shown how to identify infinite massless charge orbits using data which is not completely local but rather associated to a patch where singular divisors can intersect. In particular, this significantly extends the infinite charge orbits that were identified in \cite{Grimm:2018ohb}. It also forms a starting point towards a global understanding of the infinite towers of states associated to the monodromies in the full moduli space. 

A first point to stress is that our current definition of $\mathbf{Q}$ and $\mathbf{q}_0$ 
vitally uses the commuting $\slt$ algebras \eqref{Sl2-triples}. In particular, this fact has been exploited in 
subsection \ref{sec:general_masscomments} to show that $\mathbf{Q}$ and $\mathbf{q}_0$ have the same Hodge norm growth. 
However, the usage of the commuting $\slt$ basis containing $N_i^-$ could be just an intermediate step to show the desired results. In fact, it is an important strategy of \cite{CKS,Kerr2017} to translate the final statement back to the formulation with the $N_i$. It may be that a similar result can be shown for our constructions. Therefore, a natural candidate charge orbit is then 
\beq
   \mathbf{\tilde Q}(\mathbf{\tilde q}_0 | m_1, \ldots ,m_{n_\cE}) \equiv \text{exp}\Big(\sum_{i=1}^{n_{\cE}}  m_i N_i \Big) \mathbf{\tilde q}_0\ ,
\eeq
which is the natural analogue to \eqref{charge-orbit_gen}. In order to identify the seed charge $\mathbf{\tilde q}_0$, we would 
then require that it satisfies the massless condition \eqref{massless_q0} within the monodromy weight filtration $W\big(N_{(i)}\big)$ in order to generate an orbit that becomes massless when approaching $P$ within a growth sector. This requirement is natural due to the fact that $W\big(N_{(i)}^-\big) = W\big(N_{(i)}\big)$ as already stated in \eqref{W(N)=W(N-)}. More concretely, unpacking the filtration $W\big(N_{(i)}\big)$ with definition \eqref{def-Wi} and using the concrete Hodge-Deligne diamonds of singularity types shown in table \ref{HD:enumeration}, we see that the seed charge has to obey, for every $i = 1, \ldots, n_P - 1$:
\begin{itemize}
  \item If ${\sf Type\ A}_i = \rI$ or $\rII$, then $N_{(i)} \mathbf{\tilde q}_0 = 0$;\vspace*{-.2cm}
  \item If ${\sf Type\ A}_i = \rIII$, then there exists charge vectors $\mathbf{b}_i$ and $\mathbf{u}_i$ with $N_{(i)} \mathbf{u}_i = 0$ such that $\mathbf{\tilde q}_0 = N_{(i)} \mathbf{b}_i + \mathbf{u}_i$;\vspace*{-.2cm}
  \item If ${\sf Type\ A}_i = \rIV$, then there exists charge vectors $\mathbf{w}_i$ and $\mathbf{x}_i$ with $N_{(i)} \mathbf{w}_i = 0$ and $\big(N_{(i)}\big)^3 \mathbf{x}_i = 0$ such that $\mathbf{\tilde q}_0 = \mathbf{w}_i + N_{(i)} \mathbf{x}_i$.
\end{itemize}
Furthermore, the following conditions are imposed at position $n_P$:
\begin{itemize}
  \item If ${\sf Type\ A}_{n_P} = \rII$, then there is a charge vector $\mathbf{a}$ such that $\mathbf{\tilde q}_0 = N_{(n_P)} \mathbf{a}$;\vspace*{-.2cm}
  \item If ${\sf Type\ A}_{n_P} = \rIII$, then there is a charge vector $\mathbf{u}_{n_P}$ with $\big(N_{(n_P)}\big)^2 \mathbf{u}_{n_P} = 0$ such that $\mathbf{\tilde q}_0 = N_{(n_P)} \mathbf{u}_{n_P}$;\vspace*{-.2cm}
  \item If ${\sf Type\ A}_{n_P} = \rIV$, then there are charge vectors $\mathbf{c}$ and $\mathbf{w}_{n_P}$ with $\big(N_{(n_P)}\big)^2 \mathbf{w}_{n_P} = 0$, such that $\mathbf{\tilde q}_0 = \big(N_{(n_P)}\big)^2 \mathbf{c} + N_{(n_P)} \mathbf{w}_{n_P}$.
\end{itemize}
Finally to ensure infiniteness, we require the existence of an $N_{(j)}$ with $n_P \le j \le n_\cE$ such that $N_{(j)} \mathbf{\tilde q}_0 \ne 0$. Applying the growth theorem as before we see that the seed charge $\mathbf{\tilde q}_0$ satisfying the above properties becomes massless when approaching $P$ within a growth sector if either of the two conditions (R1), (R2) of subsection \ref{general_chargeorbit} are satisfied. Moreover the resulting $\mathbf{\tilde Q}$ is infinite by the same reasoning in subsection \ref{sec:infinteness}. We would then claim that this $\mathbf{\tilde Q}$ becomes massless when approaching $P$ within the same growth sector as $\mathbf{\tilde q}_0$. Let us stress, however, that establishing full equivalent to the results of subsections \ref{sec:general_masscomments} and \ref{multi-N-analysis}, including the explicit constructions of table \ref{tab:Chains_with_q0}, without using the commuting basis would require more work and will be left for the future.

We have discussed how the intersection points can be utilised to build the infinite distance networks in moduli space, which follow the rules of enhancement in table \ref{sing_enhancements}. If we consider such a network we can identify orbits in patches which contain type IV loci or intersections which enhance the singularity type.\footnote{Note that this implies the identification always holds in the large volume regime of the mirror.} Once such an orbit is identified, it will retain its identity along any finite distance along the singularity curve moving away from this local patch. This is because the limiting Hodge structure is defined over the full singular locus. If we move an infinite distance away, so towards a different intersection with some other infinite distance locus, then it is more difficult to track this orbit. We actually expect that the charge orbit can be `transferred' between singular divisors which intersect even when there is no enhancement of the singularity type. By this we mean that a set of charges identified by a charge orbit on one divisor has a corresponding set on the divisor which intersects it. This is supported by tracking the Hodge-Deligne diamonds from one divisor to the other through the intersection. Should we be able to track the orbit this way, we would be able to identify an infinite charge orbit over a full intersecting infinite distance network. However, we leave a detailed study of this possibility for future work. 

In \cite{Grimm:2018ohb} it was shown that the monodromy charge orbit is fully populated by BPS states as long as one of the charges corresponds to a BPS state. This was shown for the only case where such an orbit could be identified, which is for type IV singularities. In this work, we are not able to show such a connection between the charge orbit and BPS states. This is not unexpected, the argument of \cite{Grimm:2018ohb} was based on walls of marginal stability. While being away from a wall of marginal stability ensures that a BPS state remains in the spectrum, this is not a necessary condition, i.e. there are many examples of BPS states which by, charge and energy conservation alone, could decay to other BPS states. So one expects that the spectrum of BPS states has some finer underlying structure. The utilisation of the charge orbits in this work amounts to a proposal that this finer structure includes the population of charge orbits by BPS states, at least asymptotically towards infinite distance.

\section{Conclusions}
\label{sec:sum}

In this paper we studied aspects of the Swampland Distance Conjecture in the complex structure moduli space of Calabi-Yau manifolds. In this context, the set of infinite distance loci in field space can be understood both generally and precisely. We utilised the powerful mathematical tools of the orbit theorems and mixed Hodge structures to analyse infinite distance points in complete generality, so any infinite distance point in any Calabi-Yau threefold. We showed that any infinite distance point is part of a locus in moduli space to which we can associate a set of discrete topological data, its Hodge-Deligne diamond, that defines its key characteristics. We also showed how to extract this data from the monodromy, associated to axion-type shifts, about the infinite distance locus. The data can be used to completely classify infinite distance loci in the moduli space, and this classification includes an understanding of how different infinite distance loci can intersect and change their type. In this way, the different types of infinite distance loci form a rich intersecting infinite distance network. We showed that there are rules for how such intersections can occur and so for which kinds of infinite distance networks can be built. These rules and networks therefore are uncovering a new perspective on the distance conjecture where global structures in the field space are emerging. 

The intersections between different types of infinite distance loci are clearly central to this global perspective, and so naturally most of the investigation was focused on them. Within a local patch in field space containing such an intersection, we were able to reach a significant number of results regarding the nature of the infinite tower of states of the distance conjecture. More precisely, to each infinite distance locus one can associate a nilpotent matrix $N$, and when the loci intersect the different matrices commute. However, a remarkable result of \cite{CKS}, known as the general Sl(2)-orbit theorem, shows that the nilpotent matrices can further be completed into fully commuting $\slt$ algebras. This can be thought of as a type of factorisation of the infinite distance loci, and greatly simplifies the analysis of the intersections. In particular, it allows for a rather precise identification of an infinite tower of states in terms of a {\it charge orbit}. This orbit generalises the monodromy orbits presented in \cite{Grimm:2018ohb} by utilising the commuting structure of the $\slt$ algebras. Importantly, it can be generalised recursively to any number of intersecting infinite distance loci. We then established general conditions when such 
a charge orbit can define an infinite tower of states that become massless when approaching the infinite distance 
point.  More specifically, we have explicitly constructed a candidate charge orbit for any infinite distance point that has another 
infinite distance locus of higher type in its vicinity. 
This non-local construction allowed us to identify the tower of states of the distance conjecture for a more general set of infinite distance loci than was done in \cite{Grimm:2018ohb}, thereby making progress towards a complete identification of the tower of states globally on the moduli space.
 However, it is important to state that in \cite{Grimm:2018ohb}, by utilising walls of marginal stability, the monodromy orbits were shown to be populated by  actual BPS states in the spectrum. We are not able to reach such a result for the more general charge orbits introduced in this paper. We therefore leave a study of the precise relation between charge orbits and BPS states for future work.

One particularly interesting new aspect of the distance conjecture in the context of intersecting infinite distance loci is that the mass spectrum of BPS states picks up a dependence on the path of approach to the intersection. The results of \cite{CKS} allowed us to quantify this path dependence rather precisely, showing how to classify paths into different growth sectors, and to determine how the masses of the tower of states behave within each growth sector. We find the encouraging result that the particular form of the infinite charge orbit of states is such that the states become massless independently of the path of approach, within a given growth sector. This lends further evidence to the proposal in \cite{Grimm:2018ohb} that the tower of states associated to the monodromy action induces the infinite distance divergence, since there are no infinite distance paths of approach whereby the tower remains massive. 

Our results show that there is a rich structure at infinite distances in field spaces of theories of quantum gravity. While we made significant progress at uncovering some of this structure, we believe that there is much more to discover. The close ties to the existing rich and deep mathematical framework of nilpotent orbits suggest that much of this structure is general. By this we mean that many of the results can be formulated just by an association of a nilpotent matrix to an infinite distance point. Such an association is rather natural from the perspective of quantum gravity as discussed in \cite{Grimm:2018ohb}. There are two ways to motivate this. The first is that the nilpotent matrix is associated to a discrete gauge symmetry, an axion shift, which is promoted to a continuous global symmetry at infinite distance. The infinite distance and infinite tower of states can then be understood as a quantum gravity obstruction to the global symmetry limit. The second way is in the context of emergence of infinite distances, so the idea that the infinite distance is itself induced by integrating out the tower of states. Then the nilpotent matrix associated to it is a remnant of the structure of this tower. The appearance of nilpotent matrices, associated to axion transformations, was also found in \cite{Herraez:2018vae}. This ties in nicely also to the ideas of \cite{Ooguri:2018wrx} where potentials on field spaces are also controlled by the towers of states. Motivated by these results, we believe that it is a natural expectation that nilpotent elements, and the rich structure associated to them which we have been exploring in this paper, may underlie much of the universal behaviour of quantum gravity theories at large distances in field space.

While our work was motivated by the distance conjecture, the results are significant also purely as a study of Calabi-Yau moduli spaces. We have adapted the recent results on relations between polarised mixed Hodge structures \cite{Kerr2017} to the moduli space, expanded on them and developed their connection to distances in the space. We have also presented the first, to our knowledge, computation of the commuting $\slt$ triples of matrices at intersections of infinite distance loci, or from the Hodge-theoretic perspective, at degenerations of polarized mixed Hodge structures. 

Our analysis was focused on the complex structure moduli space, but we have also discussed the mirror 
dual configuration in some detail. More precisely, we have explicitly determined the monodromy matrices 
relevant in the complexified K\"ahler cone when encircling the large volume point in a higher-dimensional moduli space. We showed that by only using  the triple intersection numbers and the second Chern class of the mirror threefold one is able to classify the monodromies and the arising infinite distance singularity types in this large volume regime.
In this large volume regime we can then directly apply our findings on the charge orbit. They immediately 
imply that we have shown that to every infinite distance point in the large volume regime we 
can identify an infinite charge orbit that becomes massless at this point. Crucially the considered point does not 
have to be the large volume point itself, but rather any partial limit will also share this feature. Let us stress that we 
expect that our construction of the charge orbit is also valid relevant in string compactifications that are not directly 
the mirror to the considered Type IIB configurations \cite{CorvilainGrimmValenzuela}.
Moreover, it is interesting to point 
out that this perspective gives a new way to systematically classify allowed triple intersection numbers 
and hence allowed K\"ahler potentials. In fact, the associated polarized mixed Hodge structure incorporates 
more canonically the positivity conditions on various couplings, while the growth theorem ensures that possible 
cancellations are ruled out. It would be very interesting to systematically explore the power of this 
new perspective for questions beyond the distance conjecture. 

\vspace*{.5cm}

\noindent
{\bf Acknowledgements}

\noindent
We thank Matt Kerr and Irene Valenzuela for useful discussions.

\appendix

\section{Monodromy filtrations and mixed Hodge structures} \label{app:MWF_MHS}

In this appendix we give a short review of some further mathematical concepts relevant for this 
work. We first introduce a \textit{pure} Hodge structure and its associated Hodge filtration. 
A pure Hodge structure of weight $w$ provides a splitting of the complexification $V_\bbC = V \otimes \bbC$ of 
a rational vector space $V$ by the Hodge
decomposition 
\beq \label{Hodge-decomp_app}
   V_\bbC = \cH^{w,0}  \oplus  \cH^{w-1,1}  \oplus \ldots  \oplus  \cH^{1,w-1}  \oplus \cH^{0,w}\ ,
\eeq
with the subspaces satisfying $\cH^{p,q} = \overline{\cH^{q,p}}$ with $w=p+q$, where the complex conjugation on $V_\bbC$ 
is defined with respect to the rational vector space $V$. Using the $\cH^{p,q}$ one can also  
define a Hodge filtration as $F^p = \oplus_{i\geq p} \cH^{i,w-i}$ satisfying  
\beq \label{F-filtration}
   V_\bbC = F^0 \ \supset \ F^1 \ \supset\  \ldots\ \supset\   F^{w-1}\ \supset\   F^w = \cH^{w,0}\ ,
\eeq
such that $\cH^{p,q} = F^p \cap \bar F^q$. A \textit{polarized} pure Hodge structure requires additionally 
the existence of a bilinear form $S(\cdot, \cdot)$ on $V_\bbC$, such that the conditions  
$S(\cH^{p,q}, \cH^{r,s}) = 0$ for $p \neq s,\ q \neq r$ and $\im^{p-q} S(v,\bar v) > 0$ for any non-zero $v \in \cH^{p,q}$
are satisfied.

The crucial extra ingredient relevant to define a (limiting) mixed Hodge structure, is the 
so-called \textit{monodromy weight filtration} $W_i$. It was defined in \eqref{def-Wi} using the 
kernels and images of the nilpotent matrix $N$. The rational vector subspaces $W_{j} (N) \subset 
V$ can alternatively be defined by requiring that they form a filtration
\beq 
W_{-1}\equiv 0\ \subset\  W_0\ \subset\ W_1\ \subset\  \ldots \  \subset\ W_{2w-1}\  \subset \ W_{2w} = V\ ,
\label{filtration}
\eeq
with the properties
\begin{align}
 &  1.) \quad N W_i \subset W_{i-2} &\\
   & 2.) \quad  N^j : Gr_{w+j} \rightarrow Gr_{w-j}\ \ \text{is an isomorphism,}\quad Gr_{j} \equiv W_{j}/W_{j-1} \ .& \label{iso-prop}
\end{align} 
The quotients $Gr_i$ contain equivalence classes of elements of $W_i$ that differ by elements of $W_{i-1}$.
When $V_\bbC$ also admits a Hodge filtration $F^p$ as in \eqref{F-filtration}, we require that $N$ is compatible 
with this structure and acts on it horizontally, i.e.~$N F^p \subset F^{p-1}$.

We are now in the position to define a \textit{mixed Hodge structure} $(V,W,F)$, induced by the filtrations $W_i$ 
and $F^q$ on the vector space $V$. The defining feature of this structure is that each $Gr_{j}$ defined in \eqref{iso-prop} 
admits an induced Hodge filtration 
\beq \label{mixedHodge-filtration}
    F^p Gr_j^\bbC \equiv ( F^p  \cap W_j^\bbC )/ ( F^p  \cap W_{j-1}^\bbC)\ ,
\eeq
where $Gr_j^\bbC = Gr_j \otimes \bbC$ and $W_i^\bbC = W_i \otimes \bbC$ are the complexification.
In other words, in the notation of \eqref{Hodge-decomp_app} we spilt each $Gr_j $ into a pure Hodge structure $\cH^{p,q}$
as  
\beq \label{Grj-split}
   Gr_j  = \bigoplus_{p+q=j} \cH^{p,q} \ ,\qquad   \cH^{p,q} =  F^p Gr_j \cap \overline{F^q Gr_j}\ ,
\eeq
where we recall that $w=p+q$ is the weight of the corresponding pure Hodge structure. 
The operator $N$ is a morphism among these pure Hodge structures. 
Using the action of $N$ on $W_i$ and $F^p$, we find $N Gr_j \subset Gr_{j-2}$ and 
$N \cH^{p, q} \subset \cH^{p-1,q-1}$. Note that this induces a jump in the weight of the pure Hodge structure by $-2$, 
while the mixed Hodge structure is preserved by $N$.  

\section{Construction of the $\SLt$-splitting} \label{app:Sl2-splitting}

In this appendix we review the construction of the matrices $\delta$ and $\zeta$ that 
are used to construct a special $\bbR$-split mixed Hodge structure $(V,\hat F, W)$, first discussed in 
subsection \ref{sec:sl2-orbit}, via
\beq
    \hat{F} = e^\zeta e^{-\im\delta}F\ . 
\eeq
The mixed Hodge structure $(V,\hat{F}, W)$ is called the $\SLt$-splitting of the limiting 
mixed Hodge structure $(V,F, W)$.
Here we denote by $(V,F,W)$ a vector space $V$ with filtrations $F^p$ and $W_i$, see 
appendix \ref{app:MWF_MHS}. As in subsection \ref{sec:charact_sing}
the latter is induced by some nilpotent $N$.
Using \eqref{def-Ipq} and \eqref{Fp-Wi_split} we can determine a Deligne splitting $V_\bbC = \bigoplus I^{p, q}$ from the data $(F,W)$. 
On this splitting there is a semisimple operator $T$, called the grading operator, that acts on the subspace $\bigoplus_{p + q = l} I^{p, q}$ as 
multiplication by $l$. Let $\conj{T}$ be the complex conjugate of the grading operator $T$ defined by
\begin{equation}
  \conj{T}(v) := \conj{T(\conj{v})},
\end{equation}
for all $v \in V_\bbC$. Then $\conj{T}$ and $T$ are related by a conjugation by $e^{-2\im\delta}$
\begin{equation} \label{eqn_of_delta}
  \conj{T} = e^{-2\im\delta} T e^{2\im\delta},
\end{equation}
where the real operator $\delta$ sends every $I^{p, q}$ to its ``lower parts'':
\begin{equation} \label{where_delta_lives}
  \delta(I^{p, q}) \subs \bigoplus_{\substack{r < p\\s < q}} I^{r, s}, \text{ for all } p, q.
\end{equation}
Thus we can solve equation \eqref{eqn_of_delta} with an Ansatz satisfying \eqref{where_delta_lives} for the operator $\delta$. Furthermore $\delta$ commutes with $N$ and preserves the polarisation 
$\delta^\mathrm{T}\eta + \eta\delta = 0$. Such an operator $\delta$ is unique. Let
\begin{equation}
  \tilde{F} := e^{-\im\delta} F,
\end{equation}
and the mixed Hodge structure $(V, \tilde{F}, W)$ is $\bbR$-split. For a mathematically precise discussion we refer to Proposition 2.20 of \cite{CKS}. 

The second operator $\zeta$ further builds another $\bbR$-split mixed Hodge structure out of $(V, \tilde{F}, W)$. Its construction is indirect and we refer to section 3 and Lemma 6.60 of \cite{CKS} for the full original discussion. Also section 1 of \cite{MixedSl2} contains a good review of the $\zeta$ operator and in its Appendix the authors worked out some explicit expressions that will be used in our computation.

To find $\zeta$, we first compute a `Deligne splitting' of the operator $\delta$: Let $V_\bbC = \bigoplus \tilde{I}^{p, q}$ be the Deligne splitting of the $\bbR$-split mixed Hodge structure $(V, \tilde{F}, W)$, then this Deligne splitting induces a decomposition of $\delta$
\begin{equation}
  \delta = \sum_{p, q > 0} \delta_{-p, -q},
\end{equation}
where each component $\delta_{-p, -q}$ precisely does the following:
\begin{equation}
  \delta_{-p, -q}(\tilde{I}^{r, s}) \subs \tilde{I}^{r - p, s - q}, \text{ for all } r, s.
\end{equation}

The operator $\zeta$ admits the same kind of decomposition
\begin{equation}
  \zeta = \sum_{p, q > 0} \zeta_{-p, -q},
\end{equation}
and its relation with $\delta$ is given by the equation in Lemma 6.60 of \cite{CKS}
\begin{equation}
  e^{\im\delta} = e^\zeta\left(\sum_{k\ge 0} \frac{(-\im)^k}{k!} \mathrm{ad}_N^k(\tilde{g}_k)\right),
\end{equation}
where every $\tilde{g}_k$ is an real operator preserving the polarisation $\eta$ of the real vector space $V$ and $\mathrm{ad}_N(-) = [N, -]$ is the adjoint action. The main outcome of this formula useful for us is that, upon decomposing $\delta$ and $\zeta$ into their $(-p, -q)$ components and solving for $\zeta_{-p, -q}$, we get a polynomial in $\delta_{-p, -q}$ and the iterated commutators among various components $\delta_{-p, -q}$.

Specialising to weight-$3$ degenerating variation of Hodge structures, the possible non-vanishing components of $\zeta_{-p, -q}$ are restricted to $1 \le p, q \le 3$. Then according to the Appendix of \cite{MixedSl2}, we have the following explicit expressions
\begin{align}
  \zeta_{-1, -1} & = 0,\quad \zeta_{-1, -2} = -\frac{\im}{2} \delta_{-1, -2},\quad \zeta_{-1, -3} = -\frac{3\im}{4} \delta_{-1, -3},\quad \zeta_{-2, -2} = 0,\\
  \zeta_{-2, -3} & = -\frac{3\im}{8} \delta_{-2, -3} - \frac{1}{8}[\delta_{-1, -1}, \delta_{-1, -2}],\quad \zeta_{-3, -3} = -\frac{1}{8}[\delta_{-1, -1}, \delta_{-2, -2}],\nonumber
\end{align}
while the remaining $\zeta_{-q, -p}$ are obtained from $\zeta_{-p, -q}$ by replacing all $\im$ by $-\im$ and $\delta_{-r, -s}$ by $\delta_{-s, -r}$.
Summing all $\zeta_{-p, -q}$, we get a formula for $\zeta$ that is valid in weight-$3$ degenerating variation of Hodge structures given by
\begin{align}
  \zeta = & \frac{\im}{2}(\delta_{-2, -1} - \delta_{-1, -2}) + \frac{3\im}{4}(\delta_{-3, -1} - \delta_{-1, -3}) + \frac{3\im}{8}(\delta_{-3, -2} - \delta_{-2, -3})\nonumber\\
          & - \frac{1}{8}[\delta_{-1, -1}, \delta_{-1, -2} + \delta_{-2, -1} + \delta_{-2, -2}].
\end{align}

\section{General procedure to construct the commuting $\slt$s} \label{app:constructSl2s}

The construction of commuting $\slt$s is part of the multi-variable $\SLt$-orbit theorem in \cite{CKS}. We summarise its construction in this section for completeness.

Finding the commuting $\slt$-triples associated to the intersection $\Delta_{1, \ldots, n_\cE}$ amount to $n_\cE$-times iteration. One starts with the limiting mixed Hodge structure $(F_\infty, W^{n_\cE})$, where $F_\infty$ is the limiting Hodge filtration extracted by nilpotent orbit theorem, and $W^{n_\cE}$ is the monodromy weight filtration associated to the nilpotent cone $\sigma(N_1, \ldots, N_{n_\cE})$ generated by the monodromies $N_1, \ldots, N_{n_\cE}$, i.e.,
\begin{equation}
  W^{n_\cE} = W(N_1 + \cdots + N_{n_\cE}).
\end{equation}
The limiting mixed Hodge structure $(F_\infty, W^{n_\cE})$ will be used as the input of the first iteration of the construction. Let the index $k = n_\cE$, which will be counted downwards after each iteration.

For each iteration with index $k$, we denote the input mixed Hodge structure by $(F', W^k)$. Then one computes the $\SLt$-splitting $(F_k, W^k)$ of $(F', W^k)$. Furthermore, one finds the Deligne splitting of the mixed Hodge structure $(F_k, W^k)$
\begin{equation}
  V_\bbC = \bigoplus_{p, q} I^{p, q}_{(F_k, W^k)}.
\end{equation}
Record the semisimple grading operator $Y_{(k)}$ which acts on each subspace by multiplication
\begin{equation}
  Y_{(k)} v = (p + q - 3) v, \text{ for every } v \in I^{p, q}_{(F_k, W^k)}.
\end{equation}
And set the mixed Hodge structure $(e^{\im N_k} F_k, W^{k - 1})$ as the input of the next iteration, which carries index $k - 1$.

The loop stops once $k = 0$. In the end, we get a bunch of grading operators $Y_{(n_\cE)}, \ldots, Y_{(1)}$ associated with $\bbR$-split mixed Hodge structures $(F_{n_\cE}, W^{n_\cE}), \ldots, (F_1, W^1)$. For convenience, set $Y_{(0)} = 0$.

The next step is to find the nilpotent elements $N_i^-$ in each $\slt$-triple $(N_i^-, N_i^+, Y_i)$. Every $N_i^-$ is determined by diagonalising the adjoint action of $Y_{(i - 1)}$: Decompose $N_i$ into eigenvectors of the adjoint action of $Y_{(i - 1)}$
\begin{equation}
  N_i = \sum_\alpha N_i^\alpha,
\end{equation}
where $N_i^\alpha$ satisfies $[Y_{(i - 1)}, N_i^\alpha] = \alpha N_i^\alpha$. Then the nilnegative element is extracted $N_i^- := N_i^0$. Note that one always has $N_1^- = N_1$ since $Y_{(0)} = 0$.

The neutral elements are set to be
\begin{equation}
  Y_i = Y_{(i)} - Y_{(i - 1)}.
\end{equation}
Since $Y_{(0)} = 0$, we have $Y_0 = Y_{(0)}$.

Finally, we complete the triples by solving the equations defining an $\slt$-triple
\begin{equation}
  [Y_i, N_i^+] = 2N_i^+,\qquad [N_i^+, N_i^-] = Y_i,
\end{equation}
for the nilpositive element $N_i^+$, which is required to also preserve the polarisation
\begin{equation}
  (N_i^+)^\mathrm{T} \eta + \eta N_i^+ = 0.
\end{equation}

We have thus found the commuting $\slt$-triples $(N_i^-, N_i^+, Y_i)$ for $i = 1, \ldots, n_\cE$ according to theorem (4.20) of \cite{CKS}.

\section{An example: commuting $\slt$s in the mirror of $\mathbb{P}^{(1,1,1,6,9)}[18]$}
\label{sec:exa}

This section aims to exemplify the structures of section \ref{math_background} and \ref{sec:infinite_towers}, by analysing the periods and Hodge structure of an 
explicit Calabi-Yau threefold geometry. More precisely, we will denote by $\tilde{Y}_3$ the degree-$18$ Calabi-Yau hypersurface inside the weighted projective space $\bbP^{(1, 1, 1, 6, 9)}$ and denote by $Y_3$ its mirror. We show in detail how the associated commuting $\slt$-pair for the variation of Hodge structure on $Y_3$ arises from the study of its complex structure moduli space. We also illustrate several abstract constructions introduced in the previous sections, including the Deligne splitting \eqref{def-Ipq} and the associated canonical $\SLt$-splitting MHS in $\SLt$-orbit theorem of 
appendix \ref{app:Sl2-splitting}. Note that the geometry and the periods of the  pair $(\tilde{Y}_3, Y_3)$ have been studied in detail in \cite{Candelas:1994hw} as one of the first applications of mirror symmetry.

\subsection{Introduction to the example}

\begin{figure}[h!]
  \vspace*{.5cm}
  \begin{center} 
  \includegraphics[width=7cm]{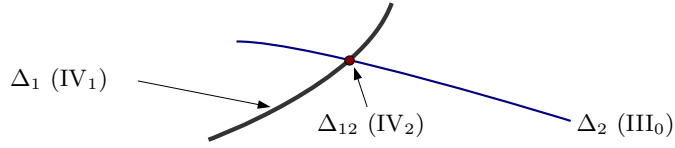} 
  \vspace*{-1cm}
  \end{center}
  \begin{picture}(0,0)
  \put(215,13){\footnotesize$\Delta_{12}\ (\rIV_2)$}
  \put(100,30){\footnotesize$\Delta_{1}\ (\rIV_1)$}
  \put(312,13){\footnotesize$\Delta_{2}\ (\rIII_0)$}
  \end{picture}
  \caption{Two singular divisors $\Delta_1$ and $\Delta_2$ intersect at the large complex structure point $\Delta_{12}$, where the corresponding types of degenerations are also labelled. The coloured divisor shows one of the possible ways of approaching the large complex structure point, namely moving along the type-$\rIII_0$ divisor towards the type-$\rIV_2$ intersection. This choice is equivalent to a choice of the ordering of the monodromies as $(N_2, N_1)$, so that we have the singularity enhancement from $\Delta_2^\circ$ to $\Delta_{12}^\circ$.} \label{limitingF_example}
\end{figure}

We focus on $2$-moduli degeneration in this section. The geometric setup is that we sit near the large complex structure point, where locally the moduli space contains two copies of punctured disk as shown in figure \ref{limitingF_example}. From the period vector around the large complex structure point, we extract the limiting Hodge filtration $F(\Delta^\circ_{12})$, whose top component $F^3(\Delta^\circ_{12})$ is generated by $\mathbf{a}_0^{(2)}$. Then $(F(\Delta^\circ_{12}), W(N_1 + N_2))$ is a limiting mixed Hodge structure. In accordance with appendix \ref{app:constructSl2s}, we denote $F_\infty := F(\Delta^\circ_{12})$ and $W^2 := W(N_1 + N_2)$ in the following discussion.

\subsubsection{The periods of $Y_3$ around the large complex structure point}
In this section we give the periods of $Y_3$ around the large complex structure point following the method described in section \ref{sec:lcs_point}. The toric and relevant topological data of $\tilde{Y}_3$ is given in section \ref{sec:charge_orbits_in_example} and we remark that the Euler characteristic of $\tilde{Y}_3$ is $\chi(\tilde{Y}_3) = -540$.

Furthermore, the generators of the Mori cone $C_1, C_2$ dual to $J_1, J_2$ are chosen to be
\begin{equation}
  C_1 = J_2 \cap J_2,\qquad C_2 = D_0 \cap J_2,
\end{equation}
so that the following K-theory basis for branes
\begin{equation}
  (\cO_{X^\circ}, \cO_{J_1}, \cO_{J_2}, \mathcal{C}_1, \mathcal{C}_2, \cO_p)
\end{equation}
yields the asymptotic period vector around the large complex structure point and the polarisation matrix:

\begin{equation}
  \mathbf{\Pi^\Omega}(t^1, t^2) =
  \begin{pmatrix}
    1\\
    t^1\\
    t^2\\
    \frac{9}{2} (t^1)^2 + 3 t^1 t^2 + \frac{1}{2} (t^2)^2 + \frac{9}{2} t^1 + \frac{1}{2} t^2 - \frac{17}{4} + \cdots\\
    \frac{3}{2} (t^1)^2 + t^1 t^2 + \frac{3}{2} t^1 - \frac{3}{2} + \cdots\\
    \frac{3}{2} (t^1)^3 + \frac{3}{2} (t^1)^2 t^2 + \frac{1}{2} t^1 (t^2)^2 - \frac{23}{4} t^1 - \frac{3}{2} t^2 - \frac{135\im\zeta(3)}{2\pi^3} + \cdots
  \end{pmatrix},
\end{equation}

\begin{equation}
  \eta =
  \begin{pmatrix}
    0  & -10 & -3 & 0  & 0 & -1\\
    10 & 0   & 1  & 1  & 0 & 0\\
    3  & -1  & 0  & 0  & 1 & 0\\
    0  & -1  & 0  & 0  & 0 & 0\\
    0  & 0   & -1 & 0  & 0 & 0\\
    1  & 0   & 0  & 0  & 0 & 0
  \end{pmatrix},
\end{equation}
where $t^i$ is the coordinate on the K\"ahler moduli space of $\tilde{Y}_3$, under the mirror map it corresponds to the coordinates $z^i$ on the complex structure moduli space of $Y_3$ via $t^i = \frac{1}{2\pi\im}\log z^i + \cdots$. The full period can be acquired by solving the Picard-Fuchs equation on the space $Y_3$ and matching the leading logarithmic behaviour of the solution with the above asymptotic period. We do not give the full instanton-corrected period vector since it is not relevant to our discussion.

The monodromy operator $T_i$ is then induced by sending $t^i \mapsto t^i - 1$:
\begin{equation}
  T_1 =
  \begin{pmatrix}
    1  & 0  & 0  & 0  & 0  & 0\\
    -1 & 1  & 0  & 0  & 0  & 0\\
    0  & 0  & 1  & 0  & 0  & 0\\
    0  & -9 & -3 & 1  & 0  & 0\\
    0  & -3 & -1 & 0  & 1  & 0\\
    0  & 9  & 2  & -1 & 0  & 1
  \end{pmatrix},\qquad
  T_2 =
  \begin{pmatrix}
    1  & 0  & 0  & 0  & 0  & 0\\
    0  & 1  & 0  & 0  & 0  & 0\\
    -1 & 0  & 1  & 0  & 0  & 0\\
    0  & -3 & -1 & 1  & 0  & 0\\
    0  & -1 & 0  & 0  & 1  & 0\\
    0  & 2  & 0  & 0  & -1 & 1
  \end{pmatrix},
\end{equation}
and they are already unipotent. Their corresponding logarithms $N_i := \log T_i$ are given by
\begin{equation}
  N_1 =
  \begin{pmatrix}
    0            & 0           & 0           & 0  & 0 & 0\\
    -1           & 0           & 0           & 0  & 0 & 0\\
    0            & 0           & 0           & 0  & 0 & 0\\
    -\frac{9}{2} & -9          & -3          & 0  & 0 & 0\\
    -\frac{3}{2} & -3          & -1          & 0  & 0 & 0\\
    \frac{3}{2}  & \frac{9}{2} & \frac{1}{2} & -1 & 0 & 0
  \end{pmatrix},\qquad
  N_2 =
  \begin{pmatrix}
    0            & 0           & 0  & 0 & 0  & 0\\
    0            & 0           & 0  & 0 & 0  & 0\\
    -1           & 0           & 0  & 0 & 0  & 0\\
    -\frac{1}{2} & -3          & -1 & 0 & 0  & 0\\
    0            & -1          & 0  & 0 & 0  & 0\\
    0            & \frac{3}{2} & 0  & 0 & -1 & 0
  \end{pmatrix}.
\end{equation}
According to the classification in table \ref{HD:enumeration}, we especially find that the degeneration types shown in figure \ref{limitingF_example}.

To find the commuting $\slt$-triples, we  need the full characterisation of the Hodge filtration $0 \subs F^3 \subs F^2 \subs F^1 \subs F^0 = V_\bbC$. According to special geometry, the period generating Hodge flags lower than $F^3$ can be obtained by taking various derivatives with respect to $t^i$. We make the following choice
\begin{equation}
  \Pi(t^1, t^2) = \left(\mathbf{\Pi^\Omega}, \partial_{t^1}\mathbf{\Pi^\Omega}, \partial_{t^2}\mathbf{\Pi^\Omega}, \frac{1}{9}\partial_{t^1}^2\mathbf{\Pi^\Omega}, \partial_{t^2}^2\mathbf{\Pi^\Omega}, \frac{1}{9}\partial_{t^1}^3\mathbf{\Pi^\Omega}\right),
\end{equation}
where the coefficient $\frac{1}{9}$ is chosen for convenience. Each entry in $\Pi(t^1, t^2)$ is understood to be a column vector, representing the Hodge basis in terms of the multi-valued integral basis $\{\gamma_i\}$. Further explanation of the period matrix representation can be found in the next subsection.

\subsection{The commuting $\slt$-pair associated to the degeneration $\rIII_0 \to \rIV_2$}

In this subsection, we compute the commuting $\slt$-pair arising from a degeneration from type $\rIII_0$ to $\rIV_2$, which amounts to an ordering $(N_2, N_1)$.
We also denote $N_{(2)} = N_2 + N_1$.

\subsubsection{Initial data: the mixed Hodge structure around the large volume point}

Let $(\gamma_5, \gamma_4, \gamma_3, \gamma_2, \gamma_1, \gamma_0)$ be the multi-valued integral basis in terms of which the Hodge basis are represented as the period matrix. Upon looping $z^i \mapsto e^{2\pi\im}z^i$ counterclockwise, they experience the monodromy transformation $(\gamma_5, \ldots, \gamma_0) \mapsto (\gamma_5, \ldots, \gamma_0)T_i$ which is equivalent to sending $t^i \mapsto t^i - 1$ in the period matrix. We first define a set of untwisted basis elements by setting
\begin{equation}
  (e_5, e_4, \ldots, e_0)_{t^1, t^2} := (\gamma_5, \gamma_4, \ldots, \gamma_0)_{t^1, t^2} e^{-t^1 N_1 - t^2 N_2}
\end{equation}
where the subscript $t^1, t^2$ reminds us that all the base vectors are $(t^1, t^2)$-dependent. The basis $\{e_i(t^1, t^2)\}$ are invariant under the monodromy transformation. Then the limiting Hodge filtration is extracted by sending $t^1, t^2 \to \im\infty$:
\begin{align}
  \Pi_\infty & = \lim_{\substack{t^1 \to \im\infty\\t^2 \to \im\infty}} e^{t^1 N_1 + t^2 N_2} \Pi(t^1, t^2)\nonumber\\
             & = \left(
                 \begin{array}{r|c|cc|cc|c}
                       & w_3                             & w_{21}        & w_{22}         & w_{11}      &  w_{12}       & w_0\\
                   \hline
                   e_5 & 1                               & 0             & 0              & 0           &  0            & 0 \\
                   e_4 & 0                               & 1             & 0              & 0           &  0            & 0 \\
                   e_3 & 0                               & 0             & 1              & 0           &  0            & 0 \\
                   e_2 & -\frac{17}{4}                   & \frac{9}{2}   & \frac{1}{2}    & 1           &  1            & 0 \\
                   e_1 & -\frac{3}{2}                    & \frac{3}{2}   & 0              & \frac{1}{3} &  0            & 0 \\
                   e_0 & -\frac{135\im \zeta(3)}{2\pi^3} & -\frac{23}{4} & -\frac{3}{2}   & 0           &  0            & 1
                 \end{array}
                 \right),
\end{align}
where the constant $\{e_i\}$ basis are now understood as the limit of the untwisted basis $\{e_i(t^1, t^2)\}$ as $t^1, t^2 \to \im\infty$.

For clarity, we explain the meaning of the period matrix: A Hodge filtration $0 \subs F^3 \subs F^2 \subs F^1 \subs F^0 = V_\bbC$ is characterised by a Hodge basis $(w_3, \ldots, w_0)$ generating the Hodge flags. In our $2$-moduli example whose Hodge numbers of the middle cohomology $H^3(Y_3, \bbC)$ are always $(1, 2, 2, 1)$, we have
\begin{align}
  F^3 & = \spanC{w_3},\quad F^2 = \spanC{w_3, w_{21}, w_{22}},\\
  F^1 & = \spanC{w_3, w_{21}, w_{22}, w_{11}, w_{12}},\quad F^0 = \spanC{w_3, w_{21}, w_{22}, w_{11}, w_{12}, w_0}. \nonumber
\end{align}
Then the period matrix representing a Hodge flag consists of column vectors expressing the Hodge basis $\{w_i\}$ in terms of the single-valued integral basis $\{e_i\}$. For example, in the above period matrix $\Pi_\infty$, the basis $w_{22} = e_3 + \frac{1}{2}e_2 - \frac{3}{2}e_0$. In the following, every operator acting on $F$ will be regarded as transforming the $\{w_i\}$ vectors, whose action is computed as right multiplication on the period matrix. While the above usage of nilpotent orbit theorem is regraded as a change of the integral basis so we have the (inverse) action of $e^{-t^1 N_1 - t^2 N_2}$ on the left. For clarity, we have labelled the column and rows in every period matrix representing the Hodge filtration in a limiting mixed Hodge structure.

We also need the monodromy weight filtration $W^2 := W\left(N_{(2)}\right)$ associated to the cone $\sigma(N_1, N_2)$. It is simply given by
\begin{equation}
  \begin{array}{rcccl}
           &   & W^2_6 & = & \spanR{e_5, e_4, e_3, e_2, e_1, e_0}\\
           &   & \cup  &   &\\
    W^2_5  & = & W^2_4 & = & \spanR{e_4, e_3, e_2, e_1, e_0}\\
           &   & \cup  &   &\\
    W^2_3  & = & W^2_2 & = & \spanR{e_2, e_1, e_0}\\
           &   & \cup  &   &\\
    W^2_1  & = & W^2_0 & = & \spanR{e_0}
  \end{array}
\end{equation}

One can check that this filtration indeed satisfies the following conditions:
\begin{align}
  N_{12}   &: W^2_i \to W^2_{i - 2} \textrm{ for every } i,\nonumber\\
  N_{12}^k &: \Gr{W^2}{3 + k} \to \Gr{W^2}{3 - k} \textrm{ is an isomorphism for every } k.\nonumber
\end{align}

From now on, it is helpful to forget the geometric origin of this limiting mixed Hodge structure and only regard it as a construction in linear algebra. To clarify: We fix a $6$-dimensional real vector space $V$ with a distinguished real basis $(e_5, \ldots, e_0)$ and two nilpotent matrices $N_1, N_2$ expressed in the $\{e_i\}$-basis. The mixed Hodge structure to work with is then $(V, F_\infty, W^2)$.

\subsubsection{First round: finding the $\SLt$-splitting of $(V, F_\infty, W^2)$}
\begin{figure}[!h]
\begin{center}
\begin{tikzpicture}[scale=1,cm={cos(45),sin(45),-sin(45),cos(45),(15,0)}]
  \draw[step = 1, gray, ultra thin] (0, 0) grid (3, 3);

  \draw[fill] (0, 0) circle[radius=0.04] node[right]{$w_0$};
  \draw[<-] (0.1, 0.1) -- (0.9, 0.9);
  \draw[fill] (1.05, 0.95) circle[radius=0.04] node[right]{$w_{12}$};
  \draw[fill] (0.95, 1.05) circle[radius=0.04] node[left]{$w_{11}$};
  \draw[<-] (1.1, 1.1) -- (1.9, 1.9);
  \draw[fill] (1.95, 2.05) circle[radius=0.04] node[left]{$w_{21}$};
  \draw[fill] (2.05, 1.95) circle[radius=0.04] node[right]{$w_{22}$};
  \draw[<-] (2.1, 2.1) -- (2.9, 2.9);
  \draw[fill] (3, 3) circle[radius=0.04] node[right]{$w_3$};
\end{tikzpicture}
\end{center}
\caption{The Hodge diamond of the mixed Hodge structure $(V, F_\infty, W^2)$, in which each dot near the $(p, q)$-site represents a base vector of the corresponding subspace $I^{p, q}_\infty$. The arrows show the action of $N_{(2)}$.} \label{fig:Hodge_diamond_IV2_of_III0IV2}
\end{figure}
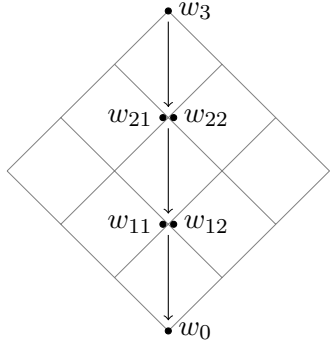

Firstly we need to find the Deligne splitting of $(V, F_\infty, W^2)$. Denote the Deligne splitting by $V_\bbC = \bigoplus I^{p, q}_\infty$ and it can be computed by directly applying the definition \eqref{def-Ipq}. The result is given in the Hodge diamond in figure \ref{fig:Hodge_diamond_IV2_of_III0IV2} and we note that the shape of the Hodge diagram clearly shows that at the large complex structure point $\Delta_{12}$ the degeneration type is $\rIV_2$.

We can further check that the splitting satisfies the conjugation property that $I_\infty^{p, q} = \conj{I_\infty^{q, p}}$ for all $p, q$ except
\begin{equation}
  I_\infty^{3, 3} = \conj{I_\infty^{3, 3}} \mod I_\infty^{0, 0},\nonumber
\end{equation}
hence the mixed Hodge structure $(F_\infty, W^2)$ is not $\bbR$-split.

The grading operator $T$ and its complex conjugate $\conj{T}$ defined in appendix \ref{app:Sl2-splitting} expressed in the Hodge basis $(w_3, \ldots, w_0)$ can be directly read out from figure \ref{fig:Hodge_diamond_IV2_of_III0IV2}
\begin{equation}
  T =
  \begin{pmatrix}
    6 & 0 & 0 & 0 & 0 & 0\\
    0 & 4 & 0 & 0 & 0 & 0\\
    0 & 0 & 4 & 0 & 0 & 0\\
    0 & 0 & 0 & 2 & 0 & 0\\
    0 & 0 & 0 & 0 & 2 & 0\\
    0 & 0 & 0 & 0 & 0 & 0
  \end{pmatrix},\qquad
  \conj{T} =
  \begin{pmatrix}
    6                            & 0 & 0 & 0 & 0 & 0\\
    0                            & 4 & 0 & 0 & 0 & 0\\
    0                            & 0 & 4 & 0 & 0 & 0\\
    0                            & 0 & 0 & 2 & 0 & 0\\
    0                            & 0 & 0 & 0 & 2 & 0\\
    \frac{810\im\zeta(3)}{\pi^3} & 0 & 0 & 0 & 0 & 0
  \end{pmatrix}.
\end{equation}

Then the $\delta$ operator written in the $\{w_i\}$ basis is solved to be
\begin{equation}
  \delta =
  \begin{pmatrix}
    0                           & 0 & 0 & 0 & 0 & 0\\
    0                           & 0 & 0 & 0 & 0 & 0\\
    0                           & 0 & 0 & 0 & 0 & 0\\
    0                           & 0 & 0 & 0 & 0 & 0\\
    0                           & 0 & 0 & 0 & 0 & 0\\
    -\frac{135\zeta(3)}{2\pi^3} & 0 & 0 & 0 & 0 & 0
  \end{pmatrix}.
\end{equation}
It is easily seen that the $\delta$ operator only has the $\delta_{-3, -3}$ component, hence the $\zeta$ operator is simply $\zeta = 0$.

Computing $F_2 = e^{-\im\delta}F_\infty$ we have found the $\SLt$-splitting $(F_2, W^2)$ of $(F_\infty, W^2)$. The filtration $F_2$ is represented by its period matrix
\begin{equation} \label{SL2splitting-F2W2}
  \Pi_2 =
  \left(
  \begin{array}{r|c|cc|cc|c}
        & w^{(2)}_3     & w^{(2)}_{21}  & w^{(2)}_{22}   & w^{(2)}_{11}  & w^{(2)}_{12} & w^{(2)}_0\\
    \hline
    e_5 & 1             & 0             & 0              & 0           & 0            & 0\\
    e_4 & 0             & 1             & 0              & 0           & 0            & 0\\
    e_3 & 0             & 0             & 1              & 0           & 0            & 0\\
    e_2 & -\frac{17}{4} & \frac{9}{2}   & \frac{1}{2}    & 1           & 1            & 0\\
    e_1 & -\frac{3}{2}  & \frac{3}{2}   & 0              & \frac{1}{3} & 0            & 0\\
    e_0 & 0             & -\frac{23}{4} & -\frac{3}{2}   & 0           & 0            & 1
  \end{array}
  \right).
\end{equation}
And it is clear that this mixed Hodge structure $(F_2, W^2)$ is $\bbR$-split

\subsubsection{The second round: finding the $\SLt$-splitting of $(F', W^1)$}
We now proceed to the second round of the computation. The starting point of this round is the mixed Hodge structure $(F', W^1)$, where $W^1 = W(N_2)$ is the monodromy weight filtration associated to $N_2$, and $F' = e^{\im N_1}F_2$. One can check that the weight filtration is now given by
\begin{equation}
  \begin{array}{rcccl}
    W^1_6  & = & W^1_5 & = & \spanR{e_5, e_4, e_3, e_2, e_1, e_0}\\
           &   & \cup  &   &\\
    W^1_4  & = & W^1_3 & = & \spanR{e_3, e_2, e_1, e_0}\\
           &   & \cup  &   &\\
    W^1_2  & = & W^1_1 & = & \spanR{e_2, e_0}\\
           &   & \cup  &   &\\
           &   & W^1_0 & = & 0
  \end{array}
\end{equation}
and the period matrix $\Pi'$ representing $F'$ is
\begin{equation}
  \Pi' =
  \left(
  \begin{array}{r|c|cc|cc|c}
        & w'_3                           & w'_{21}            & w'_{22}            & w'_{11}     & w'_{12} & w'_0\\
    \hline
    e_5 & 1                              & 0                  & 0                  & 0           & 0       & 0\\
    e_4 & \im                            & 1                  & 0                  & 0           & 0       & 0\\
    e_3 & 0                              & 0                  & 1                  & 0           & 0       & 0\\
    e_2 & -\frac{35}{4} + \frac{9\im}{2} & \frac{9}{2} + 9\im & \frac{1}{2} + 3\im & 1           & 1       & 0\\
    e_1 & -3 + \frac{3\im}{2}            & \frac{3}{2} + 3\im & \im                & \frac{1}{3} & 0       & 0\\
    e_0 & -\frac{29\im}{4}               & -\frac{41}{4}      & -3                 & \im         & \im     & 1
  \end{array}
  \right).
\end{equation}

Denote the Deligne splitting of $(F', W^1)$ by $V_\bbC = \bigoplus I'^{p, q}$. Using the formula \eqref{def-Ipq} we find the Deligne splitting of $(V, F', W^1)$ shown in figure \ref{fig:Hodge_diamond_III0_of_III0IV2}.

\begin{figure}[!h]
\begin{center}
\begin{tikzpicture}[scale=1,cm={cos(45),sin(45),-sin(45),cos(45),(15,0)}]
  \draw[step = 1, gray, ultra thin] (0, 0) grid (3, 3);
  \draw[fill] (0, 1) circle[radius=0.05] node[left]{$w'_{12} - 2\im w'_0$};
  \draw[fill] (1, 0) circle[radius=0.05] node[right]{$w'_{12}$};
  \draw[fill] (1, 2) circle[radius=0.05] node[left]{$w'_{22} - 6\im w'_{11} + 3\im w'_{12}$};
  \draw[fill] (2, 1) circle[radius=0.05] node[right]{$w'_{22}$};
  \draw[fill] (2, 3) circle[radius=0.05] node[left]{$w'_3 - 2\im w'_{21} + 3\im w'_{22}$};
  \draw[fill] (3, 2) circle[radius=0.05] node[right]{$w'_3$};
  \draw[->] (2.9, 1.9) -- (2.1, 1.1);
  \draw[->] (1.9, 2.9) -- (1.1, 2.1);
  \draw[->] (1.9, 0.9) -- (1.1, 0.1);
  \draw[->] (0.9, 1.9) -- (0.1, 1.1);
\end{tikzpicture}
\end{center}
\caption{The Hodge diamond of the mixed Hodge structure $(V, F', W^1)$, in which each dot near the $(p, q)$-site represents a base vector of the corresponding subspace $I'^{p, q}$. The arrows show the action of $N_{2}$. The diamond is clearly of type $\rIII_0$.} \label{fig:Hodge_diamond_III0_of_III0IV2}
\end{figure}

We further remark that this splitting satisfies $I'^{p, q} = \conj{I'^{q, p}}$ for all $p, q$ except
\begin{align}
  I'^{3, 2} & = \conj{I'^{2, 3}} \mod I'^{2, 1} \oplus I'^{1, 0} \oplus I'^{0, 1},\nonumber\\
  I'^{2, 1} & = \conj{I'^{1, 2}} \mod I'^{1, 0}.\nonumber
\end{align}

The Deligne splitting in figure \ref{fig:Hodge_diamond_III0_of_III0IV2} yields the following grading operator $T'$ and its complex conjugate $\conj{T}'$ expressed in the Hodge basis $(w'_3, \ldots, w'_0)$
\begin{equation}
  T' =
  \left(
  \begin{array}{cccccc}
    5 & 0  & 0 & 0  & 0 & 0\\
    0 & 5  & 0 & 0  & 0 & 0\\
    0 & -3 & 3 & 0  & 0 & 0\\
    0 & 0  & 0 & 3  & 0 & 0\\
    0 & 0  & 0 & -1 & 1 & 0\\
    0 & 0  & 0 & 0  & 0 & 1
  \end{array}
  \right),\qquad
  \conj{T}' =
  \left(
  \begin{array}{cccccc}
    5      & 0      & 0     & 0     & 0 & 0\\
    0      & 5      & 0     & 0     & 0 & 0\\
    -6\im  & -3     & 3     & 0     & 0 & 0\\
    0      & -18\im & 0     & 3     & 0 & 0\\
    18     & -18\im & -6\im & -1    & 1 & 0\\
    -24\im & -18    & 0     & -2\im & 0 & 1
  \end{array}
  \right).
\end{equation}

The operator $\delta'$ written in the Hodge basis $(w'_3, \ldots, w'_0)$ is solved to be
\begin{equation}
  \delta' =
  \left(
  \begin{array}{cccccc}
    0              & 0               & 0           & 0           & 0 & 0\\
    0              & 0               & 0           & 0           & 0 & 0\\
    \frac{3}{2}    & 0               & 0           & 0           & 0 & 0\\
    0              & \frac{9}{2}     & 0           & 0           & 0 & 0\\
    \frac{9\im}{4} & \frac{9}{4}     & \frac{3}{2} & 0           & 0 & 0\\
    3              & -\frac{9\im}{4} & 0           & \frac{1}{2} & 0 & 0
  \end{array}
  \right).
\end{equation}
This matrix does not seem to be real because we are working in the complex basis $\{w'_i\}$. If we transform it into the $(e_5, \ldots, e_0)$ basis using the period matrix $\Pi'$ then all of its entries are real numbers. Hence $\delta'$ is indeed a real operator.

Let $\tilde{F}' = e^{-\im \delta'}F'$, and we have found the first $\bbR$-split mixed Hodge structure associated with $(F', W^1)$. Let $(\tilde{w}'_3, \ldots, \tilde{w}'_0) = (w_3, \ldots, w_0)e^{-\im \delta'}$ and we have a new set of Hodge basis $\{\tilde{w}'_i\}$. The Deligne splitting of $(F', W^1)$ is the same as in figure \ref{fig:Hodge_diamond_III0_of_III0IV2} with all $w'_i$ replaced by $\tilde{w}'_i$. Then the decomposition of the operator $\delta'$ is found to be
\begin{equation}
  \delta' = \delta'_{-1, -1} + \delta'_{-2, -2} + \delta'_{-3, -1} + \delta'_{-1, -3},
\end{equation}
where $\delta'_{-p, -q}$ maps $\tilde{I}'^{r, s}$ to $\tilde{I}'^{r - p, s - q}$. The components are given by, in the $\tilde{w}'_i$ basis,
\begin{equation}
  \delta'_{-1, -1} =
  \left(
  \begin{array}{cccccc}
    0           & 0           & 0           & 0           & 0 & 0\\
    0           & 0           & 0           & 0           & 0 & 0\\
    \frac{3}{2} & 0           & 0           & 0           & 0 & 0\\
    0           & \frac{9}{2} & 0           & 0           & 0 & 0\\
    0           & 0           & \frac{3}{2} & 0           & 0 & 0\\
    0           & 0           & 0           & \frac{1}{2} & 0 & 0
  \end{array}
  \right),\qquad
  \delta'_{-2, -2} =
  \left(
  \begin{array}{cccccc}
    0              & 0               & 0 & 0 & 0 & 0\\
    0              & 0               & 0 & 0 & 0 & 0\\
    0              & 0               & 0 & 0 & 0 & 0\\
    0              & 0               & 0 & 0 & 0 & 0\\
    \frac{3\im}{4} & \frac{3}{4}     & 0 & 0 & 0 & 0\\
    0              & -\frac{3\im}{4} & 0 & 0 & 0 & 0
  \end{array}
  \right),
\end{equation}
\begin{equation}
  \delta'_{-3, -1} =
  \left(
  \begin{array}{cccccc}
    0              & 0               & 0 & 0 & 0 & 0\\
    0              & 0               & 0 & 0 & 0 & 0\\
    0              & 0               & 0 & 0 & 0 & 0\\
    0              & 0               & 0 & 0 & 0 & 0\\
    \frac{3\im}{2} & \frac{3}{4}     & 0 & 0 & 0 & 0\\
    3              & -\frac{3\im}{2} & 0 & 0 & 0 & 0
  \end{array}
  \right),\qquad
  \delta'_{-1, -3} =
  \left(
  \begin{array}{cccccc}
    0 & 0           & 0 & 0 & 0 & 0\\
    0 & 0           & 0 & 0 & 0 & 0\\
    0 & 0           & 0 & 0 & 0 & 0\\
    0 & 0           & 0 & 0 & 0 & 0\\
    0 & \frac{3}{4} & 0 & 0 & 0 & 0\\
    0 & 0           & 0 & 0 & 0 & 0
  \end{array}
  \right).
\end{equation}
Furthermore, all components are commuting with each other $[\delta'_{-p, -q}, \delta'_{-r, -s}] = 0$.

The operator $\zeta'$ given in terms of its decomposition $\zeta' = \sum \zeta'_{-p, -q}$ only has two non-vanishing components $\zeta'_{-1, -3}$ and $\zeta'_{-3, -1}$, hence, written in the $(\tilde{w}'_3, \ldots, \tilde{w}'_0)$ basis
\begin{equation}
  \zeta' = \frac{3\im}{4}(\delta'_{-3, -1} - \delta'_{-1, -3}) =
         \left(
         \begin{array}{cccccc}
           0              & 0           & 0 & 0 & 0 & 0\\
           0              & 0           & 0 & 0 & 0 & 0\\
           0              & 0           & 0 & 0 & 0 & 0\\
           0              & 0           & 0 & 0 & 0 & 0\\
           -\frac{9}{8}   & 0           & 0 & 0 & 0 & 0\\
           \frac{9\im}{4} & \frac{9}{8} & 0 & 0 & 0 & 0
         \end{array}
         \right).
\end{equation}
Finally, applying the operator $e^{\zeta'}$ to $\tilde{F}'$, we arrive at the $\SLt$-splitting $(F_1, W^1)$ associated to $(F', W^1)$. The period matrix $\Pi_1$ representing the Hodge filtration $F_1 = e^{\zeta'}e^{-\im\delta'}F'$ is
\begin{equation}
  \Pi_1 =
  \left(
  \begin{array}{r|c|cc|cc|c}
        & w^{(1)}_3                       & w^{(1)}_{21}                 & w^{(1)}_{22}                 & w^{(1)}_{11}  & w^{(1)}_{12} & w^{(1)}_0\\
    \hline
    e_5 & 1                               & 0                            & 0                            & 0             & 0            & 0\\
    e_4 & \im                             & 1                            & 0                            & 0             & 0            & 0\\
    e_3 & -\frac{3\im}{2}                 & 0                            & 1                            & 0             & 0            & 0\\
    e_2 & -\frac{17}{4} + \frac{15\im}{4} & \frac{9}{2} + \frac{9\im}{4} & \frac{1}{2} + \frac{3\im}{2} & 1             & 1            & 0\\
    e_1 & -\frac{3}{2} + \frac{3\im}{2}   & \frac{3}{2} + \frac{3\im}{2} & \im                          & \frac{1}{3}   & 0            & 0\\
    e_0 & -\frac{7\im}{2}                 & -\frac{23}{4}                & -\frac{3}{2}                 & \frac{\im}{2} & \im          & 1
  \end{array}
  \right)
\end{equation}

\subsubsection{Final output: the commuting $\slt$-pair}

With the two $\SLt$-splittings $(F_i, W^i)$ we can now compute the commuting $\slt$-pair. First we read out the semisimple grading $Y_{(i)}$ which acts on $I^{p, q}_{(F_i, W^i)}$ as multiplication by $p + q - 3$. Writing now everything in the real basis $(e_5, \ldots, e_0)$ for convenience, we have
\begin{equation}
  Y_{(1)} =
  \left(
  \begin{array}{cccccc}
    2             & 0             & 0  & 0  & 0 & 0\\
    0             & 2             & 0  & 0  & 0 & 0\\
    0             & -3            & 0  & 0  & 0 & 0\\
    -\frac{25}{2} & 12            & 1  & -2 & 3 & 0\\
    -3            & 3             & 0  & 0  & 0 & 0\\
    0             & -\frac{37}{2} & -3 & 0  & 0 & -2
  \end{array}
  \right),\qquad
  Y_{(2)} =
  \left(
  \begin{array}{cccccc}
    3   & 0   & 0  & 0  & 0  & 0\\
    0   & 1   & 0  & 0  & 0  & 0\\
    0   & 0   & 1  & 0  & 0  & 0\\
    -17 & 9   & 1  & -1 & 0  & 0\\
    -6  & 3   & 0  & 0  & -1 & 0\\
    6   & -23 & -6 & 0  & 0  & -3
  \end{array}
  \right),
\end{equation}
so the neutral elements in the $\slt$-pair are
\begin{equation} \label{sl2_neutral_III0_IV2}
  Y_1 = Y_{(1)},\quad Y_2 = Y_{(2)} - Y_{(1)}
      =
      \left(
      \begin{array}{cccccc}
        1            & 0            & 0  & 0 & 0  & 0\\
        0            & -1           & 0  & 0 & 0  & 0\\
        0            & 3            & 1  & 0 & 0  & 0\\
        -\frac{9}{2} & -3           & 0  & 1 & -3 & 0\\
        -3           & 0            & 0  & 0 & -1 & 0\\
        0            & -\frac{9}{2} & -3 & 0 & 0  & -1
      \end{array}
      \right).
\end{equation}
In addition, $N_1^- = N_2$ is already one of the nilnegative elements. We kindly remind the reader that the particular ordering $(N_2, N_1)$ of the monodromies is adopted so as to study the degeneration $\rIII_0 \to \rIV_2$.

To find the other nilnegative element $N_2^-$, we compute the decomposition of $N_1$ into the eigenvectors of the adjoint representation $[Y_{(1)}, -]$. Denote the decomposition $N_1 = \sum N_1^\alpha$, where $[Y_{(1)}, N_1^\alpha] = \alpha N_1^\alpha$ is the component corresponding to the eigenvalue $\alpha$. Bearing in mind that any component $N_1^\alpha$ must also preserve the polarisation $(N_1^\alpha)^\mathrm{T} \eta + \eta N_1^\alpha = 0$, we find that
\begin{equation}
  N_1 = N_1^{-4} + N_1^{-2} + N_1^0,
\end{equation}
where
\begin{equation}
  N_1^{-4} =
  \left(
  \begin{array}{cccccc}
    0 & 0            & 0 & 0 & 0 & 0\\
    0 & 0            & 0 & 0 & 0 & 0\\
    0 & 0            & 0 & 0 & 0 & 0\\
    0 & -\frac{9}{4} & 0 & 0 & 0 & 0\\
    0 & 0            & 0 & 0 & 0 & 0\\
    0 & 0            & 0 & 0 & 0 & 0
  \end{array}
  \right),\qquad
  N_1^{-2} =
  \left(
  \begin{array}{cccccc}
    0            & 0            & 0            & 0 & 0            & 0\\
    0            & 0            & 0            & 0 & 0            & 0\\
    -\frac{3}{2} & 0            & 0            & 0 & 0            & 0\\
    -\frac{3}{4} & -\frac{9}{2} & -\frac{3}{2} & 0 & 0            & 0\\
    0            & -\frac{3}{2} & 0            & 0 & 0            & 0\\
    0            & \frac{9}{4}  & 0            & 0 & -\frac{3}{2} & 0
  \end{array}
  \right),
\end{equation}
and the $N_1^0$ is what we need for the nilnegatives
\begin{align} \label{sl2_nilnegative_III0_IV2}
  N_1^- = N_2,\quad N_2^- = N_1^0 =
            \left(
            \begin{array}{cccccc}
              0             & 0            & 0            & 0  & 0           & 0\\
              -1            & 0            & 0            & 0  & 0           & 0\\
              \frac{3}{2}   & 0            & 0            & 0  & 0           & 0\\
              -\frac{15}{4} & -\frac{9}{4} & -\frac{3}{2} & 0  & 0           & 0\\
              -\frac{3}{2}  & -\frac{3}{2} & -1           & 0  & 0           & 0\\
              \frac{3}{2}   & \frac{9}{4}  & \frac{1}{2}  & -1 & \frac{3}{2} & 0
            \end{array}
            \right).
\end{align}

The last step is to find the nilpositive element $N_i^+$. Solving the equations
\begin{equation}
  [Y_i, N_i^+] = 2N_i^+,\quad [N_i^+, N_i^-] = Y_i,\quad (N_i^+)^\mathrm{T}\eta + \eta N_i^+ = 0,\nonumber
\end{equation}
simply yields the following unique pair of matrices
\begin{equation} \label{sl2_nilpositive_III0_IV2}
  N_1^+ =
  \left(
  \begin{array}{cccccc}
    0             & -3            & -2           & 0  & 0           & 0\\
    -3            & 3             & 0            & 0  & -2          & 0\\
    \frac{1}{2}   & 0             & 1            & -2 & 6           & 0\\
    -\frac{53}{4} & 9             & \frac{9}{2}  & -1 & -6          & -3\\
    -\frac{9}{2}  & -\frac{5}{2}  & 0            & 0  & -3          & -2\\
    \frac{33}{2}  & -\frac{69}{4} & -\frac{3}{2} & 3  & \frac{5}{2} & 0
  \end{array}
  \right),\qquad
  N_2^+ =
  \left(
  \begin{array}{cccccc}
    0            & -1           & 0            & 0 & 0            & 0\\
    0            & 0            & 0            & 0 & 0            & 0\\
    -\frac{3}{2} & \frac{3}{2}  & 0            & 0 & -1           & 0\\
    -\frac{3}{4} & -\frac{3}{4} & -\frac{3}{2} & 0 & -\frac{1}{2} & -1\\
    0            & \frac{3}{2}  & 0            & 0 & 0            & 0\\
    \frac{9}{4}  & -\frac{9}{4} & 0            & 0 & \frac{3}{2}  & 0
  \end{array}
  \right).
\end{equation}

One can finally check that the $(N_i^-, N_i^+, Y_i)$ with matrices in the $(e_5, \ldots, e_0)$ basis given by \eqref{sl2_nilnegative_III0_IV2}, \eqref{sl2_nilpositive_III0_IV2}, \eqref{sl2_neutral_III0_IV2} are indeed two sets of $\slt$-Lie algebra elements and the two sets of operators commute with each other. This completes our computation of the commuting $\slt$-pair arising from the $\rIII_0 \to \rIV_2$ degeneration in the complex structure moduli space of the Calabi-Yau threefold $Y_3$.

\subsection{The commuting $\slt$-pair associated to the degeneration $\rIV_1 \to \rIV_2$}
The other singularity locus $\Delta_1$ in the moduli space of $Y_3$ has the type $\rIV_1$. In this subsection we also work out the commuting $\slt$-pair as we move along $\Delta_1$ towards the large complex structure point of type $\rIV_2$. This amounts to switch the ordering of the monodromy cone to $(N_1, N_2)$. The computation is essentially the same as the $\rIII_0 \to \rIV_2$ degeneration, so we only list the results here without explanation.

\subsubsection{The two $\SLt$-splittings}
The starting point $(F_\infty, W^2)$ is the same as the starting point of $\rIII_0 \to \rIV_2$, hence also its $\SLt$-splitting is the same $(F_2, W^2)$ with the period matrix \eqref{SL2splitting-F2W2}. Now, we consider the limiting mixed Hodge structure $(F', W^1)$ where $F' = e^{\im N_2} F_2$ and $W^1 = W(N_1)$. The period matrix of $F'$ is now given by
\begin{equation}
  \Pi' =
  \left(
    \begin{array}{r|c|cc|cc|c}
        & w'_3                          & w'_{21}             & w'_{22}           & w'_{11}       & w'_{12} & w'_0\\
    \hline
    e_5 & 1                             & 0                   & 0                 & 0             & 0       & 0\\
    e_4 & 0                             & 1                   & 0                 & 0             & 0       & 0\\
    e_3 & \im                           & 0                   & 1                 & 0             & 0       & 0\\
    e_2 & -\frac{19}{4} + \frac{\im}{2} & \frac{9}{2} + 3\im  & \frac{1}{2} + \im & 1             & 1       & 0\\
    e_1 & -\frac{3}{2}                  & \frac{3}{2} + \im   & 0                 & \frac{1}{3}   & 0       & 0\\
    e_0 & -\frac{3\im}{2}               & -\frac{25}{4}       & -\frac{3}{2}      & \frac{\im}{3} & 0       & 1
    \end{array}
  \right),
\end{equation}
and the monodromy weight filtration $W^1$ has now the form
\begin{equation}
  \begin{array}{rcccl}
          &   & W^1_6 & = & \spanR{e_5, e_4, e_3, e_2, e_1, e_0}\\
          &   & \cup  &   &\\
    W^1_5 & = & W^1_4 & = & \spanR{e_4, e_3, e_2, e_1, e_0}\\
          &   & \cup  &   &\\
          &   & W^1_3 & = & \spanR{-e_4 + 3e_3, e_2, e_1, e_0}\\
          &   & \cup  &   &\\
          &   & W^1_2 & = & \spanR{3e_2 + e_1, e_0}\\
          &   & \cup  &   &\\
    W^1_1 & = & W^1_0 & = & \spanR{e_0}
  \end{array}
\end{equation}

So the Deligne splitting $V_\bbC = \bigoplus I'^{p, q}$ is found to be in the figure \ref{fig:Hodge_diamond_IV1_of_IV1IV2}.
\begin{figure}[!h]
\begin{center}
\begin{tikzpicture}[scale=1,cm={cos(45),sin(45),-sin(45),cos(45),(15,0)}]
  \draw[step = 1, gray, ultra thin] (0, 0) grid (3, 3);

  \draw[fill] (0, 0) circle[radius=0.05] node[right]{$w'_0$};
  \draw[<-] (0.1, 0.1) -- (0.9, 0.9);
  \draw[fill] (1, 1) circle[radius=0.05] node[right]{$w'_{11}$};
  \draw[<-] (1.1, 1.1) -- (1.9, 1.9);
  \draw[fill] (1, 2) circle[radius=0.05] node[left]{$-w'_{21} + 3w'_{22} + 6\im w'_{11} - 6\im w'_{12}$};
  \draw[fill] (2, 1) circle[radius=0.05] node[right]{$-w'_{21} + 3w'_{22}$};
  \draw[fill] (2, 2) circle[radius=0.05] node[right]{$w'_{21}$};
  \draw[<-] (2.1, 2.1) -- (2.9, 2.9);
  \draw[fill] (3, 3) circle[radius=0.05] node[right]{$w'_3$};
\end{tikzpicture}
\end{center}
\caption{The Hodge diamond of the mixed Hodge structure $(V, F', W^1)$, in which each dot near the $(p, q)$-site represents a base vector of the corresponding subspace $I'^{p, q}$. The arrows show the action of $N_{1}$. The diamond is clearly of type $\rIV_1$.} \label{fig:Hodge_diamond_IV1_of_IV1IV2}
\end{figure}

This structure is again far from $\bbR$-split, and we can check that $I^{p, q} = \conj{I^{q, p}}$ for all $p, q$ except
\begin{align}
  I'^{3, 3} & = \conj{I'^{3, 3}} \mod I'^{2, 2} \oplus I'^{1, 2} \oplus I'^{2, 1} \oplus I'^{1, 1},\\
  I'^{2, 2} & = \conj{I'^{2, 2}} \mod I'^{1, 1} \oplus I'^{0, 0},\nonumber\\
  I'^{2, 1} & = \conj{I'^{1, 2}} \mod I'^{0, 0},\nonumber\\
  I'^{1, 1} & = \conj{I'^{1, 1}} \mod I'^{0, 0}.\nonumber
\end{align}

Reading out the grading and solving for $\delta'$, we find, in the $(w'_3, \ldots, w'_0)$ basis
\begin{equation}
  \delta' = \left(
    \begin{array}{cccccc}
      0           & 0 & 0              & 0           & 0 & 0\\
      0           & 0 & 0              & 0           & 0 & 0\\
      1           & 0 & 0              & 0           & 0 & 0\\
      \im         & 3 & 1              & 0           & 0 & 0\\
      -\im        & 0 & 0              & 0           & 0 & 0\\
      \frac{2}{9} & 0 & -\frac{\im}{3} & \frac{1}{3} & 0 & 0
    \end{array}
  \right),
\end{equation}
which consists of real elements once we transform back to the $\{e_i\}$ basis.

The operator $\delta'$ now admits the following Deligne splitting
\begin{equation}
  \delta' = \delta'_{-3, -3} + \delta'_{-2, -1} + \delta'_{-1, -1} + \delta'_{-1, -2},
\end{equation}
where various components are given by
\begin{align}
  \delta'_{-3, -3} & = \left(
    \begin{array}{cccccc}
      0           & 0 & 0 & 0 & 0 & 0\\
      0           & 0 & 0 & 0 & 0 & 0\\
      0           & 0 & 0 & 0 & 0 & 0\\
      0           & 0 & 0 & 0 & 0 & 0\\
      0           & 0 & 0 & 0 & 0 & 0\\
      \frac{2}{9} & 0 & 0 & 0 & 0 & 0
    \end{array}
    \right),\qquad
  \delta'_{-2, -1} = \left(
    \begin{array}{cccccc}
      0            & 0 & 0              & 0 & 0            & 0\\
      -\frac{1}{6} & 0 & 0              & 0 & 0            & 0\\
      \frac{1}{2}  & 0 & 0              & 0 & 0            & 0\\
      \im          & 0 & 0              & 0 & 0            & 0\\
      -\im         & 0 & 0              & 0 & 0            & 0\\
      0            & 0 & -\frac{\im}{3} & 0 & -\frac{1}{6} & 0
    \end{array}
    \right),\nonumber\\
  \delta'_{-1, -1} & = \left(
    \begin{array}{cccccc}
      0           & 0 & 0 & 0           & 0           & 0\\
      \frac{1}{3} & 0 & 0 & 0           & 0           & 0\\
      0           & 0 & 0 & 0           & 0           & 0\\
      0           & 3 & 1 & 0           & 0           & 0\\
      0           & 0 & 0 & 0           & 0           & 0\\
      0           & 0 & 0 & \frac{1}{3} & \frac{1}{3} & 0
    \end{array}
    \right),\qquad
  \delta'_{-1, -2} = \left(
    \begin{array}{cccccc}
      0            & 0 & 0 & 0 & 0            & 0\\
      -\frac{1}{6} & 0 & 0 & 0 & 0            & 0\\
      \frac{1}{2}  & 0 & 0 & 0 & 0            & 0\\
      0            & 0 & 0 & 0 & 0            & 0\\
      0            & 0 & 0 & 0 & 0            & 0\\
      0            & 0 & 0 & 0 & -\frac{1}{6} & 0
    \end{array}
    \right),
\end{align}
with the only non-vanishing commutator
\begin{equation}
  [\delta'_{-2, -1}, \delta'_{-1, -2}] = -\frac{3\im}{2}\delta'_{-3, -3}.
\end{equation}

Then the only non-vanishing components of $\zeta'$ are $\zeta'_{-1, -2}$ and $\zeta'_{-2, -1}$, hence they sum to the $\zeta'$ operator
\begin{equation}
  \zeta' = \frac{\im}{2}(\delta'_{-2, -1} - \delta'_{-1, -2})
         = \left(
            \begin{array}{cccccc}
              0            & 0 & 0           & 0 & 0 & 0\\
              0            & 0 & 0           & 0 & 0 & 0\\
              0            & 0 & 0           & 0 & 0 & 0\\
              -\frac{1}{2} & 0 & 0           & 0 & 0 & 0\\
              \frac{1}{2}  & 0 & 0           & 0 & 0 & 0\\
              0            & 0 & \frac{1}{6} & 0 & 0 & 0
            \end{array}
            \right),
\end{equation}
where the matrix is written in the $\{w'\}$ basis.

The Hodge filtration $F_1 = e^{\zeta'}e^{-\im \delta'}F_2$ is given by its period matrix $\Pi_1$
\begin{equation}
  \Pi_1 = \left(
        \begin{array}{r|c|cc|cc|c}
                & w^{(1)}_3     & w^{(1)}_{21}  & w^{(1)}_{22}   & w^{(1)}_{11}  & w^{(1)}_{12} & w^{(1)}_0\\
          \hline
          e'_5  & 1             & 0             & 0              & 0             & 0            & 0\\
          e'_4  & 0             & 1             & 0              & 0             & 0            & 0\\
          e'_3  & 0             & 0             & 1              & 0             & 0            & 0\\
          e'_2  & -\frac{17}{4} & \frac{9}{2}   & \frac{1}{2}    & 1             & 1            & 0\\
          e'_1  & -\frac{3}{2}  & \frac{3}{2}   & -\frac{\im}{3} & \frac{1}{3}   & 0            & 0\\
          e'_0  & 0             & -\frac{23}{4} & -\frac{3}{2}   & 0             & 0            & 1
        \end{array}
        \right),
\end{equation}
thus we have arrived at the $\SLt$-splitting $(F_1, W^1)$ of $(F', W^1)$.

\subsubsection{The commuting $\slt$-pair}
For convenience we express everything in the $(e_5, \ldots, e_0)$ basis in this subsection. We read out the grading elements
\begin{equation}
  Y_{(1)} =
  \left(
  \begin{array}{cccccc}
    3 & 0 & 0 & 0 & 0 & 0 \\
    0 & 1 & \frac{1}{3} & 0 & 0 & 0 \\
    0 & 0 & 0 & 0 & 0 & 0 \\
    -17 & 9 & 2 & -1 & 0 & 0 \\
    -\frac{71}{12} & 3 & \frac{2}{3} & -\frac{1}{3} & 0 & 0 \\
    0 & -23 & -\frac{77}{12} & 0 & 0 & -3 \\
  \end{array}
  \right),\qquad
  Y_{(2)} =
  \left(
  \begin{array}{cccccc}
    3 & 0 & 0 & 0 & 0 & 0 \\
    0 & 1 & 0 & 0 & 0 & 0 \\
    0 & 0 & 1 & 0 & 0 & 0 \\
    -17 & 9 & 1 & -1 & 0 & 0 \\
    -6 & 3 & 0 & 0 & -1 & 0 \\
    0 & -23 & -6 & 0 & 0 & -3 \\
  \end{array}
  \right),
\end{equation}
so that the neutral elements are given by
\begin{equation}
  Y_1 = Y_{(1)},\quad Y_2 = Y_{(2)} - Y_{(1)} =
    \left(
    \begin{array}{cccccc}
     0 & 0 & 0 & 0 & 0 & 0 \\
     0 & 0 & -\frac{1}{3} & 0 & 0 & 0 \\
     0 & 0 & 1 & 0 & 0 & 0 \\
     0 & 0 & -1 & 0 & 0 & 0 \\
     -\frac{1}{12} & 0 & -\frac{2}{3} & \frac{1}{3} & -1 & 0 \\
     0 & 0 & \frac{5}{12} & 0 & 0 & 0 \\
    \end{array}
    \right).
\end{equation}

Decompose $N_2 = \sum N_2^\alpha$ into the eigen-components of the action $[Y_{(1)}, -]$ and we have
\begin{equation}
  N_2 = N_2^{-3} + N_2^{-2} + N_2^0,
\end{equation}
where
\begin{equation}
  N_2^{-3} =
  \left(
  \begin{array}{cccccc}
    0 & 0 & 0 & 0 & 0 & 0 \\
    \frac{1}{3} & 0 & 0 & 0 & 0 & 0 \\
    -1 & 0 & 0 & 0 & 0 & 0 \\
    1 & 0 & 0 & 0 & 0 & 0 \\
    \frac{1}{2} & 0 & 0 & 0 & 0 & 0 \\
    -\frac{1}{2} & 0 & -\frac{1}{6} & \frac{1}{3} & -1 & 0 \\
  \end{array}
  \right),\qquad
  N_2^{-2} =
  \left(
  \begin{array}{cccccc}
    0 & 0 & 0 & 0 & 0 & 0 \\
    -\frac{1}{3} & 0 & 0 & 0 & 0 & 0 \\
    0 & 0 & 0 & 0 & 0 & 0 \\
    -\frac{3}{2} & -3 & -1 & 0 & 0 & 0 \\
    -\frac{1}{2} & -1 & -\frac{1}{3} & 0 & 0 & 0 \\
    \frac{1}{2} & \frac{3}{2} & \frac{1}{6} & -\frac{1}{3} & 0 & 0 \\
  \end{array}
  \right),
\end{equation}
and the remaining $N_2^0$ together with $N_1$ constitute the nilnegative elements
\begin{equation}
  N_1^- = N_1,\quad N_2^- = N_2^0 =
      \left(
      \begin{array}{cccccc}
        0 & 0 & 0 & 0 & 0 & 0 \\
        0 & 0 & 0 & 0 & 0 & 0 \\
        0 & 0 & 0 & 0 & 0 & 0 \\
        0 & 0 & 0 & 0 & 0 & 0 \\
        0 & 0 & \frac{1}{3} & 0 & 0 & 0 \\
        0 & 0 & 0 & 0 & 0 & 0 \\
      \end{array}
      \right).
\end{equation}

Finally we can complete the $(N_i^-, Y_i)$ into the complete $\slt$-triples $(N_i^-, N_i^+, Y_i)$ with nilpositive elements
\begin{equation}
  N_1^+ =
  \left(
  \begin{array}{cccccc}
    0 & -3 & -1 & 0 & 0 & 0 \\
    -\frac{17}{9} & 2 & \frac{2}{9} & -\frac{4}{9} & 0 & 0 \\
    0 & 0 & 0 & 0 & 0 & 0 \\
    -\frac{17}{2} & \frac{9}{2} & \frac{3}{4} & -2 & 0 & -3 \\
    -\frac{17}{6} & \frac{7}{4} & \frac{1}{3} & -\frac{2}{3} & 0 & -1 \\
    \frac{391}{36} & -\frac{23}{2} & -\frac{23}{18} & \frac{23}{9} & 0 & 0 \\
  \end{array}
  \right),\qquad
  N_2^+ =
  \left(
  \begin{array}{cccccc}
    0 & 0 & 0 & 0 & 0 & 0 \\
    -\frac{1}{12} & 0 & -\frac{1}{6} & \frac{1}{3} & -1 & 0 \\
    \frac{1}{4} & 0 & \frac{1}{2} & -1 & 3 & 0 \\
    -\frac{1}{4} & 0 & -\frac{1}{2} & 1 & -3 & 0 \\
    -\frac{1}{8} & 0 & -\frac{1}{4} & \frac{1}{2} & -\frac{3}{2} & 0 \\
    \frac{5}{48} & 0 & \frac{5}{24} & -\frac{5}{12} & \frac{5}{4} & 0 \\
  \end{array}
  \right).
\end{equation}

This completes our computation of the commuting $\slt$-pair for the degeneration from $\rIV_1$ to $\rIV_2$.

\section{Deriving the polarised relations}
\label{sec:derpol}
In this section we summarise the definition of polarised relation proposed in \cite{Kerr2017} and exemplify the derivation of the relation $\rIII_c \to \rIV_{\hat{d}}$ in table \ref{sing_enhancements}.

For the ease of notation, we follow \cite{Kerr2017} to consider an entire Hodge-Deligne diamond at once. Given a Hodge-Deligne diamond consisting of Hodge-Deligne numbers $\{i^{p, q}\}$, we can define an integer-valued function $\HD(p, q) := i^{p, q}$ on the lattice $\bbZ \times \bbZ$. On the other hand, we define a Hodge-Deligne diamond of a variation of weight-$w$ Hodge structure polarised by $N$ with Hodge numbers $(h^{w, 0}, h^{w - 1, 1}, \ldots, h^{0, w})$ to be any integer-valued function $\HD(p, q)$ on the lattice $\bbZ \times \bbZ$ such that
\begin{equation}
  \sum^{w}_{q = 0} \HD(p, q) = h^{p, w - p}, \textrm{ for all } p,
\end{equation}
and satisfying the usual symmetry properties
\begin{align}
  \HD(p, q) = \HD(q, p) = \HD(w - q, w - p), & \text{ for all } p, q,\\
  \HD(p - 1, q - 1) \le \HD(p, q),           & \text{ for } p + q \le w.
\end{align}

In this fashion the sum $\HD = \HD_1 + \HD_2$ of two Hodge-Deligne diamonds $\HD_1$ and $\HD_2$ is naturally defined pointwise
\begin{equation}
  \HD(p, q) := \HD_1(p, q) + \HD_2(p, q).
\end{equation}
And also the shifted Hodge-Deligne diamond $\HD[a]$ of $\HD$ is defined to be
\begin{equation}
  \HD[a](p, q) = \HD(p + a, q + q).
\end{equation}

Now it comes to the enhancement relation \cite{Kerr2017}. Let $(F_1, N_1)$ and $(F_2, N_2)$ be two nilpotent orbits 
with limiting mixed Hodge structures $(F_1, W(N_1))$ and $(F_2, W(N_2))$. Denote $\HD(F_1, N_1)$ and $\HD(F_2, N_2)$ respectively their Hodge-Deligne diamonds. Considering a possible degeneration relation $(F_1, N_1) \to (F_2, N_2)$ there is the following equivalent condition:

Every primitive subspace $P^k(N_1)$ ($3 \le k \le 6$) of $(F_1, W(N_1))$ carries a weight-$k$ Hodge structure $P^k(N_1) = \bigoplus_{p + q = k} P^{p, q}(N_1)$. Denote its Hodge numbers by $j^{p, q}_1 := \dim_\bbC P^{p, q}(N_1)$. Let $\HD(F'_k, N'_k)$ be a Hodge-Deligne diamond of the variation of weight-$k$ Hodge structure polarised by $S(\cdot, N_1^k \cdot)$ on $P^k(N_1)$ with Hodge numbers $(j^{k, 0}, j^{k - 1, 1}, \ldots, j^{0, k})$ where $S$ is the polarisation bilinear form \eqref{def-S}. If one can decompose $\HD(F_2, N_2)$ as
\begin{equation}
  \HD(F_2, N_2) = \sum_{\substack{3 \le k \le 6\\0 \le a \le k - 3}} \HD(F'_k, N'_k)[a],
\end{equation}
where $\HD(F'_k, N'_k)[a]$ is the shifted Hodge-Deligne diamond defined above, then the degeneration relation
\begin{equation}
  (F_1, N_1) \to (F_2, N_2)
\end{equation}
holds. The converse is also true.

We refer the reader to \cite{Kerr2017} for details.

We exemplify the above condition on the relation $\rIII_c \to \rIV_{\hat{d}}$. Firstly we list the primitive Hodge-Deligne sub-diamond of $\rIII_c$ containing only primitive Hodge-Deligne numbers $j^{p, q}_1$ and the Hodge-Deligne diamond of $\rIV_{\hat{d}}$:
\begin{center}
  \begin{tabular}{c c}
  \begin{tikzpicture}[scale=1,cm={cos(45),sin(45),-sin(45),cos(45),(15,0)}]
    \draw[step = 1, gray, ultra thin] (0, 0) grid (3, 3);

    \draw[fill] (1, 2) circle[radius=0.03] node[above=2pt] {$c' - 1$};
    \draw[fill] (2, 2) circle[radius=0.03] node[above=2pt] {$c$};
    \draw[fill] (2, 1) circle[radius=0.03] node[above=2pt] {$c' - 1$};
    \draw[fill] (2, 3) circle[radius=0.03];
    \draw[fill] (3, 2) circle[radius=0.03];

    \draw[dashed, gray, ultra thin] (-0.5, 3.5) -- (3.5, -0.5) node[right]{$k = 3$};
    \draw[dashed, gray, ultra thin] (0.5, 3.5) -- (3.5, 0.5) node[right]{$k = 4$};
    \draw[dashed, gray, ultra thin] (1.5, 3.5) -- (3.5, 1.5) node[right]{$k = 5$};
  \end{tikzpicture} &
  \begin{tikzpicture}[scale=1,cm={cos(45),sin(45),-sin(45),cos(45),(15,0)}]
    \draw[step = 1, gray, ultra thin] (0, 0) grid (3, 3);

    \draw[fill] (0, 0) circle[radius=0.03];
    \draw[fill] (1, 2) circle[radius=0.03] node[above=2pt] {$\hat{d'}$};
    \draw[fill] (1, 1) circle[radius=0.03] node[above=2pt] {$\hat{d}$};
    \draw[fill] (2, 2) circle[radius=0.03] node[above=2pt] {$\hat{d}$};
    \draw[fill] (2, 1) circle[radius=0.03] node[above=2pt] {$\hat{d'}$};
    \draw[fill] (3, 3) circle[radius=0.03];

    \draw[dashed, gray, ultra thin, opacity=0] (-0.5, 3.5) -- (3.5, -0.5);
    \draw[dashed, gray, ultra thin, opacity=0] (0.5, 3.5) -- (3.5, 0.5);
    \draw[dashed, gray, ultra thin, opacity=0] (1.5, 3.5) -- (3.5, 1.5);
  \end{tikzpicture}\\
  $\rIII_c^{\mathrm{prim}}$ & $\rIV_{\hat{d}}$
  \end{tabular}
\end{center}

For the relation $\rIII_c \to \rIV_{\hat{d}}$ to hold, we need to find three Hodge-Deligne diamonds with Hodge numbers $(0, c' - 1, c' - 1, 0)$, $(0, 0, c, 0, 0)$ and $(0, 0, 1, 1, 0, 0)$ that sums (with proper shifts) to $\rIV_{\hat{d}}$. The following are three such Hodge-Deligne diamonds

\begin{center}
\begin{tabular}{c c c}
  \begin{tikzpicture}[scale=0.7,cm={cos(45),sin(45),-sin(45),cos(45),(15,0)}]
    \draw[step = 1, gray, ultra thin] (0, 0) grid (3, 3);

    \draw[fill] (1, 2) circle[radius=0.03] node[left] {$c' - r - 1$};
    \draw[fill] (1, 1) circle[radius=0.03] node[below] {$r$};
    \draw[fill] (2, 2) circle[radius=0.03] node[above] {$r$};
    \draw[fill] (2, 1) circle[radius=0.03] node[right] {$c' - r - 1$};
  \end{tikzpicture} & 
  \begin{tikzpicture}[scale=0.7,cm={cos(45),sin(45),-sin(45),cos(45),(15,0)}]
    \draw[dotted, step = 1, gray] (0, 0) grid (4, 4);
    \draw[fill, white] (0, 0) rectangle (3, 3);
    \draw[step = 1, gray, ultra thin] (0, 0) grid (3, 3);

    \draw[fill] (2, 2) circle[radius=0.03] node[above=2pt] {$c$};
  \end{tikzpicture} &
  \begin{tikzpicture}[scale=0.7,cm={cos(45),sin(45),-sin(45),cos(45),(15,0)}]
    \draw[dotted, step = 1, gray] (0, 0) grid (5, 5);
    \draw[fill, white] (0, 0) rectangle (3, 3);
    \draw[step = 1, gray, ultra thin] (0, 0) grid (3, 3);

    \draw[fill] (2, 2) circle[radius=0.03];
    \draw[fill] (3, 3) circle[radius=0.03];
  \end{tikzpicture}
  \\
  $\HD(F'_3, N'_3)$ with $r \ge 0$ & $\HD(F'_4, N'_4)$ & $\HD(F'_5, N'_5)$\\
  for $(0, c' - 1, c' - 1, 0)$. & for $(0, 0, c, 0, 0)$. & for $(0, 0, 1, 1, 0, 0)$.
\end{tabular}
\end{center}

Then we consider the sum
\begin{equation}
   \HD(F'_3, N'_3) + \HD(F'_4, N'_4) + \HD(F'_4, N'_4)[1] + \HD(F'_5, N'_5) + \HD(F'_5, N'_5)[1] + \HD(F'_5, N'_5)[2]
\end{equation}
which can be depicted as
\begin{center}
\begin{tikzpicture}[scale=1,cm={cos(45),sin(45),-sin(45),cos(45),(15,0)}]
  \draw[step = 1, gray, ultra thin] (0, 0) grid (3, 3);

  \draw[fill] (0, 0) circle[radius=0.03];
  \draw[fill] (1, 2) circle[radius=0.03] node[left=2pt] {$c' - r - 1$};
  \draw[fill] (1, 1) circle[radius=0.03] node[below=2pt] {$c + r + 2$};
  \draw[fill] (2, 2) circle[radius=0.03] node[above=2pt] {$c + r + 2$};
  \draw[fill] (2, 1) circle[radius=0.03] node[right=2pt] {$c' - r - 1$};
  \draw[fill] (3, 3) circle[radius=0.03];
\end{tikzpicture}
\end{center}
and we expect that this diamond agrees with the Hodge-Deligne diamond of $\rIV_{\hat{d}}$. This is possible if the condition
\begin{equation}
  \hat{d} = c + r + 2,
\end{equation}
together with the usual $r \ge 0$, $0 \le c \le h^{2, 1} - 2$ and $1 \le \hat{d} \le h^{2, 1}$ are satisfied. Hence we have derived the condition $c + 2 \le \hat{d}$ for the relation $\rIII_c \to \rIV_{\hat{d}}$ in table \ref{sing_enhancements}.


\bibliographystyle{JHEP}
\bibliography{refs-swamp.bib}

\providecommand{\href}[2]{#2}\begingroup\raggedright\begin{thebibliography}{10}

\bibitem{Ooguri:2006in}
H.~Ooguri and C.~Vafa, {\it {On the Geometry of the String Landscape and the
  Swampland}},  {\em Nucl. Phys.} {\bf B766} (2007) 21--33,
  [\href{http://arxiv.org/abs/hep-th/0605264}{{\tt hep-th/0605264}}].

\bibitem{Cecotti:2015wqa}
S.~Cecotti, {\em {Supersymmetric Field Theories}}.
\newblock Cambridge University Press, 2015.

\bibitem{Baume:2016psm}
F.~Baume and E.~Palti, {\it {Backreacted Axion Field Ranges in String Theory}},
   {\em JHEP} {\bf 08} (2016) 043, [\href{http://arxiv.org/abs/1602.06517}{{\tt
  arXiv:1602.06517}}].

\bibitem{Klaewer:2016kiy}
D.~Klaewer and E.~Palti, {\it {Super-Planckian Spatial Field Variations and
  Quantum Gravity}},  \href{http://arxiv.org/abs/1610.00010}{{\tt
  arXiv:1610.00010}}.

\bibitem{Palti:2015xra}
E.~Palti, {\it {On Natural Inflation and Moduli Stabilisation in String
  Theory}},  {\em JHEP} {\bf 10} (2015) 188,
  [\href{http://arxiv.org/abs/1508.00009}{{\tt arXiv:1508.00009}}].

\bibitem{Valenzuela:2016yny}
I.~Valenzuela, {\it {Backreaction Issues in Axion Monodromy and Minkowski
  $4$-forms}},  {\em JHEP} {\bf 06} (2017) 098,
  [\href{http://arxiv.org/abs/1611.00394}{{\tt arXiv:1611.00394}}].

\bibitem{Blumenhagen:2017cxt}
R.~Blumenhagen, I.~Valenzuela, and F.~Wolf, {\it {The Swampland Conjecture and
  F-term Axion Monodromy Inflation}},  {\em JHEP} {\bf 07} (2017) 145,
  [\href{http://arxiv.org/abs/1703.05776}{{\tt arXiv:1703.05776}}].

\bibitem{Palti:2017elp}
E.~Palti, {\it {The Weak Gravity Conjecture and Scalar Fields}},  {\em JHEP}
  {\bf 08} (2017) 034, [\href{http://arxiv.org/abs/1705.04328}{{\tt
  arXiv:1705.04328}}].

\bibitem{Lust:2017wrl}
D.~Lust and E.~Palti, {\it {Scalar Fields, Hierarchical UV/IR Mixing and The
  Weak Gravity Conjecture}},  {\em JHEP} {\bf 02} (2018) 040,
  [\href{http://arxiv.org/abs/1709.01790}{{\tt arXiv:1709.01790}}].

\bibitem{Hebecker:2017lxm}
A.~Hebecker, P.~Henkenjohann, and L.~T. Witkowski, {\it {Flat Monodromies and a
  Moduli Space Size Conjecture}},  {\em JHEP} {\bf 12} (2017) 033,
  [\href{http://arxiv.org/abs/1708.06761}{{\tt arXiv:1708.06761}}].

\bibitem{Cicoli:2018tcq}
M.~Cicoli, D.~Ciupke, C.~Mayrhofer, and P.~Shukla, {\it {A Geometrical Upper
  Bound on the Inflaton Range}},  \href{http://arxiv.org/abs/1801.05434}{{\tt
  arXiv:1801.05434}}.

\bibitem{Grimm:2018ohb}
T.~W. Grimm, E.~Palti, and I.~Valenzuela, {\it {Infinite Distances in Field
  Space and Massless Towers of States}},  {\em JHEP} {\bf 08} (2018) 143,
  [\href{http://arxiv.org/abs/1802.08264}{{\tt arXiv:1802.08264}}].

\bibitem{Heidenreich:2018kpg}
B.~Heidenreich, M.~Reece, and T.~Rudelius, {\it {Emergence of Weak Coupling at
  Large Distance in Quantum Gravity}},  {\em Phys. Rev. Lett.} {\bf 121}
  (2018), no.~5 051601, [\href{http://arxiv.org/abs/1802.08698}{{\tt
  arXiv:1802.08698}}].

\bibitem{Blumenhagen:2018nts}
R.~Blumenhagen, D.~Kläwer, L.~Schlechter, and F.~Wolf, {\it {The Refined
  Swampland Distance Conjecture in Calabi-Yau Moduli Spaces}},  {\em JHEP} {\bf
  06} (2018) 052, [\href{http://arxiv.org/abs/1803.04989}{{\tt
  arXiv:1803.04989}}].

\bibitem{Landete:2018kqf}
A.~Landete and G.~Shiu, {\it {Mass Hierarchies and Dynamical Field Range}},
  {\em Phys. Rev.} {\bf D98} (2018), no.~6 066012,
  [\href{http://arxiv.org/abs/1806.01874}{{\tt arXiv:1806.01874}}].

\bibitem{Lee:2018urn}
S.-J. Lee, W.~Lerche, and T.~Weigand, {\it {Tensionless Strings and the Weak
  Gravity Conjecture}},  \href{http://arxiv.org/abs/1808.05958}{{\tt
  arXiv:1808.05958}}.

\bibitem{Lee:2018spm}
S.-J. Lee, W.~Lerche, and T.~Weigand, {\it {A Stringy Test of the Scalar Weak
  Gravity Conjecture}},  \href{http://arxiv.org/abs/1810.05169}{{\tt
  arXiv:1810.05169}}.

\bibitem{Blumenhagen:2018hsh}
R.~Blumenhagen, {\it {Large Field Inflation/Quintessence and the Refined
  Swampland Distance Conjecture}},  {\em PoS} {\bf CORFU2017} (2018) 175,
  [\href{http://arxiv.org/abs/1804.10504}{{\tt arXiv:1804.10504}}].

\bibitem{Ooguri:2018wrx}
H.~Ooguri, E.~Palti, G.~Shiu, and C.~Vafa, {\it {Distance and de Sitter
  Conjectures on the Swampland}},  \href{http://arxiv.org/abs/1810.05506}{{\tt
  arXiv:1810.05506}}.

\bibitem{Obied:2018sgi}
G.~Obied, H.~Ooguri, L.~Spodyneiko, and C.~Vafa, {\it {De Sitter Space and the
  Swampland}},  \href{http://arxiv.org/abs/1806.08362}{{\tt arXiv:1806.08362}}.

\bibitem{Garg:2018reu}
S.~K. Garg and C.~Krishnan, {\it {Bounds on Slow Roll and the de Sitter
  Swampland}},  \href{http://arxiv.org/abs/1807.05193}{{\tt 1807.05193}}.

\bibitem{Candelas:1993dm}
P.~Candelas, X.~De~La~Ossa, A.~Font, S.~H. Katz, and D.~R. Morrison, {\it
  {Mirror Symmetry for Two Parameter Models. 1}},  {\em Nucl. Phys.} {\bf B416}
  (1994) 481--538, [\href{http://arxiv.org/abs/hep-th/9308083}{{\tt
  hep-th/9308083}}]. [AMS/IP Stud. Adv. Math.1,483(1996)].

\bibitem{Candelas:1994hw}
P.~Candelas, A.~Font, S.~H. Katz, and D.~R. Morrison, {\it {Mirror Symmetry for
  Two Parameter Models. 2}},  {\em Nucl. Phys.} {\bf B429} (1994) 626--674,
  [\href{http://arxiv.org/abs/hep-th/9403187}{{\tt hep-th/9403187}}].

\bibitem{CKS}
E.~Cattani, A.~Kaplan, and W.~Schmid, {\it {Degeneration of Hodge Structures}},
   {\em Ann. of Math.} {\bf 123} (1986), no.~3 457--535.

\bibitem{Kerr2017}
M.~Kerr, G.~Pearlstein, and C.~Robles, {\it {Polarized Relations on Horizontal
  $\mathrm{SL}(2)$s}},  \href{http://arxiv.org/abs/1705.03117}{{\tt
  arXiv:1705.03117}}.

\bibitem{Andrianopoli:1996cm}
L.~Andrianopoli, M.~Bertolini, A.~Ceresole, R.~D'Auria, S.~Ferrara, P.~Fre, and
  T.~Magri, {\it {$N=2$ Supergravity and $N=2$ Super Yang-Mills Theory on
  General Scalar Manifolds: Symplectic Covariance, Gaugings and the Momentum
  Map}},  {\em J. Geom. Phys.} {\bf 23} (1997) 111--189,
  [\href{http://arxiv.org/abs/hep-th/9605032}{{\tt hep-th/9605032}}].

\bibitem{Craps:1997gp}
B.~Craps, F.~Roose, W.~Troost, and A.~Van~Proeyen, {\it {What is Special
  K\"ahler Geometry?}},  {\em Nucl. Phys.} {\bf B503} (1997) 565--613,
  [\href{http://arxiv.org/abs/hep-th/9703082}{{\tt hep-th/9703082}}].

\bibitem{Alim:2012gq}
M.~Alim, {\it {Lectures on Mirror Symmetry and Topological String Theory}},
  \href{http://arxiv.org/abs/1207.0496}{{\tt arXiv:1207.0496}}.

\bibitem{Viehweg}
E.~Viehweg, {\em Quasi-projective Moduli for Polarized Manifolds}.
\newblock Ergebnisse der Mathematik und ihrer Grenzgebiete. 3. Folge / A Series
  of Modern Surveys in Mathematics. Springer Berlin Heidelberg, 1995.

\bibitem{Hironaka}
H.~Hironaka, {\it {Resolution of Singularities of an Algebraic Variety Over a
  Field of Characteristic Zero: I}},  {\em Ann. of Math.} {\bf 79} (1964),
  no.~1 109--203.

\bibitem{landman}
A.~Landman, {\it {On the Picard-Lefschetz Transformation for Algebraic
  Manifolds Acquiring General Singularities}},  {\em Trans. Amer. Math. Soc.}
  {\bf 181} (1973) 89--126.

\bibitem{Schmid}
W.~Schmid, {\it {Variation of Hodge Structure: The Singularities of the Period
  Mapping}},  {\em Invent. Math.} {\bf 22} (1973) 211--320.

\bibitem{CattaniKaplan}
E.~Cattani and A.~Kaplan, {\it {Polarized Mixed Hodge Structures and the Local
  Monodromy of a Variation of Hodge Structure}},  {\em Invent. Math.} {\bf 67}
  (1982) 101--116.

\bibitem{Kashiwara}
M.~Kashiwara, {\it {The Asymptotic Behavior of a Variation of Polarized Hodge
  Structure}},  {\em Publ. RIMS, Kyoto Univ.} {\bf 21} (1985), no.~4 853--875.

\bibitem{Djokovic1982}
D.~\v{Z}. Djokovi\'{c}, {\it {Closures of Conjugacy Classes in Classical Real
  Linear Lie Groups. II}},  {\em Trans. Amer. Math. Soc.} {\bf 270} (1982),
  no.~1 217--252.

\bibitem{Collingwood1993}
D.~H. Collingwood and W.~M. McGovern, {\em {Nilpotent Orbits in Semisimple Lie
  Algebras}}.
\newblock Van Nostrand Reinhold, 1993.

\bibitem{Wang1997}
C.-L. Wang, {\it {On the Incompleteness of the Weil-Petersson Metric along
  Degenerations of Calabi-Yau Manifolds}},  {\em Math. Res. Lett.} {\bf 4}
  (1997), no.~1 157--171.

\bibitem{Wang2015}
C.-L. Wang, {\it {Aspects on Calabi-Yau Moduli}},  in {\em Proceeding of
  Uniformizations, Riemann--Hilbert correspondence, Calabi--Yau manifolds, and
  Picard--Fuchs equations}, Mittag--Leffler Institute, 2015.

\bibitem{Morrison1993}
D.~R. Morrison, {\it {Compactifications of Moduli Spaces Inspired by Mirror
  Symmetry}},  {\em Journ{\'{e}}es de G{\'{e}}om{\'{e}}trie Alg{\'{e}}brique
  d'Orsay (Juillet 1992), Ast{\'{e}}risque} {\bf 218} (Apr, 1993) 243--271,
  [\href{http://arxiv.org/abs/alg-geom/9304007}{{\tt alg-geom/9304007}}].

\bibitem{Candelas1991a}
P.~Candelas, X.~C. {De La Ossa}, P.~S. Green, and L.~Parkes, {\it {A Pair of
  Calabi-Yau Manifolds as an Exactly Soluble Superconformal Theory}},  {\em
  Nucl. Phys. B} {\bf 359} (Jul, 1991) 21--74.

\bibitem{Cota2018}
C.~F. Cota, A.~Klemm, and T.~Schimannek, {\it {Modular Amplitudes and
  Flux-Superpotentials on Elliptic Calabi-Yau Fourfolds}},  {\em JHEP} {\bf
  2018} (Jan, 2018) 86, [\href{http://arxiv.org/abs/1709.02820}{{\tt
  arXiv:1709.02820}}].

\bibitem{Gerhardus2016}
A.~Gerhardus and H.~Jockers, {\it {Quantum Periods of Calabi–Yau Fourfolds}},
   {\em Nucl. Phys. B} {\bf 913} (Dec, 2016) 425--474,
  [\href{http://arxiv.org/abs/1604.05325}{{\tt arXiv:1604.05325}}].

\bibitem{Bloch:2016izu}
S.~Bloch, M.~Kerr, and P.~Vanhove, {\it {Local Mirror Symmetry and the Sunset
  Feynman Integral}},  {\em Adv. Theor. Math. Phys.} {\bf 21} (2017)
  1373--1453, [\href{http://arxiv.org/abs/1601.08181}{{\tt arXiv:1601.08181}}].

\bibitem{Iritani2011}
H.~Iritani, {\it {Quantum Cohomology and Periods}},  {\em Ann. l'Institut
  Fourier} {\bf 61} (2011) 2909--2958,
  [\href{http://arxiv.org/abs/1101.4512}{{\tt arXiv:1101.4512}}].

\bibitem{Katzarkov2008}
L.~Katzarkov, M.~Kontsevich, and T.~Pantev, {\it {Hodge Theoretic Aspects of
  Mirror Symmetry}},  \href{http://arxiv.org/abs/0806.0107}{{\tt
  arXiv:0806.0107}}.

\bibitem{Wall1966}
C.~T.~C. Wall, {\it {Classification Problems in Differential Topology. V On
  Certain 6-Manifolds}},  {\em Invent. Math.} {\bf 1} (1966) 355--374.

\bibitem{Candelas1991}
P.~Candelas and X.~C. de~la Ossa, {\it {Moduli Space of Calabi-Yau Manifolds}},
   {\em Nucl. Phys. B} {\bf 355} (May, 1991) 455--481.

\bibitem{Herraez:2018vae}
A.~Herraez, L.~E. Ibanez, F.~Marchesano, and G.~Zoccarato, {\it {The Type IIA
  Flux Potential, $4$-forms and Freed-Witten Anomalies}},  {\em JHEP} {\bf 09}
  (2018) 018, [\href{http://arxiv.org/abs/1802.05771}{{\tt arXiv:1802.05771}}].

\bibitem{CorvilainGrimmValenzuela}
P.~Corvilain, T.~W. Grimm, and I.~Valenzuela, to appear.

\bibitem{MixedSl2}
K.~{Kato}, C.~{Nakayama}, and S.~{Usui}, {\it {$\mathrm{SL}(2)$-Orbit Theorem
  for Degeneration of Mixed Hodge Structure}},  {\em {J. Algebr. Geom.}} {\bf
  17} (2008), no.~3 401--479.

\end{thebibliography}\endgroup
\end{document}